\documentstyle[12pt]{report}
\newcommand{\D}{\displaystyle}
\newcommand{\T}{\textstyle}

\begin{document}
\title{
Light Front Quantization}
\author{
Matthias Burkardt\\
Institute for Nuclear Theory\\
University of Washington, Box 351550\\
Seattle, WA 98195-1550
\\[4.ex]
to appear in:\\[2.ex]
Advances in Nuclear Physics
}
\date{}
\maketitle
\begin{abstract}
An introductory overview on Light-Front quantization, with
some emphasis on recent achievements, is given.
Light-Front quantization is the most promising and physical tool
to study deep inelastic scattering on the basis of quark gluon
degrees of freedom. The simplified vacuum structure (nontrivial
vacuum effects can only appear in zero-mode degrees of freedom)
and the physical basis allows for a description of hadrons that
stays close to intuition. Recent progress has ben made in
understanding the connection between effective LF Hamiltonians
and nontrivial vacuum condesates. Discrete Light-Cone
Quantization, the transverse lattice and Light-Front Tamm-Dancoff
(in combination with renormalization group techniques) are the
main tools for exploring LF-Hamiltonians nonperturbatively.
\end{abstract}

\tableofcontents
\chapter{Introduction}
\section{Deep Inelastic Scattering}
More than twenty years after quantum chromodynamics (QCD)
was introduced as a microscopic theory of strong interactions, very little
is known about its solutions. At least in principle, it should
be possible to describe the interaction of nucleons with
external probes using quark and gluon degrees of freedom
on the basis of QCD.
So far, however, the extreme complexity of this theory has
slowed any progress in this direction considerably.

Deep inelastic lepton-nucleon scattering (DIS) provides
access to quark and gluon degrees of freedom in nucleons
and nuclei. In these experiments one shoots high energy
leptons (e.g. electrons) at a hadronic target
(usually protons or nuclei) and measures the energy and
momentum transfer to the target by detecting the
final state lepton (Fig. \ref{fig:dis}).
\begin{figure}
\unitlength1.cm
\begin{picture}(14,6)(-1.5,-8.)
\includegraphics{dis.ps}
\end{picture}
\caption{Inclusive process $e^-+N \rightarrow e^{-\prime}+X$,
where $X$ is an unidentified hadronic state.
}
\label{fig:dis}
\end{figure}
The hadronic final state $X$ is not measured
(usually the nucleon is destroyed in these reactions and the
hadronic final state consists of many particles).
Because of the extremely large momentum transfer to the
target (typical momentum transfers in DIS experiments are
several $GeV/c$ or more),
the inclusive cross sections are dominated by single
particle response functions along the light-cone.
To illustrate this let us use the optical theorem which
relates the differential lepton nucleon cross section
to the imaginary part of the forward Compton amplitude
\cite{yn:qcd} (Fig.\ref{fig:compt}).
One finds
\begin{equation}
\frac{d^2\sigma}{d\Omega dE^\prime}
= \frac{\alpha^2}{q^4} \left( \frac{E^\prime}{E}
\right) l^{\mu \nu}
\frac{\Im T_{\mu \nu}}{2\pi}
\end{equation}
where $E, E^\prime$ are the energies of the initial and final lepton.
\mbox{$q=k-k^\prime$}
is the four momentum transfer of the lepton on the
target and \mbox{$l_{\mu\nu}=2k_\mu k^\prime_\nu + 2k_\nu k^\prime_\mu +
q^2g_{\mu \nu}$} is the leptonic tensor. The hadronic
tensor
\begin{equation}
T_{\mu \nu}(P,q) = \frac{i}{2M_N} \sum_S\int \frac{d^4x}{2\pi}
e^{iq\cdot x}
\langle P,S |T\left(J_\mu(x) J_\nu(0)\right)|P,S \rangle
\label{eq:htensor}
\end{equation}
($S$ is the spin of the target proton)
contains all the information about the parton substructure
of the target proton.

In the Bjorken limit ($Q^2 \equiv -q^2 \rightarrow \infty$,
$P\cdot q \rightarrow \infty$, $x_{Bj} = Q^2/2P\cdot q$ fixed),
deep inelastic structure functions exhibit Bjorken scaling:
up to kinematical coefficients, the hadronic tensor
(\ref{eq:htensor}) depends only on $x_{Bj}$ but no longer
on $Q^2$ (within perturbatively calculable logarithmic
corrections). In order to understand this result, it is
convenient to introduce {\it light-front} variables
$a_\mp = a^\pm = \left( a^0 \pm a^3\right)/\sqrt{2}$ so that
the scalar product reads
$a \cdot b = a_+b^+ + a_- b^- -{\vec a}_\perp
{\vec b}_\perp =a_+b_- + a_- b_+ -{\vec a}_\perp
{\vec b}_\perp$. Furthermore let us choose a frame where
${\vec q}_\perp =0$. The Bjorken limit corresponds to
$p^\mu$ and $q_-$ fixed, while $q_+\rightarrow \infty$.
Bjorken scaling is equivalent to the statement that
the structure functions become independent of $q_+$ in this limit (again up
to trivial kinematic coefficients).
In this limit, the integrand in Eq.(\ref{eq:htensor}) contains the rapidly
oscillating factor
$\exp (iq_+x^+)$, which kills all contributions to the
integral except those where the integrand is singular
\cite{jaffe}. Due to causality, the integrand must vanish
for $x^2=2x^+x^- -{\vec x}^2_\perp <0$ and the current
product is singular at $x^+=0$, ${\vec x}_\perp =0$.
The leading singularity can be obtained from the
operator product expansion by contracting two fermion
operators in the product
$T\left( J_\mu(x) J_\nu(0) \right) \equiv
T\left( \bar{\psi}(x) \gamma_\mu \psi(x)
\bar{\psi}(0) \gamma_\nu \psi(0) \right)$, yielding
a nonlocal term bilinear in the fermion field multiplying
a free (asymptotic freedom!) fermion propagator from $0$ to $x$
which gives rise to the abovementioned singularity structure
\cite{ch:gau}.
The $x^+={\vec x}_\perp =0$ dominance in the integral has two consequences.
First it explains Bjorken scaling, because $q_+$ enters the
hadronic tensor only via the term $x^+q_+$ in the exponent
and for $x^+=0$ the $q_+$ dependence drops out. Second,
and this is very important for practical calculations,
the parton distributions, i.e. the Bjorken scaled
structure functions, can be expressed in terms
of correlation functions along the {\it light-front}
space direction $x^-$.
\begin{figure}
\unitlength1.cm
\begin{picture}(15,7)(1.7,-11.5)
\includegraphics{compt.ps}
\end{picture}
\caption{Inclusive lepton nucleon cross section expressed
in terms of the imaginary part of the forward Compton amplitude.
For $Q^2=-q^2\rightarrow \infty$ only the `handbag diagram'
(both photons couple to the same quark) survives. The `crossed
diagram' (the two photons couple to different quarks) is
suppressed because of wavefunction effects.
}
\label{fig:compt}
\end{figure}
For example, for the spin
averaged parton distribution one obtains
\begin{equation}
2P_-f(x_{Bj}) = \int \frac{dx^-}{2\pi} \langle P |
\bar{\psi}(0)\gamma_- \psi(x^-)|P \rangle \exp (iP_-x^- x_{Bj}),
\label{eq:parton}
\end{equation}

The physical origin of this result can be understood as
follows. Consider again the virtual forward Compton
amplitude (Fig. \ref{fig:compt}).
In principle, the photons in the first and second interaction
in Fig. \ref{fig:compt} can couple to the same as well as to different
quarks in the target. However, the hadronic wavefunction can
only absorb momenta which are of the order of the
QCD-scale ($\Lambda_{QCD}\approx 200 MeV$).
Therefore, in the limit of large momentum transfer, only
such diagrams survive where the two photons in Fig. \ref{fig:compt}
couple to the same quark. All other diagrams have
large momenta flowing through the wavefunction or they
involve extra hard gluon exchanges which results in
their suppression at large $Q^2$.
The large momentum transfer is also important because of
asymptotic freedom. Since $\alpha_S(Q^2)\sim 1/\log
\left(Q^2/\Lambda_{QCD}^2\right)$,
the running coupling constant of QCD,
goes to zero for large $Q^2$, all interactions of
the struck quark can be neglected and it propagates
essentially without interaction
between the two
photon-vertices. Furthermore, since the momentum
transfer is much larger than the masses of the quarks
in the target, the struck quarks propagation between
becomes ultra-relativistic, i.e. it moves
exceedingly close to the light cone $x^2=0$.
Due to the high-energy nature of the scattering, the
relativistic structure function is a LF correlation
\cite{rj:70,ji:com}. Already at this point it should
be clear that LF-coordinates play a distinguished
role in the analysis of DIS experiments ---
a point which will become much more obvious after
we have introduced some of the formal ideas of
LF quantization.

\section{Advantages of Light-Front Coordinates}
LF quantization is very similar to canonical equal
time (ET) quantization \cite{di:49} (here we closely follow Ref.
\cite{kent}).
Both are Hamiltonian formulations of
field theory, where one specifies the fields on a
particular initial surface. The evolution of the fields
off the initial surface is determined by the
Lagrangian equations of motion. The main difference
is the choice of the initial surface, $x^0=0$ for
ET and $x^+=0$ for the LF respectively.
In both frameworks states are expanded in terms of fields
(and their derivatives) on this surface. Therefore,
the same physical state may have very different
wavefunctions\footnote{By ``wavefunction'' we mean here
the collection of all Fock space amplitudes.}
in the ET and LF approaches because fields at $x^0=0$
provide a different basis for expanding a state than
fields at $x^+=0$. The reason is that the microscopic
degrees of freedom --- field amplitudes at $x^0=0$
versus field amplitudes at $x^+=0$ --- are in general
quite different from each other in the two formalisms.

This has important consequences for the practical
calculation of parton distributions (\ref{eq:parton})
which are real time response functions in the equal
time formalism.
\footnote{The arguments of $\bar{\psi}$ and $\psi$ in
Eq.(\ref{eq:parton}) have different time components!}
In order to evaluate Eq.(\ref{eq:parton}) one needs to
know not only the ground state wavefunction of
the target, but also matrix elements to excited states.
In contrast, in the framework of LF quantization,
parton distributions are correlation functions at equal
LF-time $x^+$, i.e.
{\it within} the initial surface $x^+=0$ and can thus
be expressed directly in terms of ground state
wavefunctions (As a reminder: ET wavefunctions and
LF wavefunctions are in general different objects).
In the LF framework, parton distributions $f(x_{Bj}$) can
be easily calculated and have a very simple physical
interpretation as single particle momentum densities,
where $x_{Bj}$ measures the fraction of
momentum carried by the hadron
\footnote{In DIS with nonrelativistic kinematics
(e.g. thermal neutron scattering off liquid $^4$He)
one also observes scaling and the structure functions can be
expressed in terms of single particle response functions.
However, due to the different kinematics, nonrelativistic
structure functions at large momentum transfer
are dominated by Fourier transforms of
equal time response functions, i.e. ordinary momentum
distributions.}
\begin{equation}
x_{Bj} = \frac{p_-^{parton}}{P_-^{hadron}}.
\end{equation}

Although DIS is probably the most prominent example for
practical applications of LF coordinates, they prove useful in many other
places as well. For example,
LF coordinates have been used in the context current algebra
sum rules in particle physics \cite{fu:inf}.
Another prominent example is form factors, where moments
of the wave function along the LF determine the asymptotic
falloff at large momentum transfer \cite{br:lep}.
More recently, LF quantization found applications
in inclusive decays of heavy quarks \cite{bj:hq,mb:hq,wz:hq}.

{}From the purely theoretical point of view, various advantages
of LF quantization derive from properties of the ten generators
of the Poincar\'e group (translations $P^\mu$,
rotations ${\vec L}$ and boosts ${\vec K}$) \cite{di:49,kent}.
Those generators which leave the initial surface
invariant (${\vec P}$ and ${\vec L}$ for ET and
$P_-$, ${\vec P}_\perp$, $L_3$ and ${\vec K}$ for LF)
are ``simple'' in the sense that they have very simple
representations in terms of the fields (typically just
sums of single particle operators). The other generators, which include
the ``Hamiltonians'' ($P_0$, which is conjugate
to $x^0$ in ET and $P_+$, which is conjugate to the LF-time
$x^+$ in LF quantization) contain interactions among the
fields and are typically very complicated.
Generators which leave the initial surface invariant are also
called {\it kinematic} generators, while the others are called
{\it dynamic} generators. Obviously it is advantageous to have as
many of the ten generators kinematic as possible. There are
seven kinematic generators on the LF but only six in ET quantization.

The fact that $P_-$, the generator of $x^-$ translations, is
kinematic (obviously it leaves $x^+=0$ invariant!)
and positive has striking
consequences for the LF vacuum\cite{kent}. For free fields $p^2=m^2$ implies
for the LF energy $p_+ = \left(m^2 + {\vec p}_\perp \right)/2p_-$.
Hence positive energy excitations have positive $p_-$. After the
usual reinterpretation of the negative energy states this implies
that $p_-$ for a single particle is positive (which makes sense,
considering that $p_- =\left(p_0-p_3\right)/\sqrt{2}$).
$P_-$ being kinematic
means that it is given by the sum of single particle $p_-$.
Combined with the positivity of $p_-$ this implies that the
Fock vacuum (no particle excitations) is the unique state
with $P_-=0$. All other states have positive $P_-$.
Hence, even in the presence of interactions,
the LF Fock vacuum does not mix with any other state and is
therefore an exact eigenstate of the LF Hamiltonian $P_+$
(which commutes with $P_-$). If one further assumes parity
invariance of the ground state this implies that the Fock
vacuum must be the exact ground state of the fully interacting LF quantum
field theory.
\footnote{Practical calculations show that typical LF
Hamiltonians are either unbounded from below
or their ground state is indeed the Fock vacuum.}
In sharp contrast to other
formulations of field theory, the LF-vacuum is trivial!
This implies a tremendous technical advantage but also raises
the question whether nonperturbative LF-field theory is
equivalent to conventional field theory, where nonperturbative
effects usually result in a highly nontrivial vacuum structure.
This very deep issue will be discussed in more detail in Chapter
\ref{vac}.

Dirac was the first who had the idea to formulate field theory in
LF-coordinates \cite{di:49}.\footnote{Later, a similar framework
was developed independently on the basis of a Lorentz frame
(``the infinite momentum frame'') that moves with
$v \rightarrow c$ \cite{fu:inf,su:inf,we:69,ks:qed,bj:71,brs:73}.}
In this remarkable work
(almost 20 years before scaling was discovered in deep
inelastic lepton nucleon scattering !) he has shown that it should
in principle be possible to formulate a consistent quantum
theory on the LF. This work laid the basis for all further
developments, but left many details open. The main issues are
the structure of the vacuum, renormalization and practical
algorithms for solution.

\section{Outline}
There are many similarities between the formal steps
in ET quantization and LF quantization. In Chapter
\ref{canoni} we will explain the basic steps in constructing
LF Hamiltonians and give examples for scalar fields,
fermions and gauge fields.
The vacuum on the LF is very controversial. On the one hand
simple kinematical arguments seem to show that in LF field theory
the vacuum of
interacting field theories is the same as the free field
theory vacuum (all interactions turned off). In QCD we know
that chiral symmetry is spontaneously broken. It is up to now
unclear whether a LF Hamiltonian, with its trivial vacuum,
is capable of describing this physics. We will elaborate on
this point in Chapter \ref{vac}.
Renormalization is an issue because the LF-approach to field theory
is not manifestly covariant. Thus UV-divergences
(which occur on the LF as they do everywhere in quantum field theory)
are not necessarily the same for all Lorentz components of
a particular operator under consideration. Clearly this requires
a more complex counterterm structure to render the theory finite
and to restore Lorentz invariance for physical observables
(see Chapter \ref{ren}). Despite certain technical simplifications,
field theory on the LF is {\it a priori} still an enormously
complex many body problem. In particular in QCD one knows
from DIS experiments that the nucleon consists not only of
the three valence quarks, but that sea quark pairs and gluons
are a significant, if not dominant, component of the
nucleon's LF wavefunction, i.e. one should not expect that
the LF wavefunctions of ground state hadrons in QCD are simple.
Recent attempts to cast LF bound state problems into a form
that can be solved on a computer will be described in
Chapter \ref{num}.

\chapter{Canonical Quantization}
\label{canoni}
\section{Quantization in Light-Front Coordinates}
In this chapter, the formal steps
for quantization on the light-front are presented. For pedagogical
reasons this will be done by comparing with conventional quantization
(with $x^0$ as ``time''). On the one hand this shows that the basic
steps in the quantization procedure in LF and in ET
formalism are in fact very similar. More importantly, however,
we will thus be able to
highlight the essential differences between these two approaches
to quantum field theory more easily.

In the context of canonical quantization one usually starts from
the action
\begin{equation}
S = \int d^4x {\cal L}.
\end{equation}
(${\cal L}={\cal L}(\phi, \partial_\mu \phi)$)
After selecting a time direction
$\tau$ \footnote{Here $\tau$ may stand for ordinary time $x^0$ as well as
for LF time $x^+=\left(x^0+x^3\right)/\sqrt{2}$ or any other
(not space-like) direction.}
one forms the momenta which are canonically conjugate to
$\phi$
\begin{equation}
\Pi (x)= \frac{ \delta {\cal L}}{\delta \partial_\tau \phi}
\end{equation}
and postulates canonical commutation relations
between fields and corresponding momenta at equal ``time'' $\tau$
(Table \ref{tab:can}).
\footnote{The canonical quantization procedure in the ET
formulation can for example be found in Ref. \cite{bj:rel}.
The  rules for canonical
LF-quantization have been taken from Refs. \cite{ch:73,yan:sd2}.}
\begin{table}
\begin{tabular}{c|c}
normal coordinates & light-front \\[1.5ex]
\hline
\multicolumn{2}{c}{coordinates}\\[1.5ex]
$\begin{array}{ll}
x^0 & \mbox{time} \\ x^1,x^2,x^3 & \mbox{space}
\end{array} $ &
$ \begin{array}{ll}
x^+ = \frac{\T x^0+x^3}{\T \sqrt{2}} & \mbox{time} \\
x^- = \frac{\T x^0-x^3}{\T \sqrt{2}}, x^1, x^2  & \mbox{space}
\end{array} $ \\[1.5ex]
\multicolumn{2}{c}{scalar product}\\[1.5ex]
$a \cdot b = $ & $a \cdot b = $ \\
$a^0b^0-a^1b^1-a^2b^2-a^3b^3 $ & $a^+b^-+a^-b^+-a^1b^1-a^2b^2 $ \\
$= a^0b^0-{\vec a}{\vec b} $ & $= a^+b^-+a^-b^+-{\vec a}_\perp
{\vec b}_\perp $\\[1.5ex]
\multicolumn{2}{c}{Lagrangian density}\\[1.5ex]
${\cal L} = \frac{1}{2} \left(\partial_0 \phi\right)^2
-\frac{1}{2}\left(
\stackrel{\rightarrow}{\nabla} \phi \right)^2 -V(\phi)
$ & ${\cal L} = \partial_+\phi \partial_-\phi
-\frac{1}{2}\left(
\stackrel{\rightarrow}{\nabla}_\perp \phi \right)^2 -V(\phi) $ \\[1.5ex]
\multicolumn{2}{c}{conjugate momenta}\\[1.5ex]
$\pi = \frac{\T \delta{\cal L}}{\T\delta\partial_0\varphi} =
 \partial_0\varphi$ &
$\pi = \frac{\T \delta{\cal L}}{\T\delta\partial_+\varphi} =
 \partial_-\varphi$ \\[1.5ex]
\multicolumn{2}{c}{canonical commutation relations}\\[1.5ex]
$ [\pi(\vec{x},t),\varphi(\vec{y},t)] $ &
$ [\pi(x^-,x_{\perp},x^+), \varphi(y^-,y_{\perp},x^+)] $ \\
$ = -i\delta^3(\vec{x}-\vec{y}) $ &
$ = -\frac{i}{2} \delta(x^- -y^-)
\delta^2({\vec x}_{\perp}-{\vec y}_{\perp}) $
\\[1.5ex]
\multicolumn{2}{c}{Hamilton operator}\\[1.5ex]
$P^0 = {\D\int} d^3x\; \cal{H}(\varphi,\pi) $ &
$P_+ = {\D\int} dx^- {\D\int} d^2x_{\perp}\; {\cal H}(\varphi,\pi) $ \\
${\cal H} = \pi \partial_0 \varphi - {\cal L} $ &
${\cal H} = \pi\partial_+\varphi - {\cal L} $ \\[1.5ex]
\multicolumn{2}{c}{momentum operator}\\[1.5ex]
$\vec{P} = {\D\int} d^3x\; \pi \vec{\bigtriangledown}\varphi $ &
$P_- = {\D\int} dx^-d^2x_{\perp}\; \pi \partial_-\varphi $ \\
& ${\vec P}_\perp = {\D\int} dx^-d^2x_{\perp}\;
\pi {\vec \partial}_{\perp}\varphi $ \\
\multicolumn{2}{c}{eigenvalue equation}\\[1.5ex]
$P^0 |\psi_n\!> = E_n |\psi_n\!> $ &
$P_+ |\psi_n\!> = P_{+n} |\psi_n\!> $ \\[1.5ex]
$\vec{P}$ fixed & $P_-, {\vec P}_{\perp}$ fixed \\[1.5ex]
\multicolumn{2}{c}{hadron masses} \\[1.5ex]
$M_n^2 = E_n^2 - \vec{P}^2 $ &
$M_n^2 = 2 P_{+n}P_- - {\vec P}_{\perp}^2 $
\end{tabular}
\caption{canonical quantization in ordinary coordinates and on
the light-front}
\label{tab:can}
\end{table}

In the next step one constructs the Hamilton operator and the other
components of the momentum vector. Thus one has completely specified the
dynamics and can start solving the equations of motion.
Typically, one either makes some variational ansatz or a Fock space
expansion.
In the latter approach one writes the hadron wave function
as a sum over components with a fixed number of elementary quanta
(for example in QCD: $q\bar{q}$, $q\bar{q}q\bar{q}$, $q\bar{q}g$, e.t.c.).
The expansion coefficients, i.e. the wavefunctions for the corresponding
Fock space sector are used as variational parameters. They are determined
by making the expectation value of the energy stationary
with respect to variations in the wavefunction. Typically the variation is
done
for fixed momentum.\footnote{On
the LF this is very important because
$P_+ \propto 1/P_-$, i.e. unrestricted variation
($P_-$ allowed to vary) results in $P_-\rightarrow \infty$.}
This whole procedure results in coupled integral equations
for the Fock space components. In general they have to be solved
numerically. In practical calculations, since one cannot
include infinitely many Fock components, one has to introduce
some {\it ad hoc} cutoff in the Fock space. Thus it is very important to
demonstrate that physical observables do not depend on how many
Fock components are included.

Until one selects the canonically conjugate momenta and postulates equal
$\tau$ commutation relations,
i.e. at the level of the classical Lagrangian,
the transition from ET to the LF consists of a mere rewriting. After
quantization, the independent degrees
of freedom consist of the fields and their conjugate
momenta on the initial surface ($x^0=0$ for ET and
$x^+=0$ for LF). Thus different degrees of freedom are
employed to expand physical states in the ET
and in the LF approach. Of course, after solving the equations of motion,
physical observables must not depend on the choice of quantization plane.
However, it may turn out that one approach is more efficient
(e.g. faster numerical convergence) than the other or more elegant and
more easy to interpret physically. In general, this will of course
depend on the details of the interaction. An extreme example is
$\mbox{QCD}_{1+1}(N_C \rightarrow \infty)$. In the ET approach
\cite{ba:qcd,wi:vak,ne:nor} one first has to solve coupled, nonlinear
integral
equations with a singular kernel to obtain the Hartree-Fock
solution for the vacuum. Then, in order to calculate meson masses,
one has to solve the two body equation in this background, which
amounts to solving another set of coupled (linear) integral equations with
singular kernel.
In the LF-approach \cite{th:qcd,ei:str} all one has to
do is solve one linear integral equation with singular kernel. The numerical
results for the meson spectrum are in extremely good agreement between
the two approaches, but numerically the LF calculation is more than one
order of
magnitude faster! In this case the simplification arises because the
LF-vacuum
is trivial --- a point which will be elaborated in more detail below as well
as in Chapter \ref{vac}.

Which approach is preferable may, however, also depend on the observables in
which one is interested.
The most prominent example is deep inelastic scattering.
As discussed in the introduction, parton distributions are much more easily
accessible on the LF than in usual coordinates.

\section{$\varepsilon$-Coordinates on Finite Light-Front \mbox{Intervals}}
\label{eps}
One issue one may be worried about is the question of equivalence
between the LF approach to field theories and other approaches.
On the LF one imposes commutation relations at equal LF-time, i.e.
between two space-time points that are connected by a light-like
distance. Thus it is {\it a priori} not clear whether the initial
value problem with initial conditions on a null plane
is well defined \cite{ro:ini,smu} and whether there arise
any conflicts with causality on the LF. The situation becomes
particularly worrisome when one introduces a ``box'' in the
longitudinal $x^-$ direction (to keep IR-singularities
under control) and  imposes periodic or quasiperiodic
boundary conditions at the ends of the box --- i.e. one imposes
boundary conditions between points that may be causally related.
One way to address this issue in a well defined way
is to define the LF via a limiting procedure by starting
from a spacelike quantization surface and carefully rotating
this surface until one has `reached' the LF (note: although there
are some similarities, this should not be confused with
a Lorentz boost to infinite momentum \cite{we:69,brs:73}).

In order to be able to control infrared singularities, let
us formulate the dynamics on a finite LF interval with extension
$L$ in the $x^-$ direction.
\footnote{To simplify the notation, only 1+1 dimensional examples
will be discussed in this section.}
On a finite interval, boundary
conditions have to be specified, e.g.
$\phi(x^-+L,x^+) = \phi(x^-,x^+)$. However, if one is working
on the LF, imposing boundary conditions means relating
fields at points that are separated by a light-like distance ---
obviously one may run into trouble with causality at this point.
To avoid this dilemma, Lenz et al. \cite{le:ap}
have introduced $\varepsilon$-coordinates
which are defined as follows,
\footnote{See also Ref.\cite{fr:eps}.
A slightly different approach, where both $x^+$ and $x^-$
are rotated away from the light-cone, has been studied
in Ref.\cite{ho:vac}.}
\begin{eqnarray}
x^-_\varepsilon &=& x^- \nonumber\\
x^+_\varepsilon &=& x^+ + \frac{\varepsilon}{L} x^-.
\end{eqnarray}
Now points at opposite ends of the interval (with coordinates
$(x^-_\varepsilon +L,x^+_\varepsilon)$ and
$(x^-_\varepsilon,x^+_\varepsilon)$ are separated
by a spacelike distance $ds^2 = -2\varepsilon L$ and no conflict
with causality arises from imposing boundary conditions.
In $\varepsilon$-coordinates the scalar product is given by
\begin{equation}
A\cdot B = A_+B_- + A_- B_+ + 2 \frac{\varepsilon}{L} A_+B_+
\end{equation}
and thus the Lagrangian density (for the rest of this section,
the subscript $\varepsilon$ will be dropped to simplify the
notation) for $\phi^4_{1+1}$ reads
\begin{equation}
{\cal L} = \partial_+\phi\left(\partial_-\phi+\frac{\varepsilon}{L}
\partial_+\phi\right) - \frac{m^2}{2}\phi^2
-\frac{\lambda}{4!} \phi^4.
\label{eq:leps}
\end{equation}
Since ${\cal L}$
is quadratic in $\partial_+\phi$, quantization in
$\varepsilon$-coordinates is straightforward (as in usual
coordinates). One finds \cite{le:ap}
\begin{equation}
\Pi = \frac{\delta {\cal L}}{\delta \partial_+\phi}
= \partial_-\phi + \frac{2\varepsilon}{L} \partial_+\phi
\label{eq:momfi}
\end{equation}
\begin{equation}
\left[ \Pi(x^-,x^+), \phi(y^-,x^+) \right] = -i\delta(x^--y^-)
\label{eq:quaneps}
\end{equation}
and
\begin{equation}
H = \int dx^- \frac{L}{4\varepsilon}
\left( \Pi - \partial_-\phi\right)^2 + \frac{m^2}{2}\phi^2
+\frac{\lambda}{4!} \phi^4.
\label{eq:heps}
\end{equation}
In these coordinates, the free dispersion relation
($\lambda=0$) is given by
\begin{equation}
p_+(n) = \frac{L}{2\varepsilon} \left( -p_-(n)\pm \sqrt{p_-(n)^2 +
\frac{2\varepsilon}{L}m^2 }\right),
\label{eq:epsdisp}
\end{equation}
where $p_-(n)=\frac{2\pi n}{L}$ as usual in a box with periodic
boundary conditions.
Later, we will also need the normal mode expansion of
the fields with periodic boundary conditions
\begin{eqnarray}
\phi(x^-) &=& \sum_n \frac{1}{2\sqrt{\omega_n}}
\left[
a_n e^{-ip_-(n)x^-} + a_n^\dagger e^{ip_-(n)x^-}\right]
\nonumber\\
\Pi(x^-) &=& \sum_n \frac{-i\sqrt{\omega_n}}{L}
\left[
a_n e^{-ip_-(n)x^-} - a_n^\dagger e^{ip_-(n)x^-}\right],
\label{eq:nmode}
\end{eqnarray}
where
$\omega_n = L\sqrt{p_-(n)^2 + \frac{2\varepsilon}{L}m^2}$
and
the $a$, $a^\dagger$ satisfy the usual
commutation relations, e.g
\begin{equation}
\left[ a_m, a_n^\dagger \right] = \delta_{m,n}.
\end{equation}
The most significant difference between the dispersion relation
in $\varepsilon$-coordinates (\ref{eq:epsdisp}) and the
dispersion relation on the LF \mbox{($p_+ = \frac{m^2}{2p_-}$)}
is the appearance of two solutions of $p_+$ for each $p_-$
in $\varepsilon$-coordinates, while the dispersion relation
on the LF yields just one solution for each $p_-$
(Figure \ref{fig:disp}).
\begin{figure}
\unitlength1.cm
\begin{picture}(14,7)(-1.,1.5)
\includegraphics{disp.ps}
\end{picture}
\caption{Free dispersion relation in $\varepsilon$-coordinates
versus the dispersion relation in the LF limit.
}
\label{fig:disp}
\end{figure}
For positive energy ($p_+>0$) modes, the LF momentum $p_-$
is positive whereas the momentum $p_-$ in
$\varepsilon$-coordinates can be both positive and negative.
This has importance consequences for the vacuum structure which
will be discussed in Chapter \ref{vac}.

In the limit $\frac{\varepsilon}{L} \rightarrow 0$
($L$ fixed) the LF is recovered:
\begin{equation}
\Pi \stackrel{\frac{\varepsilon}{L}\rightarrow 0}{\longrightarrow}
\partial_- \phi
\end{equation}
\begin{equation}
H\stackrel{\frac{\varepsilon}{L}\rightarrow 0}{\longrightarrow}
\int dx^-
\frac{m^2}{2}\phi^2
+\frac{\lambda}{4!} \phi^4.
\end{equation}

For all nonzero $\varepsilon$, the relation between the
momenta and the fields (\ref{eq:momfi}) contains the time
derivative of the fields and the fields are quantized
as usual (\ref{eq:quaneps}). However, for $\varepsilon=0$,
Eq.(\ref{eq:momfi}) becomes a constraint equation, and the
Dirac-Bergmann algorithm (see Appendix \ref{dirac}) yields
$\left[ \partial_-\phi(x^-,x^+), \phi(y^-,x^+) \right] = \frac{i}{2}
\delta(x^--y^-)$.

It should be noted, that the order
of limits does matter, i.e. it is important
whether one takes the LF limit \mbox{($\frac{\varepsilon}{L}\rightarrow 0$)}
first or the continuum limit \mbox{($L\rightarrow \infty$)}.
This will be discussed in detail in Chapter \ref{vac}.

\section{Examples for Canonical Light-Front Hamiltonians}
\subsection{Scalar Fields}
\label{ex:scalar}
Self-interacting Scalar fields in the LF framework have been
discussed in Refs.\cite{ch:73,yan:sd2}.
In order to keep the discussion as general as possible, we will
work in $D_\perp$ transverse dimensions, where $D_\perp =0,1,2$.
For a polynomial interaction
\footnote{In $3+1$ dimensions, renormalizability restricts the
interaction to 4th order polynomials, but in $2+1$ or $1+1$
dimensions higher order polynomials are conceivable
(6th order and $\infty$ order respectively).},
\begin{equation}
{\cal L}= \frac{1}{2}\partial_\mu \phi \partial^\mu \phi
- \frac{m^2}{2}\phi^2-{\cal L}^{int},
\label{eq:lascal}
\end{equation}
where ${\cal L}^{int}$ is a polynomial in $\phi$,
the momenta conjugate to $\phi$ are
\begin{equation}
\Pi = \partial_-\phi
\end{equation}
with commutation relations
\begin{equation}
\left[ \Pi(x^-,{\vec x}_\perp,x^+),\phi(y^-,{\vec y}_\perp,x^+)
\right] = -\frac{i}{2} \delta(x^--y^-)
\delta({\vec x}_\perp-{\vec y}_\perp).
\label{eq:comscal}
\end{equation}
Note that this implies nonlocal commutation relations for the field
$\phi$, e.g.
\begin{equation}
\left[ \phi(x^-,{\vec x}_\perp,x^+),\phi(y^-,{\vec y}_\perp,x^+)
\right] = -\frac{i}{4} \varepsilon(x^--y^-)\delta({\vec x}_\perp-{\vec
y}_\perp),
\end{equation}
where $\varepsilon(x) = 1$ for $x>0$ and $\varepsilon(x) = -1$ for $x<0$.
The Hamiltonian density
($P_+= \int dx^- d^{D_\perp} x_\perp {\cal H} $)
is obtained from Eq.(\ref{eq:lascal}) via a Legendre transformation
\begin{eqnarray}
{\cal H} &=& \Pi \partial_+\phi - {\cal L}\nonumber\\
&=& \frac{1}{2} \left( {\vec \nabla}_\perp \phi \right)^2
+ \frac{m^2}{2}\phi^2+{\cal L}^{int}.
\end{eqnarray}
The commutation relations (\ref{eq:comscal}) are easily satisfied
if we make a mode expansion
\begin{equation}
\phi(x) =
\int_0^{\infty}\frac{dk_-}{\sqrt{4\pi k_-}}
\int \frac{ d^{D_\perp}k_\perp }{(2\pi)^{D_\perp/2}}
\left[ a_{k_-{\vec k}_\perp}e^{-ikx}
+a^\dagger_{k_-{\vec k}_\perp}e^{ikx} \right]
\end{equation}
where $a_{k_-{\vec k}_\perp}$, $a^\dagger_{k_-{\vec k}_\perp}$
satisfy the usual boson commutation relations, e.g.
\begin{equation}
\left[a_{k_-{\vec k}_\perp},a^\dagger_{q_-{\vec q}_\perp}\right]
=\delta(k_--q_-) \delta({\vec k}_\perp -{\vec q}_\perp).
\end{equation}
Longitudinal and transverse momentum operators contain no
interaction terms
\begin{eqnarray}
P_- &=& \int dx^- d^{D_\perp}x_\perp \Pi \partial_-\phi
= \int_0^{\infty} dk_-
\int d^{D_\perp}k_\perp k_-a^\dagger_{k_-{\vec k}_\perp}
a_{k_-{\vec k}_\perp}
\nonumber\\
{\vec P}_\perp &=& \int dx^- d^{D_\perp}
x_\perp \Pi {\vec \nabla}_\perp\phi
= \int_0^{\infty} dk_-
\int d^{D_\perp}k_\perp {\vec k}_{\perp}
a^\dagger_{k_-{\vec k}_\perp} a_{k_-{\vec k}_\perp}
\end{eqnarray}
where normal ordering terms have been dropped.
Most of the numerical studies of self-interacting scalar fields
have been done in $1+1$ dimensions \cite{hari,mb:sg} using
discrete light-cone quantization (Section \ref{dlcq}).
A more recent work employs Monte Carlo techniques
to solve $\phi^4$-theory in $2+1$ dimensions \cite{mb:lfepmc}.

Complex scalar fields can always be reduced to real scalar fields
by working in a Cartesian basis
$\Phi = \frac{1}{\sqrt{2}}\left(\phi_1+i\phi_2\right)$ and thus
need not be discussed here.

\subsection{Fermions with Yukawa Interactions }
\label{ex:ferm}
To keep the discussion as general as possible we assume
an interaction of the form $\bar{\psi}\Gamma \psi \phi$, where
$\phi$ is either scalar or pseudoscalar and $\Gamma$ is either
$1$ or $i\gamma_5$
\begin{equation}
{\cal L} = \bar{\psi}\left(i\not \! \partial - M -g\Gamma \phi \right)
\psi + \frac{1}{2}\left(\partial_\mu \phi \partial^\mu \phi
-m^2 \phi^2 \right).
\label{eq:lagsd}
\end{equation}
One novel feature compared to normal coordinates and compared to
self-interacting scalar fields on the LF is the fact that not all
components of $\psi$ are independent dynamical degrees of freedom.
To see this, let us introduce projection matrices
${\cal P}^{(\pm)} = \frac{1}{2} \gamma^\mp\gamma^\pm$ where
$\gamma^\pm = \left( \gamma^0 \pm \gamma^3 \right)/\sqrt{2}$.
Note that $\gamma^+\gamma^+=\gamma^- \gamma^-=0$ implies
${\cal P}^{(+)}{\cal P}^{(-)}={\cal P}^{(-)}{\cal P}^{(+)}=0$. These
projection matrices
can be used to decompose the fermion spinors into dynamical and
non-dynamical components
$\psi = \psi_{(+)} + \psi_{(-)}$ ,where
$\psi_{(\pm)} \equiv {\cal P}^{(\pm)}\psi$. The Lagrangian does not contain
a LF-time derivative ($\partial_+$) of $\psi_{(-)}$
\begin{eqnarray}
{\cal L} &=& \sqrt{2} \psi_{(+)}^\dagger i\partial_+\psi_{(+)}
+ \sqrt{2} \psi_{(-)}^\dagger i\partial_-\psi_{(-)}
- \psi_{(+)}^\dagger \left(i{\vec \alpha}_\perp {\vec \partial}_\perp+
\gamma^0{\cal M}\right)\psi_{(-)}\nonumber\\
& &- \psi_{(-)}^\dagger \left(i{\vec \alpha}_\perp {\vec \partial}_\perp+
\gamma^0{\cal M}\right)\psi_{(+)}
+ \frac{1}{2}\left(\partial_\mu \phi \partial^\mu \phi
-m^2 \phi^2\right),
\label{eq:lpm}
\end{eqnarray}
where ${\vec \alpha}_\perp = \gamma^0 {\vec \gamma}_\perp$ and
${\cal M}(x) = M + g\Gamma \phi(x)$. Thus the Euler-Lagrange
equation for $\psi_{(-)}$ is a constraint equation
\begin{equation}
\sqrt{2}i\partial_-\psi_{(-)}=\left(i{\vec \alpha}_\perp {\vec
\partial}_\perp+
\gamma^0 {\cal M}\right)\psi_{(+)}.
\label{eq:psiconstr}
\end{equation}
It is therefore necessary to
eliminate the dependent degrees of freedom ($\psi_{(-)}$) before
quantizing the theory.
Here we proceed by solving Eq.(\ref{eq:psiconstr})
and inserting the solution back in the Lagrangian (\ref{eq:lpm}), yielding
\cite{yan:sd2}
\footnote{Another option would be to use Dirac-Bergmann quantization
(Appendix \ref{dirac}). Up to possible differences
in the zero-mode sector, the result is the same.}
\begin{eqnarray}
{\cal L}_{(+)}&=& \sqrt{2} \psi_{(+)}^\dagger \partial_+
\psi_{(+)}+\frac{1}{2}\left(\partial_\mu \phi \partial^\mu \phi
-m^2 \phi^2 \right)\nonumber\\
& &-\frac{1}{\sqrt{2}}
\psi_{(+)}^\dagger \left(i{\vec \alpha}_\perp {\vec \partial}_\perp+
\gamma^0{\cal M}\right)\frac{1}{i\partial_-}
\left(i{\vec \alpha}_\perp {\vec \partial}_\perp
+ \gamma^0{\cal M}\right)\psi_{(+)}.
\end{eqnarray}
The ambiguities associated with the inversion of the differential
operator will be discussed in Section \ref{fzm}. Here we just define
\begin{equation}
\left( \frac{1}{\partial_-} f \right)(x^-,{\vec x}_\perp)
= \frac{1}{2}\int_{-\infty}^\infty dy^- \varepsilon(x^--y^-) f(y^-,{\vec
x}_\perp)
{}.
\end{equation}
The rest of the quantization procedure is now straightforward.
The Hamiltonian is given by
\begin{equation}
P^- = \int dx^- d^{D_\perp}x_\perp {\cal H}
\end{equation}
where
\begin{eqnarray}
{\cal H} &=&\frac{1}{\sqrt{2}}
\psi_{(+)}^\dagger \left(i{\vec \alpha}_\perp
{\vec \partial}_\perp+ \gamma^0{\cal M}\right)\frac{1}{i\partial_-}
\left(i{\vec \alpha}_\perp {\vec \partial}_\perp+
\gamma^0{\cal M}\right)\psi_{(+)}\nonumber\\
& &+\frac{1}{2} \left[\left( {\vec \nabla}_\perp \phi \right)^2
+m^2\phi^2\right]
\label{eq:hyuk}
\end{eqnarray}
Note that Eq.(\ref{eq:hyuk}) contains four-point interactions
of the form
$\psi_{(+)}^\dagger \phi \left( i\partial_- \right)^{-1}\phi\psi_{(+)}$
which were not present in the original Lagrangian (\ref{eq:lagsd}).
Note also, that the fermion mass $M$ enters the Hamiltonian density
(\ref{eq:hyuk}) in two different places: in the kinetic term for
the dynamical fermion field,
${\cal H}_{kin} \propto M^2 \psi_{(+)}^\dagger\left( i\partial_-
\right)^{-1}\psi_{(+)}$,
as well as in the three point vertex,
$M g \psi_{(+)}^\dagger\left( i\partial_- \right)^{-1} \Gamma
\phi\psi_{(+)} + h.c.$.
In Chapter \ref{ren} we will find that in general
these two masses are renormalized differently.

The scalar field $\phi$ is quantized as in Section (\ref{ex:scalar}).
For the fermions, one imposes anti-commutation relations only for
the independent component $\psi_{(+)}$ \cite{yan:sd2}
\begin{equation}
\left\{ \psi_{(+)}(x),\psi_{(+)}^\dagger(y)\right\}_{x^+=y^+}
=\frac{1}{\sqrt{2}} {\cal P}^{(+)} \delta(x^--y^-)
\delta({\vec x}_\perp -{\vec y}_\perp )
\label{eq:anti}
\end{equation}
with $\left\{ \psi_{(+)}^\dagger(x),\psi_{(+)}^\dagger(y)\right\}_{x^+=y^+}$
and
$\left\{ \psi_{(+)}(x),\psi_{(+)}(y)\right\}_{x^+=y^+}$ both vanishing.

For practical calculations it is very useful to make a mode expansion.
Let $u(p,s)$, $v(p,s)$ be the usual particle and antiparticle spinors,
satisfying $\left(\not \! p-M \right)u(p,s)=0$ and
$\left(\not \! p+M \right)v(p,s)=0$, where $s$ labels the spin.
The normalization is fixed such that
\begin{eqnarray}
\sqrt{2} u^\dagger_{(+)}(p,s^\prime) u_{(+)}(p,s) &=&
\bar{u}(p,s^\prime)\gamma_-u(p,s)=
2p_- \delta_{ss^\prime}\nonumber\\
\sqrt{2} v^\dagger_{(+)}(p,s^\prime) v_{(+)}(p,s) &=&
\bar{v}(p,s^\prime)\gamma_-v(p,s)=
2p_- \delta_{ss^\prime}.
\end{eqnarray}
For $\psi_{(+)}$ we make a plane wave ansatz
\begin{equation}
\psi_{(+)}(x) =
\int \frac{d^{D_\perp}p_\perp}{(2\pi)^{D_\perp/2} }
\int_0^\infty \frac{dp_-}{\sqrt{4\pi p_-}}\sum_s
\left[ b_{p_-{\vec p}_\perp s} u_{(+)}(p,s) e^{-ipx}
+ d^\dagger_{p_-{\vec p}_\perp s} v_{(+)}(p,s) e^{ipx}\right].
\label{eq:pwd}
\end{equation}
One can easily verify that $\psi_{(+)}$ in Eq.(\ref{eq:pwd})
satisfies the anti-commutation relations above (\ref{eq:anti}),
provided
\begin{eqnarray}
\left\{ b^\dagger_{p_-{\vec p}_\perp r},b_{q_-{\vec q}_\perp s} \right\}
&=& \delta(p_--q_-) \delta({\vec p}_\perp -{\vec q}_\perp) \delta_{rs}
\nonumber\\
\left\{ d^\dagger_{p_-{\vec p}_\perp r},d_{q_-{\vec q}_\perp s} \right\}
&=& \delta(p_--q_-) \delta({\vec p}_\perp -{\vec q}_\perp) \delta_{rs}
\end{eqnarray}
with all other anti-commutators vanishing.

Nonperturbative numerical works on the LF Hamiltonian with
Yukawa interactions (\ref{eq:hyuk}) have been restricted to
DLCQ calculations in 1+1 dimensions \cite{pa:dlcq} as well
as to 1+1 dimensional \cite{hari:yuk}
3+1 dimensional \cite{wo:93} calculations which use Tamm-Dancoff
truncations to fermion and at most two bosons or
fermion, antifermion and at most one boson.

\subsection{QED and QCD}
\label{ex:gauge}
Before one can canonically quantize a gauge field theory,
one must fix the gauge --- otherwise one has to deal
with the infinite degeneracy associated with the gauge
symmetry.

In the context of LF quantization one usually picks the
LF-gauge $A^+=A_-=0$. There are several (related) reasons for
this choice. In QED, the constraint equation for the
``bad'' spinor component reads
\begin{equation}
\sqrt{2} i D_-\psi_{(-)} \equiv \sqrt{2} \left(i\partial_- -eA_-\right)
\psi_{(-)} = \left( i {\vec \alpha}_\perp {\vec D}_\perp
+\gamma^0 M \right)\psi_{(+)}.
\end{equation}
The solution to this constraint equation,
\begin{equation}
\psi_{(-)}=\frac{1}{\sqrt{2}}\left(i\partial_- -eA_-\right)^{-1}
\left( i {\vec \alpha}_\perp {\vec D}_\perp
+\gamma^0 M \right)\psi_{(+)},
\label{eq:psielim}
\end{equation}
contains $A_-$ in the denominator and, unless one chooses
$A_-=0$ gauge, one thus obtains terms which have $A_-$ in the
denominator in the LF-Hamiltonian. In other words, in any
gauge other than the LF-gauge the canonical LF-Hamiltonian
always contains all powers of $A_-$ (after expanding the
geometric series) appearing in the interactions.

In QCD one faces the additional problem that
${\vec \Pi}_\perp$, the momentum conjugate to ${\vec A}_\perp$,
satisfies a nonlinear constraint equation if $A_-\neq 0$
(Section \ref{gaugezero}). Another reason to pick LF gauge is
that $A_-=0$ is invariant under the kinematic Lorentz
symmetries of the LF, i.e. under all transformations
that leave the plane $x^+=0$ invariant.
It is for these reasons that the LF gauge has been commonly
used for canonical LF quantization of gauge field theories,
and will also be used here
--- despite all the difficulties which are inherent
to the LF and axial gauges
\cite{ml:reg,ba:reg,wu:int,le:qm,le:qed,le:qcd}.

Even after fixing the gauge, not all degrees of freedom
are dynamical (similar to $\psi_{(-)}$, their time derivative
does not enter the Lagrangian).
Before we can proceed with the canonical quantization
we first have to eliminate these dependent variables by solving
those equations of motion which are constraint equations.
For $\psi_{(-)}$ we use Eq.(\ref{eq:psielim}) (note: $A_-=0$)
and proceed similar to the example of the Yukawa theory.
Since the time derivative of $A_+$ does not enter the
Lagrangian, $A_+$ has to be eliminated as well, by solving
its constraint equation (obtained by varying the Lagrangian
density with respect to $A_+$)
\begin{equation}
\partial_-^2 A_+ = \partial_- {\vec \nabla}_\perp {\vec A}_\perp
-j^+,
\end{equation}
where $j^+=\sqrt{2}e\psi^\dagger_{(+)} \psi_{(+)}$, in QED and
\begin{equation}
\partial_-^2 A_{+a} = \partial_-
{\vec \nabla}_\perp  {\vec A}_{\perp a}+
gf^{abc}{\vec A}_{\perp b}\partial_-{\vec A}_{\perp c} - j^+_a,
\end{equation}
where $j^+_a =\sqrt{2}g\psi^\dagger_{(+)\alpha } \psi_{(+)\beta }
\frac{\lambda_a^{\alpha \beta}}{2}$, in QCD. After inserting $A_+$
back into the Lagrangian one can proceed
with the quantization as usual. One finds
\begin{eqnarray}
{\cal H}_{QED} &=& \frac{1}{\sqrt{2}}
\psi_{(+)}^\dagger
\left( i {\vec \alpha}_\perp {\vec D}_\perp + \gamma^0M\right)
\frac{1}{i\partial_-}
\left( i {\vec \alpha}_\perp {\vec D}_\perp + \gamma^0M\right)
\psi_{(+)}
\nonumber\\
& &-\frac{1}{2}
\left(\partial_- {\vec \nabla}_\perp {\vec A}_\perp
-j^+\right) \frac{1}{\partial_-^2}
\left(\partial_- {\vec \nabla}_\perp {\vec A}_\perp
-j^+\right),
\label{eq:hqed}
\end{eqnarray}
and
\begin{eqnarray}
{\cal H}_{QCD} &=& \frac{1}{\sqrt{2}}
\psi_{(+)}^\dagger
\left( i {\vec \alpha}_\perp {\vec D}_\perp + \gamma^0M\right)
\frac{1}{i\partial_-}
\left( i {\vec \alpha}_\perp {\vec D}_\perp + \gamma^0M\right)
\psi_{(+)}
\nonumber\\
& &-\frac{1}{2}
\left({\vec D}_\perp \partial_-{\vec A}_{\perp a}-j^+_a\right)
\frac{1}{\partial_-^2}
\left({\vec D}_\perp \partial_-{\vec A}_{\perp a}-j^+_a \right)
\label{eq:hqcd}
\end{eqnarray}
where
${\vec D}_\perp \partial_-{\vec A}_{\perp a}=
 {\vec \nabla}_\perp \partial_-{\vec A}_{\perp a}
+gf^{abc}{\vec A}_{\perp b}\partial_-{\vec A}_{\perp c}$.
The commutation relations are similar to the ones in Yukawa theory
\begin{eqnarray}
\mbox{QED:}\ \ \ & &
\left\{ \psi_{(+)}(x),\psi_{(+)}^\dagger(y)\right\}_{x^+=y^+}
=\frac{1}{\sqrt{2}} {\cal P}^{(+)} \delta(x^--y^-)
\delta({\vec x}_\perp -{\vec y}_\perp )
\nonumber\\
& &
\left[ \partial_- A_{\perp i}(x), A_{\perp j}(y)\right]_{x^+=y^+}
= - \frac{i}{2}\delta (x^--y^-)
\delta ({\vec x}_\perp -{\vec y}_\perp )\delta_{ij}
\nonumber\\
\mbox{QCD:}\ \ \ & &
\left\{ \psi_{(+)\alpha}(x),
\psi_{(+)\beta}^\dagger(y)\right\}_{x^+=y^+}
=\frac{1}{\sqrt{2}} {\cal P}^{(+)} \delta(x^--y^-)
\delta({\vec x}_\perp -{\vec y}_\perp )\delta_{\alpha \beta}
\nonumber\\
& &
\left[ \partial_- A_{\perp a i}(x), A_{\perp b j}(y)
\right]_{x^+=y^+} = - \frac{i}{2}\delta (x^--y^-)
\delta ({\vec x}_\perp -{\vec y}_\perp )\delta_{ij}\delta_{ab}.
\nonumber\\
\end{eqnarray}
Similar to the approach to scalar field theories and Yukawa
theories, one may now attempt to solve the above Hamiltonians
by making a mode expansion and using matrix diagonalization.
In $1+1$ dimension this method was very successful
\cite{pa:qed,el:qed,ho:sea,mb:phd,mb:deu,mb:rb}.
In $3+1$ dimensions, this approach suffers from a
fundamental problem \footnote{Besides numerical difficulties
which will be discussed in Section \ref{dlcq}.}:
charged particles are subject to a linear, confining interaction
--- which is present even in ${\cal H}_{QED}$. For gauge
invariant amplitudes (all intermediate states included, which
contribute to a given order of the coupling) this linear
potential is canceled by infrared singular couplings of
charges to the $\perp$-components of the gauge field.
However, in most practical calculations, drastic truncations
of the Fock space are used to keep the dimension of the
Hamiltonian matrix within practical limits \cite{tang,kaluza}.
This approximation results in incomplete cancelations of
IR singularities and IR divergences result. Partly responsible
for this disaster is an improper treatment of
zero-modes and incomplete gauge fixing. If one integrates
the Maxwell equation for $F^{\mu+}$ over $x^-$ one finds
\cite{kalli}
\begin{equation}
-{\vec \partial}_\perp^2 \int dx^- A_-(x^+,x^-,x_\perp)
=\int dx^- j^+(x^+,x^-,x_\perp),
\label{eq:glaw}
\end{equation}
i.e. in general, when $\int dx^- j^+(x^+,x^-,x_\perp) \neq 0$,
the ``gauge'' $A_-=0$ is inconsistent with the
equations of motion. On a finite interval, with periodic
boundary condition, this becomes clearer because then
a Wilson loop ``around the torus'',
$\exp\left(ie\oint dx^- A_-(x^+,x^-,x_\perp)\right)$, is a gauge
invariant quantity. The closest one can get to the LF gauge
is $\partial_- A_-=0$.
In this gauge one can now investigate the problem of
incomplete gauge fixing.
The gauge $\partial_-A_-=0$ still leaves the freedom of
$x^-$-independent gauge transformations
$A_\mu \rightarrow A_\mu^\prime = A_\mu + \partial_\mu \chi$
where $\partial_-^2 \chi =0$
(or $\partial_- \chi=0$ if we restrict ourselves to periodic
$\chi$) \cite{kalli}. In such an incompletely gauge fixed
situations, not all degrees of freedom are physical and
approximations may result in inconsistencies.
A typical example is the residual or transverse
Gau\ss ' law (\ref{eq:glaw}), which
is a constraint on the physical Hilbert space. Such constraints
must either be imposed on the states or one can also use
them to eliminate ``unphysical'' degrees of freedom
(here ${\vec \partial}^2_\perp \int dx^- A^+$). The abovementioned,
incomplete
cancelation of IR singularities in the Tamm-Dancoff
approximation occurs because the transverse Gau\ss ' law
(\ref{eq:glaw}) is violated.
A more thorough discussion on this subject and possible caveats
can be found in Refs. \cite{le:qm,kalli}.

\chapter{The Light-Front Vacuum}
\label{vac}
\section{The Physical Picture}
In the Fock space expansion one starts from the vacuum as the ground
state and constructs physical hadrons by successive application
of creation operators.
In an interacting theory the vacuum is in general an extremely
complicated state and not known {\it a priori}. Thus, in general,
a Fock space expansion is not practical because one does not
know the physical vacuum (i.e. the ground state of the
Hamiltonian). In normal coordinates, particularly
in the Hamiltonian formulation, this is a serious obstacle
for numerical calculations.
As is illustrated in Table \ref{tab:vac}, the LF formulation
provides a dramatic simplification at this point.
\begin{table}
\begin{center}
\begin{tabular*}{14.4cm}[t]{@{\extracolsep{\fill}}c|c}
normal coordinates & light-front \\
\rule{7.2cm}{0.cm}&\rule{7.2cm}{0.cm}\\
\hline
\multicolumn{2}{c}{free theory}\\[1.5ex]
\multicolumn{2}{c}{
\setlength{\unitlength}{0.8mm}
\special{em:linewidth 0.4pt}
\def\empoint##1{\special{em:point ##1}}
\def\emline##1##2{\special{em:line ##1,##2}}
\begin{picture}(160,65)
\put( 0, 0){\begin{picture}(80,60)( 0, 0)
\put(5,0){\line(1,0){50}}
\put(30,-3){\line(0,1){50}}
\put(60,0){\makebox(0,0){$P_z$}}
\put(75,0){\line(0,1){60}}
\put(30,55){\makebox(0,0){$P^0 = \sqrt{m^2+{\vec{P}}^2}$}}
  \put( 5,27){\empoint{3}}
  \put(10,22){\empoint{4}}
  \put(15,18){\empoint{5}}
  \put(20,14){\empoint{6}}
  \put(25,11){\empoint{7}}
  \put(30,10){\empoint{8}}
  \put(35,11){\empoint{9}}
  \put(40,14){\empoint{10}}
  \put(45,18){\empoint{11}}
  \put(50,22){\empoint{12}}
  \put(55,27){\empoint{13}}
  \emline{3}{4} \emline{4}{5}
  \emline{5}{6} \emline{6}{7} \emline{7}{8} \emline{8}{9} \emline{9}{10}
  \emline{10}{11} \emline{11}{12} \emline{12}{13}
\end{picture}}
\put(80, 0){\begin{picture}(80,60)
\put(5,0){\line(1,0){50}}
\put(10,-3){\line(0,1){50}}
\put(60,0){\makebox(0,0){$P_-$}}
\put(10,55){\makebox(0,0)[l]{$P_+ =
\frac{\T m^2+{\vec P}_{\perp}^2}{\T 2 P_-}$}}
\put(15,49){\empoint{31}}
\put(20,25){\empoint{32}}
\put(25,17){\empoint{33}}
\put(30,13){\empoint{34}}
\put(35,10){\empoint{35}}
\put(40,8){\empoint{36}}
\put(45,7){\empoint{37}}
\put(50,6){\empoint{38}}
\put(55,6){\empoint{39}}
\emline{31}{32}
\emline{32}{33}
\emline{33}{34} \emline{34}{35}
\emline{35}{36} \emline{36}{37} \emline{37}{38} \emline{38}{39}
\end{picture}}
\end{picture}
}\\[1.cm]
$ P^0 = {\D\sum\limits_{\vec{k}}} a^\dagger_{\vec{k}} a_{\vec{k}} \sqrt{m^2
+ \vec{k}^2 } $ &
$ P_+ = {\D\sum\limits_{k_-,{\vec k}_{\perp}}}
a^\dagger_{k_-\!,{\vec k}_{\perp}} a_{k_-\!,{\vec k}_{\perp}}
 \frac{ m^2 + {\vec k}_{\perp}^2 }{ 2 k_-} $ \\[1.5ex]
\multicolumn{2}{c}{vacuum (free theory)}\\[1.5ex]
$\D a_{\vec{k}}|0\rangle = 0 $ & $\D a_{k_-\!,k_\perp}|0
\rangle = 0 $\\[1.5ex]
\multicolumn{2}{c}{vacuum (interacting theory)}\\[1.5ex]
many states with $\vec{P}=0$ & $k_- \ge 0$ \\
(e.\,g.\ $a_{\vec{k}}^\dagger
 a_{-\vec{k}}^\dagger|0\rangle$) &
$\hookrightarrow$ only pure zero-mode \\
 & excitations have $P_-=0$\\[1.5ex]
$\hookrightarrow$ $|\tilde{0}\!>$ very complex &
$\hookrightarrow$ $|\tilde{0}\!>$ can only contain \\
 & zero-mode excitations
\end{tabular*}
\end{center}
\caption{Zero Modes and the Vacuum}
\label{tab:vac}
\end{table}
While all components of the momentum in normal coordinates can
be positive as well as negative, the longitudinal LF momentum
$P_-$ is always positive. In free field theory (in normal
coordinates as well as on the LF) the vacuum is the state which
is annihilated by all annihilation operators $a_k$.
In general, in an interacting theory, excited states (excited with
respect to the free Hamiltonian) mix with the trivial vacuum
(i.e. the free field theory vacuum) state
resulting in a complicated physical vacuum.
Of course, there are certain selection rules and only states with
the same quantum numbers as the trivial vacuum can mix with
this state; for example, states with the same momentum as
the free vacuum (${\vec P}=0$ in normal coordinates,
$P_-=0$, ${\vec P}_\perp =0$ on the LF).
In normal coordinates this has no deep consequences because there
are many excited states which have zero momentum. On the LF
the situation is completely different. Except for pure zero-mode
excitations, i.e. states where only the zero-mode
(the mode with $k_-=0$) is excited, all excited states have
positive longitudinal momentum $P_-$. Thus only these pure zero-mode
excitations can mix with the trivial LF vacuum.
Thus with the exception of the zero-modes the physical LF vacuum
(i.e. the ground state) of an interacting
field theory must be trivial.\footnote{Cases where the
LF Hamiltonian has no ground state will be discussed below.}

Of course, this cannot mean that the vacuum is entirely
trivial. Otherwise it seems impossible to describe
many interesting problems which are related to spontaneous
symmetry breaking within the LF formalism. For example one knows
that chiral symmetry is spontaneously broken in QCD
and that this is responsible for the relatively small mass of
the pions --- which play an important role in strong interaction
phenomena at low energies. What it means is that one has
reduced the problem of finding the LF vacuum to the problem
of understanding the dynamics of these zero-modes.

First this sounds just like merely shifting the problem
about the structure of the vacuum from nonzero-modes
to zero-modes. However, as the
free dispersion relation on the LF,
\begin{equation}
k_+=\frac{m^2+{\vec k}^2_{\perp}}{2k_-},
\end{equation}
indicates, zero-modes are high energy modes!
Hence it should, at least in principle, be possible
to eliminate these zero-modes systematically
giving rise to an effective LF field theory
\cite{le:ap}.

Before we embark on theoretically analyzing
zero-modes, it should be emphasized that zero-modes
may have experimentally measurable implications.
This is discussed in Refs.\cite{mb:nag,mb:delta}.

\section{Examples for Zero Modes}
Usually, in the context of LF quantization,
fields that do not depend on $x^-$ are called zero-modes
(regardless whether they depend on ${\vec x}_\perp$ or not).
However, for practical purposes, the following classification scheme
seems to be particularly useful \cite{kalli}:
If one denotes
\begin{equation}
\langle f \rangle_o \equiv \frac{1}{2L} \int_{-L}^L dx^- f(x^-,{\vec
x}_\perp),
\end{equation}
then
\begin{equation}
\langle f \rangle \equiv \frac{1}{(2L_\perp)^2} \int d^2x_\perp \langle f
\rangle_o =\frac{1}{2L(2L_\perp)^2} \int d^2x_\perp\int_{-L}^L dx^-
f(x^-,{\vec x}_\perp)
\end{equation}
is called the {\it global zero-mode}, while
\begin{equation}
\stackrel{o}{f} \equiv \langle f \rangle_o
-\langle f \rangle
\end{equation}
is called {\it proper zero-mode}. The ``rest'', i.e.
\begin{equation}
\stackrel{n}{f} \equiv  f
-\langle f \rangle_0
\end{equation}
is called the {\it normal mode} part of $f$.
The motivation for this distinction arises primarily from
the fact that usually only the global zero-mode can
develop a vacuum expectation value but also since
proper and global zero-modes have a very different dynamics.

Zero modes occur in various contexts and it is not yet entirely
clear to what extend the various zero-mode effects, which will be
discussed below, are connected.
\subsection{Constant Scalar Fields}
In $\phi^4$ theory,
\begin{equation}
{\cal L} = \frac{1}{2}\partial_\mu \phi\partial^\mu \phi
-\frac{m^2}{2} \phi^2 - \frac{\lambda}{4!}\phi^4,
\label{eq:lphi4}
\end{equation}
if one chooses the ``wrong'' sign for the mass
($m^2<0$, $\lambda >0$), spontaneous
symmetry breaking occurs already at the classical level.
The field $\phi$ develops a vacuum expectation value
and the symmetry $\phi \rightarrow -\phi$ is
spontaneously broken. At least in $1+1$ and $2+1$ dimensions, with
appropriate values for the renormalized mass,
a similar behavior is observed in the quantum version.
Clearly, such a scenario requires a zero-mode. In the case
of $\phi^4$ theory, one may imagine that a redefinition
\begin{equation}
\phi \rightarrow \tilde{\phi}+\langle 0|\phi| 0\rangle
\label{eq:phishift}
\end{equation}
eliminates the VEV of the global zero-mode \cite{hari}.
However, this does not mean that one has eliminated
the zero-modes. In fact, by integrating the
equations of motion over $x^-$, one finds
\begin{equation}
0=m^2\langle \phi \rangle  +\frac{\lambda}{3!}\frac{1}{2L}\int_{-L}^L dx^-
\phi^3 \ ,
\label{eq:czm}
\end{equation}
which relates the zero-mode part of the field to the
normal mode part. Clearly, this nonlinear operator identity
implies that (for finite $L$), a mere shift of the
scalar field is not sufficient to completely eliminate
the zero-mode.
Instead, two main classes of approaches are being used
get the zero-modes under control.
In DLCQ one attempts to
solve the zero-mode constraint equation [Eq.(\ref{eq:czm})] using
various approximation or expansion schemes \cite{pisa:1,pisa:2,pisa:3}.
Due to nonlinear effects and operator ordering
ambiguities, solving Eq.(\ref{eq:czm}) becomes a nontrivial endeavor
 \cite{pisa:1,pisa:2,pisa:3,dave:symm,hksw:92}.
In the other approach (the effective LF-Hamiltonian approach, which
will be discussed in detail in Section 3.3) one makes use of the fact that
zero-modes freeze out for $L \rightarrow \infty$. Instead of
keeping zero-modes explicitly, one allows for an effective
Hamiltonian, which should account for their effects on
normal modes in the large volume limit \cite{mb:sg}.

So far it is not
known whether either one of these approaches to LF-quantization
(explicit zero-modes and effective LF-Hamiltonian)
leads to a consistent formulation
of $\phi^4$ theory in the broken phase. It is also not known
to what extend the particle spectrum in the equal time
formulation agrees with the spectrum on the LF.
Since the broken phase of
$\phi^4$ in 1+1 dimensions has a rather rich spectrum:
mesons, solitons \footnote{Often excluded by boundary conditions
on the fields.}, bound states and
scattering states in the soliton-antisoliton sector
\footnote{In general not excluded
by boundary conditions.}, this seems to be an ideal test case
for the various approaches to scalar zero-modes on the LF.
So far, all works on $\phi^4$ on the LF have concentrated
on demonstrating that spontaneous symmetry breaking occurs
and on reproducing the numerical value of the critical coupling
constant from ET quantization.\footnote{In view of the
nontrivial renormalization effects on the LF (see Chapter
\ref{ren}), comparing critical coupling constants on the
LF with those from ET quantization is very treacherous.}

One of the most striking consequences of spontaneous
symmetry breaking in $\phi^4_{1+1}$ is the emergence of
solitons. While most LF workers choose boundary conditions
that make it impossible to study solitons,
soliton-antisoliton scattering states are often still
possible. These states often have a very clear signature
\cite{mb:sg} and one can easily determine their threshold.
Considering the extensive literature on LF-$\phi^4_{1+1}$
(see e.g. Refs. \cite{pisa:1,pisa:2,pisa:3} and references therein),
it is surprising that solitons have been ignored so far.

\subsection{Fermionic Zero Modes}
\label{fzm}
Consider the free Dirac equation
\begin{equation}
0=\left( i \gamma_\mu \partial^\mu -M\right) \psi
= \left( i\gamma_-\partial_+ + i\gamma_+ \partial_-
 -i {\vec \gamma}_\perp {\vec \partial}_\perp -M\right)\psi.
\label{eq:Dirac}
\end{equation}
Multiplying Eq. (\ref{eq:Dirac}) with ${\cal P}^{(+)}$, where
${\cal P}^{(\pm)} = \gamma^\mp \gamma^\pm /2$ are the
projection matrices introduced in Section \ref{ex:ferm}, one
obtains
\begin{equation}
i\gamma^-\partial_-\psi^{(-)} = \left(
i {\vec \gamma}_\perp {\vec \partial}_\perp +M\right)\psi^{(+)}
\label{eq:fconstr}
\end{equation}
with $\psi^{(\pm)} = {\cal P}^{(\pm)}\psi$.
Clearly Eq.(\ref{eq:fconstr}) is a constraint equation
and one must eliminate $\psi^{(-)}$ before one can
canonically quantize the theory
(the kinetic term in the fermionic Lagrangian does
not contain a LF-time derivative of $\psi^{(-)}$).
For all modes but the
zero-modes this is straightforward. However, Eq.(\ref{eq:fconstr})
does not determine the $x^-$-independent components of
$\psi^{(-)}$. In other words, because of possible
``integration constants'', there is some ambiguity
in defining the inverse of the differential operator
$\partial_-$.

For scalar fields, the time derivative is always accompanied
by a space derivative (kinetic term: $\phi \partial_+ \partial_- \phi$).
Therefore, the zero-mode for scalar fields is not a dynamical degree
of freedom, since its time derivative does not enter the Lagrangian.
For Dirac fields this is different, since there the
Lagrangian is linear in the derivatives, and the fermionic
zero-mode is a dynamical degree of freedom.
Little is known in this case beyond perturbation theory
(see e.g. Refs. \cite{mb:al2,smu}).

\subsection{Gauge Field Zero-Modes}
\label{gaugezero}
For practical reasons one would like to work in
the LF gauge $A_-=0$ when quantizing gauge fields
on the LF. The reason is that, only in the LF gauge are
canonical field momenta simple.
For example, in QCD, the kinetic term for the gauge field
in the Lagrangian,
$-\frac{1}{4}F_{\mu\nu}F^{\mu\nu}$ contains terms like
$
\left[ D_+, {\vec A}_\perp\right]
\left[ D_-, {\vec A}_\perp\right]
$, i.e. in general, the term multiplying
$\partial_+{\vec A}_\perp$ contains interactions.
As usual in LF coordinates, the canonically conjugate momentum
satisfies a constraint equation
\begin{equation}
{\vec \Pi}_\perp =
\frac{\delta {\cal L}}{\delta \partial_+ {\vec A}_\perp}
= \partial_-{\vec A}_\perp -ig\left[ A_-,{\vec A}_\perp \right]
{}.
\label{eq:gfmom}
\end{equation}
Only in the LF gauge is the constraint equation for
${\vec \Pi}_\perp$ linear in the fields,
and one obtains simple commutation relations between the fields.

The problem with the LF gauge, as with axial gauges in
general, has to do with infrared singularities,
particularly in the nonabelian case. In order to arrive at a
well defined formulation of the theory, it is often very helpful
to formulate the theory in a finite `box' with periodic boundary
conditions (i.e. a torus). That way, it is generally easier to
keep track of surface terms that appear in formal manipulations
which include integrations by parts.

If one starts from an arbitrary gauge field configuration on
a torus, it is in general not possible to reach the LF gauge
(or spatial axial gauges) by means of a gauge transformation
\cite{manton,kalli}.
This can be easily shown by considering the Wilson loop
around the torus in the $x^-$ direction:
\mbox{$W = P \exp(ig\oint dx^- A_-)$}. This is a gauge invariant quantity
and thus does not change under a gauge transformation. If
it were possible to reach the LF gauge, $A_-=0$, be means of
a gauge transformation this would mean transforming $W$ to $1$,
which is a contradiction. It turns out that on a torus, the
closest one can get to the LF gauge is $\partial_- A_-=0$, i.e.
the zero-modes for $A_-$ remain and, due to their relation
to the Wilson loop around the torus, they have a gauge invariant meaning
\cite{kalli}.
They are dynamical degrees
of freedom (their $\partial_+$ derivative enters the Lagrangian).
The zero-modes of $a^i$ behave very similar to a scalar field,
in the sense that their time
derivative does not enter the Lagrangian and hence they are not
dynamical degrees of freedom.
Recently, Kalloniatis, Robertson and collaborators \cite{kalli}
have developed a systematic scheme to disentangle and resolve
the various zero-mode problems that appear in QED and QCD.
For example, projecting the QED Maxwell equations onto the
proper zero-mode sector, they obtain:
\begin{eqnarray}
-\partial_\perp^2 \stackrel{o}{A}^+ &=& g\stackrel{o}{J}^+
\label{eq:mw1}\\
 -2\partial_+^2 \stackrel{o}{A}^--\partial_\perp^2 \stackrel{o}{A}^-
 -2\partial_i\partial_+ \stackrel{o}{A}^i
&=& g\partial_\perp^2 \stackrel{o}{J}^-
\label{eq:mw2}\\
 -\partial_\perp^2 \stackrel{o}{A}^+ +\partial_i\partial_+ \stackrel{o}{A}^+
 +\partial_i\partial_j \stackrel{o}{A}^j
&=& g\partial_\perp^2 \stackrel{o}{J}^j\ ,
\label{eq:mw3}
\end{eqnarray}
where $J^\mu$ is the fermionic current operator.
The first of these equations (\ref{eq:mw1}) is a constraint equation
and can be used
to eliminate the proper zero-mode of $A^+$ in terms of the current $J^+$
\footnote{In the charge neutral sector, the global zero-mode of $J^+$
vanishes
and thus the inverse
Laplace is well defined.}
\begin{equation}
\stackrel{o}{A}^+ = -g\frac{1}{\partial_\perp^2} \stackrel{o}{J}^+\ ,
\end{equation}
which again demonstrates that $A^+=0$ is in general not consistent
with the equations of motion.

Further simplification can be obtained by taking the (transverse) divergence
of Eq.(\ref{eq:mw3}), yielding
\begin{equation}
\partial_+\stackrel{o}{A}^+ = g \frac{1}{\partial_\perp^2} \partial_i
\stackrel{o}{J}^i\ .
\end{equation}
Inserting this back into Eq.(\ref{eq:mw3}), one finds
\begin{equation}
-\partial_\perp^2 \left( \delta_{ij} -\frac{\partial_i \partial_j}
{\partial_\perp^2}\right) \stackrel{o}{A}^j
=g\left( \delta_{ij} -\frac{\partial_i \partial_j}
{\partial_\perp^2}\right) \stackrel{o}{J}^j,
\label{eq:tproj}
\end{equation}
which can be used to eliminate the transverse projection
of the proper zero-mode of $A^j$.
Note that so far we have not yet completely fixed the gauge,
since $\partial_-A^+$ sill leaves the freedom of purely transverse
gauge transformations, $A^\mu \rightarrow A^\mu + \partial^\mu \Omega$,
where $\Omega = \Omega(x^+,x_\perp)$. One can use this residual
gauge freedom to set $\partial_i \stackrel{o}{A}^i=0$. In combination
with Eq.(\ref{eq:tproj}), this completely
determines the proper zero-mode of $A^i$.
Up to this point, it seems that the zero-modes
in QED pose no real problems in the LF formulation.

The real problems in this formalism arise when one tries to implement
these results in a quantum formulation that includes fermions.
This can be seen when one
inserts the solution for
$\int dx^-A_-$ back into the Lagrangian, yielding a
four Fermi interaction term of the form
\begin{equation}
\Delta {\cal L}\propto \frac{1}{2L}\stackrel{o}{J}^+
\frac{1}{\partial_\perp^2}
\stackrel{o}{J}^-.
\end{equation}
Similarly, inserting the solution for $\stackrel{o}{A}^i$
into the Lagrangian yields
\begin{equation}
\Delta {\cal L}\propto \frac{1}{2L}\stackrel{o}{J} ^i
\frac{\delta_{ij} -\frac{\partial_i \partial_j}
{\partial_\perp^2}}{\partial_\perp^2}
\stackrel{o}{J}^j.
\end{equation}
The presence of such terms, which contain
the ``bad'' current $j_+ =\sqrt{2}e \psi^\dagger_{(-)}\psi_{(-)}$
leads to nonlinear constraint equations for $\psi_{(-)}$.
Because of the difficulties in solving this nonlinear
constraint equation, it has so far not been possible
to write down the LF Hamiltonian for QED or QCD in terms of
physical degrees of freedom and
including all zero-modes, in closed form.
Only perturbative expressions for the Hamiltonian in
terms of physical degrees of freedom have been found so far
\cite{kalli}. Similar problems arise in the DLCQ formulation
of QCD with additional complications arising
from the difficulties in quantizing the gauge field when
$A^+ \not \! = 0$., arising from the nonlinear constraint
relation between fields and their canonical momenta (\ref{eq:gfmom}).

{}From the practitioner's point-of-view, it would be helpful
to know to what extend this elaborate machinery is
actually necessary if one is interested only in the large
volume limit.
On a finite interval, gauge field zero-modes clearly
play an important role. For example, they are essential
to generate the correct potential for a heavy
quark-antiquark pair in 1+1 dimensions on a circle in
Coulomb gauge
\cite{be:eng,kpp94}. However, in the latter example, zero-mode
effects for color singlet states
disappear in the limit of a large interval.
Unfortunately, it is not clear whether this result
carries through to higher dimensional gauge theories.

\subsection{Perturbative Zero-Modes}
The zero-modes discussed are either connected
to purely nonperturbative effects (like in the case
of spontaneous symmetry breaking for scalar fields)
or seem to be at least connected with nonperturbative
physics (like infrared singular long range effects
for gauge fields). There are, however, plenty of
examples where zero-mode effects appear already
on the level of perturbation theory. Examples include
disconnected vacuum diagrams \cite{ma:zero},
``generalized tadpoles'' for self-interacting scalar fields
\cite{gr:sg,mb:sg} as well as ``rainbow diagrams'' for the
fermion self-energy \cite{mb:al1,mb:al2}. These examples
will be discussed in more detail in Chapter \ref{ren}.

In perturbation theory in LF gauge the gauge field
propagator
\begin{equation}
D_{\mu \nu}(k) = \frac{ g_{\mu \nu} -
\frac{k_\mu n_\nu + k_\nu n_\mu}{k \cdot n} }{k^2 +i\varepsilon}
\end{equation}
($n \cdot A = A_-$) becomes singular as $k_- \rightarrow 0$.
There exist several ``prescriptions'' to handle this singularity.
The most useful prescription for perturbative calculations
is the Mandelstam-Leibbrandt (ML) prescription \cite{ml:reg}, where
one replaces
\begin{equation}
\frac{1}{k \cdot n} \equiv \frac{1}{k_-}
\stackrel{\mbox{ML}}{\longrightarrow}
\frac{1}{k_- +i\varepsilon \mbox{sign}(k_+)} =
\frac{k_+}{k_+k_- + i\varepsilon} .
\end{equation}
The crucial property of this prescription is that the
pole structure is similar to the one of a typical Feynman
propagator, with poles in the second and fourth quadrant of the
complex $k_0$-plane, and thus allows to perform a Wick rotation.
This is not the case for the principal value (PV) prescription
\begin{equation}
\frac{1}{k_-} \stackrel{\mbox{PV}}{\longrightarrow}
\frac{1}{2} \left(
\frac{1}{k_-+i\varepsilon}+\frac{1}{k_--i\varepsilon}\right)
\end{equation}
with poles in the first and fourth quadrant.

One of the major disadvantages of the ML prescription is the fact
that it introduces additional energy ($k_+$) dependencies in the
propagator, which cannot be generated by a canonical LF Hamiltonian
\cite{smu2}.
However, recently the ML prescription has been successfully
implemented in a LF Bethe-Salpeter approach to bound states
\cite{so:bs}. Conversely, in
$\mbox{QCD}_{1+1}(N_C \rightarrow \infty)$ the
ML prescription \cite{wu:int} yielded a spectrum that disagreed
with the canonical LF approach \cite{th:qcd} as well as with the
result from equal time quantization \cite{wi:vak}. More recently,
light-like Wilson loops in 1+1 dimensions have been calculated,  using
various prescriptions for gauge field propagator \cite{ba:2d}, and
it was found that only the principal value prescription yields
the exact area law one expects for gauge fields in 1+1 dimensions
(on noncompact manifolds).

\section{Zero Modes and the Vacuum in
\mbox{$\varepsilon$-Coordinates}}
\label{zereps}
\subsection{General Considerations}
For a free particle $p_+= \frac{L}{2\varepsilon}
\left(-p_-+\sqrt{p_-^2+2\varepsilon m^2/L}\right)$
and $p_-$ is no longer restricted to positive values
(Fig. \ref{fig:disp}). Therefore, for all finite values
of $\varepsilon /L$, the vacuum in $\varepsilon$-coordinates
is nontrivial.
Since $\varepsilon$-coordinates
(see Section \ref{eps}) provide a controlled and
well defined approach to the LF, it seems very natural
to employ this framework for studying the LF vacuum.

Let us first consider the canonical LF limit ($L$ fixed,
$\varepsilon \rightarrow 0$). In this case it is
straightforward to derive an effective LF-Hamiltonian from
the $\varepsilon$-Hamiltonian \cite{le:ap}
(for a related work, see Refs. \cite{pnp,fields}). For finite
L the momenta are discrete. Without interactions the
energy of the zero-mode ($p_+(0) = m \sqrt{\frac{L}{2\varepsilon}}$)
and the energy of modes with negative momenta
($p_+(-n) \approx \frac{2\pi n}{\varepsilon}$) diverge as
$\varepsilon \rightarrow 0$, while the energy of all
positive momentum modes ($p_-(n) \approx \frac{m^2 L}{4\pi n}$)
remains finite. For interacting fields there will be some slight
quantitative changes, but the general picture should remain
the same: zero-modes and negative momentum modes are
high energy modes ---
separated from positive momentum modes by an energy  gap
of ${\cal O}\left(\sqrt{\frac{1}{\varepsilon}}\right)$
and ${\cal O}\left(\frac{1}{\varepsilon}\right)$
respectively.
Thus although $p_-\leq 0$ modes may acquire
nontrivial occupations, $p_->0$ modes have too
little energy to cause any excitations within the
$p_-\leq 0$ sector for $\varepsilon \rightarrow 0$:
the $p_-\leq 0$ modes freeze out and can be replaced
by their vacuum expectation value (VEV).

At this point it seems that we have succeeded in deriving
a nontrivial effective LF-Hamiltonian. Unfortunately, we
arrived at this result by approaching the LF in such a way
that the invariant length of the interval ($\propto L\varepsilon$)
approaches zero, i.e. as discussed in Ref. \cite{le:ap},
the effective theory that we have obtained is not necessarily
equivalent to the original covariant theory. This can be
easily illustrated by means of a perturbative example.
Consider a simple tadpole with a mass insertion (to make
it convergent) in $1+1$ dimensions
\begin{equation}
\tilde{\Sigma} = \int \frac{d^2k}{(2\pi)^2}
\frac{1}{\left(k^2-m^2+i0\right)^2} = \frac{i}{4\pi m^2}.
\label{eq:sigcov}
\end{equation}
On a finite interval (with $\varepsilon$ coordinates) one
obtains instead ($k_-(n) = \frac{2\pi}{L}n$)
\begin{eqnarray}
\tilde{\Sigma} &=& \frac{1}{L}\sum_{k_-}\int \frac{dk_+}{2\pi}
\frac{1}{\left(\frac{2\varepsilon}{L}k_+^2 +2k_+k_- -m^2 + i0
\right)^2}\nonumber\\
&=& \frac{i}{4\sqrt{2\varepsilon L}}\sum_{n=-\infty}^{\infty}
\left[ \frac{(2\pi n)^2}{2\varepsilon L}+m^2\right]^{-3/2}
\label{eq:sigeps}.
\end{eqnarray}
Clearly, in order to recover the continuum result
(\ref{eq:sigcov}) one must take limits in such a way that
the invariant length if the interval becomes infinite.
If one takes the LF limit first ($\varepsilon \rightarrow 0$,
$L$ fixed), one obtains a divergent contribution from the
zero-mode.

A different result is obtained if one performs the
continuum limit first ($L\rightarrow \infty$,
$\varepsilon/L$ fixed). Since this corresponds to
$\varepsilon L \rightarrow \infty$ no problems with perturbation
theory arises. However, since the spectrum is now
continuous, there is no mass gap and the derivation of the
effective Hamiltonian for $\varepsilon/L\rightarrow 0$
becomes more complicated.
Nevertheless, it is still possible: first, note that the
momentum scale of the continuum Hamiltonian is
$m^2 \varepsilon/L$ since
momentum dependent terms in the continuum Hamiltonian appear only in
the kinetic term $\propto \frac{L}{2\varepsilon}\left[
-k_-+\sqrt{k_-^2+m^22\varepsilon/L}\right]$
and in vertex factors $\propto \left(k_-^2+m^22\varepsilon/L
\right)$. Thus the typical momentum scale in the vacuum is given by
$p_-^{vac} = {\cal O}\left( \sqrt{\frac{\varepsilon}{L}}m\right)$.
Similarly the energy scale for vacuum excitations
(zero total $P_-$) is of the order
${\cal O}\left( \sqrt{\frac{L}{\varepsilon}}m\right)$.
Suppose one is interested in the effective Hamiltonian for
a physical particle of total momentum $p_-^{tot}$
moving in the vacuum. If $\frac{\varepsilon}{L} \ll 1$ then
there is almost no overlap between the wave function of the
vacuum $p_-^{vac} = {\cal O}\left( \sqrt{\frac{\varepsilon}{L}}m\right)$
and the wave function of the partons in the particle $p_-^{parton}$
because the parton wavefunction (calculated
for example with a typical LF Hamiltonian) vanishes for small
momenta. Thus one can introduce an energy gap {\it by hand}
without affecting the dynamics in the limit
$\frac{\varepsilon}{L}\rightarrow 0$: for example, by selecting
cutoffs $\Lambda_1$ and $\Lambda_2$ such that
\begin{equation}
m\sqrt{\frac{\varepsilon}{L}} \ll \Lambda_1 \ll \Lambda_2
\ll p_-^{tot}
\label{eq:superineq}
\end{equation}
and removing all modes with $\Lambda_1 < k_-<\Lambda_2$.
First, this gives rise to a mass gap and one can argue
that the modes with $k_-<\Lambda_1$ remain frozen
(energy scale $k_+ > \frac{m^2}{2\Lambda_1}$)
when excitations with $k_- > \Lambda_2$ are present
(energy scale $k_+ < \frac{m^2}{2\Lambda_2}$):
in second order perturbation theory, the energy shift
for modes with $k_- > \Lambda_2$ due
to excitations of $n$ modes with $k_- < \Lambda_1$ is given by
\begin{equation}
\Delta^{(2)} E \propto\frac {\left| \Lambda_1^{-n/2} \right|^2 }
{-\frac{1}{\Lambda_1}} \Lambda_1^n = -\Lambda_1
\stackrel{\Lambda_1\rightarrow 0}{\longrightarrow}0
\label{eq:2nd}
\end{equation}
Here $\Lambda_1^{-n/2}$ is a vertex factor, arising from
the factor $\frac{1}{\sqrt{\omega_n}}$ in the expansion
of the fields $\phi$ (Eq.(\ref{eq:nmode})),
the factor $\Lambda_1^n$ is the phase space factor for
$n$ modes with $k_- ={\cal O}( \Lambda_1)$ and states with
$k_- < \Lambda_1$ excitations are off-shell by at least
$\frac{1}{\Lambda_1}$.

Since $k_- < \Lambda_1$ excitations are suppressed,
the effective LF-Hamiltonian for the modes with
$k_- >\Lambda_2$ contains the $k_-<\Lambda_1$ modes
only via their VEV (which may be nontrivial!)
\begin{equation}
V^{eff}_{k_->\Lambda_2} =
\langle 0_{k_-<\Lambda_1} | V | 0_{k_-<\Lambda_1}\rangle.
\label{eq:veffeps}
\end{equation}
The crucial point is that
the parton distribution calculated with such an
effective LF Hamiltonian vanishes for small momenta
in the above
superrenormalizable example
\footnote{Roughly speaking, the
LF kinetic energy $T$, which one can calculate from the parton
momentum distribution $f(k_-)$, using
$T=m^2 \int_0^{P_-^{tot}} dk_- \frac{f(k_-)}{2k_-}$, has to remain finite.}.
Thus as long as $\Lambda_2$
is small enough compared to the total momentum of the particle $p$,
the parton distribution vanishes already
for momenta much larger than $\Lambda_2$ and the presence of the
cutoff does not affect the parton dynamics.
Since the VEVs are nearly independent of $\frac{\varepsilon}{L}$,
so is the effective
Hamiltonian. Thus the suppression of the parton distribution
due to the kinetic energy sets in at a value
$cp^{tot}_-$, where $c$ is nearly independent of $\frac{\varepsilon}{L}$.
Thus, for $\frac{\varepsilon}{L}\rightarrow 0$,
$\Lambda_2$ can easily be chosen smaller than $cp_-^{tot}$
while Eq.(\ref{eq:superineq}) remains satisfied. In other words,
$\Lambda_2$ can be chosen such that the parton dynamics is
independent of the exact position of the cutoff.
Similarly, since vacuum momenta are restricted to
\mbox{$p_-^{vac} = {\cal O}\left(
\sqrt{\frac{\varepsilon}{L}}m\right)\ll \Lambda_1$},
the presence of the cutoff does not affect the
dynamics of the vacuum either, i.e., the numerical value
of the VEVs which enter Eq.(\ref{eq:veffeps}) is independent
of the cutoff.
\begin{figure}
\unitlength1.cm
\begin{picture}(14,5)(1,-8.)
\includegraphics{veff.ps}
\end{picture}
\caption{
Schematic occupation of modes in the presence of
a particle with momentum $p_-$ for
$\varepsilon/L \ll 1$ (i.e. ``close to the LF'').
The modes near $k_-=0$ are already present in the
vacuum and are dynamically restricted to
$k_- = {\cal O} \left( m (\varepsilon /L)^{(1/2)}\right)$
The ``parton distribution'', i.e. the modes which are
occupied in the presence of the particle but not in the
vacuum, vanish at small $k_-$ at a momentum scale which
remains finite as $\varepsilon /L\rightarrow 0$.
The presence of the cutoffs has almost no effect on the
dynamics.
}
\label{fig:veff}
\end{figure}

In the 2nd order perturbation theory argument above we made
use of $\Lambda_1 \ll \Lambda_2$ to make sure that the energy
denominator in Eq.(\ref{eq:2nd}) is of the order
${\cal O}\left( \frac{1}{\Lambda_1}\right)$. This is actually not
necessary, since the occupation of these modes is anyway
{\it dynamically} suppressed for $k_-<cp_-$ and as long as
\mbox{$\Lambda_1 \ll cp_-$}, the energy denominator will automatically
be ${\cal O}\left( \frac{1}{\Lambda_1}\right)$ or smaller.

Thus we can actually let $\Lambda_2 \rightarrow \Lambda_1$,
i.e. remove the cutoff, without altering the conclusion.
Introducing a mass gap was helpful in deriving an effective
Hamiltonian for modes with $p_-\gg m\sqrt{\frac{\varepsilon}{L}}$.
However, since the solutions of the effective Hamiltonian
vanish at small $p_-$ anyway, there is no need for a cutoff:
a region void of excitations between
$m\sqrt{\frac{\varepsilon}{L}}$ and $cp_-$ develops dynamically
(Figure \ref{fig:veff})
and this is sufficient to derive an effective Hamiltonian.

In the end, the following result is obtained. Suppose we started
from some polynomial interaction
\begin{equation}
{\cal L}^{int} = -\sum_n \lambda_n \frac{\phi^n}{n!}.
\end{equation}
Then, using Eq.(\ref{eq:veffeps}) (after some combinatorics)
the effective interaction, which enters
the LF Hamiltonian for $p_->0$ modes in the limit
$\frac{\varepsilon}{L}\rightarrow 0$ is given by
\begin{equation}
{\cal L}^{int}_{eff} = -\sum_n \lambda_n
\sum_{k=0}^n \frac{\phi^{(n-k)}}{(n-k)!}
\langle0| \frac{\phi^k}{k!}|0\rangle
\label{eq:leffeps2}
\end{equation}
(in order to obtain this result one also uses that, after
normal ordering, the $p_->0$ modes do not contribute
to the VEVs). Eq.(\ref{eq:leffeps2}) is a remarkable
result. It states that nontrivial vacuum effects enter
the LF-Hamiltonian only via effective interactions.
The effective coupling constants depend on the vacuum
condensates which, in general, cannot be obtained directly from a LF
calculation
\footnote{However, there are exceptions where one can use sum rules
or consistency conditions to determine the effective couplings
iteratively. Examples will be discussed in the following section.}.
They must be considered as renormalization parameters of the
LF theory. Eq.(\ref{eq:leffeps2}) is also valid in situations
where spontaneous symmetry breaking occurs. For example
in $\phi^4$ theory, $\langle 0|\phi |0\rangle$ may become
nonzero and a $\phi^3$ interaction will thus appear
in the effective Lagrangian.
\footnote{
It should be emphasized that we did not make any mean field
assumptions, such as $\langle 0|\phi^k|0\rangle=
\langle 0|\phi|0\rangle^k$, in order to arrive at this result.}
However, note that only a {\it finite} number of condensates
is necessary to specify the effective LF Hamiltonian:
if $N$ is the highest power of $\phi$ entering the canonical
LF Hamiltonian then only condensates $\langle 0|\phi^k|0\rangle $
with $k<N$ need to be considered.

At several points in the above discussion it was important
that the theory is free of divergences (up to a finite
number of diagrams
which can always be subtracted before applying
above argumentation). First this was important to insure
that the momentum scale in the vacuum is finite. Secondly,
it was important because only in the absence of
divergences one can apply the kinetic energy argument
to prove that the parton momentum distribution vanishes for small
parton momenta. Therefore one must be very careful when
generalizing the above results to higher dimensional field theories.
Eq.(\ref{eq:leffeps2}) can be used in renormalizable field theories
only {\it after} a cutoff has been imposed. This is for example the
case for the transverse lattice which will be discussed
in Section \ref{lfepmc}.

\subsection{A simple Example for the
limit $L\rightarrow \infty$,
$\varepsilon / L \rightarrow 0$}
The appearance of the gap as $L\rightarrow \infty$ (first) and
$\varepsilon/L\rightarrow 0$
is best understood by studying a concrete example.
Ideally this implies considering some nontrivial interacting
field theory and calculating the occupation of the modes
nonperturbatively for various $\varepsilon/L$ and
$L \rightarrow \infty$. However, even in integrable models,
such as the sine-Gordon model, the occupation of the
modes is not known exactly! Since numerical calculations at
small but finite $\varepsilon/L$ and $L\rightarrow \infty$
are very complicated --- particularly if one is interested
in momenta of the order of $m\sqrt{\varepsilon/L}$ ---
and we will proceed by studying a perturbative example.
Due to the fact that the appearance of the gap is mostly a
consequence of dimensional analysis, this will be sufficient
to highlight the essential physics of the limit
$\varepsilon/L\rightarrow 0$, $L \rightarrow \infty$
($L \rightarrow \infty$ first). The example which we will
consider is a scalar field theory in $1+1$ dimensions with
polynomial self-interactions
\begin{equation}
{\cal L} = \frac{1}{2} \partial_\mu \phi \partial^\mu \phi
-\frac{m^2}{2}\phi^2 -\frac{\lambda_4}{4!}\phi^4
-\frac{\lambda_6}{6!}\phi^6 .
\end{equation}
In $\varepsilon$-coordinates the Hamiltonian for this
model reads after normal ordering
\begin{equation}
H = \sum_n p_+(k_n) a_{k_n}^\dagger a_{k_n}
+\int_0^L dx^- \left[\frac{\lambda_4}{4!}:\phi^4:
+\frac{\lambda_6}{6!}:\phi^6:\right],
\end{equation}
where the same notation as in Section \ref{eps} has
been used. In lowest (zeroth) order in $\lambda_4$ and
$\lambda_6$ the vacuum is the Fock vacuum, defined by
$a_{k_n} |0\rangle =0$. This changes of course for
nonvanishing couplings.
For example, in second order perturbation theory in
$\lambda_4$ one finds for the occupation of states in the
vacuum
\begin{eqnarray}
\rho_0(k,\frac{\varepsilon}{L},L) &\equiv&
\langle \tilde{0}| a_k^\dagger a_k |\tilde{0}\rangle
\nonumber\\
&=& \frac{\lambda_4^2}{3!} L^2 \sum_{k_2,\ k_3}
\frac{1}{\left[ p_+(k) + p_+(k_2) + p_+(k_3) + p_+(k_4) \right]^2}
\nonumber\\
& &\quad \quad \quad \quad\times
\frac{1}{2\omega (k)}\frac{1}{2\omega (k_2)}
\frac{1}{2\omega (k_3)}\frac{1}{2\omega (k_4)},
\end{eqnarray}
where $\omega(q)=L\sqrt{q^2+2\varepsilon m^2/L}$,
$p_+(q)=\left( -q+\sqrt{q^2+2\varepsilon m^2/L}\right)
L/2\varepsilon $
and $k_4=-k-k_2-k_3$. A similar expression is found for
the ${\cal O}(\lambda_6^2)$-term, which will be omitted for simplicity. In
the limit $L \rightarrow \infty$ one thus finds
\begin{eqnarray}
\rho_0(k,\frac{\varepsilon}{L}) &\equiv&
\lim_{L \rightarrow \infty} \frac{L}{2\pi} \frac{1}{\sqrt{2\varepsilon L
m^2}}
\rho_0(k,\frac{\varepsilon}{L},L)
\nonumber\\
&=& \frac{\lambda_4^2}{96 \pi m^5}\sqrt{\frac{L}{2\varepsilon}}
\hat{\rho}_0 \left(k \sqrt{\frac{L}{2\varepsilon m^2}}\right),
\label{eq:rho0}
\end{eqnarray}
where
\begin{equation}
\hat{\rho}_0(z) = \int_{-\infty}^\infty \frac{dz_2}{2\pi}
\int_{-\infty}^\infty \frac{dz_3}{2\pi}
\frac{\hat{\omega}(z)^{-1}\hat{\omega}(z_2)^{-1}
\hat{\omega}(z_3)^{-1}\hat{\omega}(z_4)^{-1}
}{\left[ \hat{\omega}(z)+\hat{\omega}(z_2)+\hat{\omega}(z_3)+
\hat{\omega}(z_4)\right]^2}
\end{equation}
with $\hat{\omega}(z)=\sqrt{z^2+1}$ and $z_4=-z-z_2-z_3$.
The factor $L/2\pi$ arises from going from discrete to
continuous momentum $k$ and we divided by the invariant
length of the interval because the occupation in the
vacuum trivially scales like the invariant length.

Most importantly, the momentum scale in the occupation
of the vacuum is set by $m\sqrt{2\varepsilon}/L$.
The momentum density in the vacuum is sharply peaked around
$k=0$ with width ${\cal O}(\sqrt{2\varepsilon /L})$ and height
${\cal O}(\sqrt{L/2\varepsilon})$, i.e. it resembles a
$\delta$-function as $\varepsilon/L \rightarrow 0$.

Let us now consider a state with momentum P, where P is taken
independent of $L$ or $\varepsilon$. To lowest order
\begin{equation}
|P\rangle = a_P^\dagger |0\rangle
\end{equation}
and thus
\begin{equation}
\rho_P(k,\frac{\varepsilon}{L},L) \equiv \langle P|a_k^\dagger a_k
|P\rangle = \delta_{k,P} + {\cal O}(\lambda^2).
\label{eq:0order}
\end{equation}
Three classes of corrections contribute to $\rho_P$:
insertions in
disconnected vacuum diagrams (Fig.\ref{fig:rhodiag}a)
[yielding again Eq.(\ref{eq:rho0})], insertions in tadpoles
(Fig.\ref{fig:rhodiag}b) and the rest, i.e. insertions
in non-tadpole connected corrections
(Fig.\ref{fig:rhodiag}c).

\begin{figure}
\unitlength1.cm
\begin{picture}(14,5)(1,-7.5)
\includegraphics{rhopert.ps}
\end{picture}
\caption{${\cal O}(\lambda^2)$-corrections to the mode
density in the presence of a particle with momentum $P$.
a.) disconnected corrections, b.) insertions into
generalized tadpoles
(i.e. diagrams where a subgraph is connected with the
rest of the diagram at one point only) and c.) non-tadpole
connected corrections.}
\label{fig:rhodiag}
\end{figure}
The tadpole term yields
\begin{eqnarray}
\tilde{\rho}_P^{tadpole}(k,\sqrt{\frac{\varepsilon}{L}}) &\equiv&
\lim_{L\rightarrow \infty} \frac{L}{2\pi}
\rho_P^{tadpole}(k,\sqrt{\frac{\varepsilon}{L}},L)
\nonumber\\
&=&\frac{\lambda_4 \lambda_6}{96 \pi m^4 P}
\hat{\rho}_0 \left(k \sqrt{\frac{L}{2\varepsilon m^2}}\right),
\label{eq:rhotad}
\end{eqnarray}
and for the non-tadpole, connected term one finds
\begin{eqnarray}
\tilde{\rho}_P^{nt}(k,\sqrt{\frac{\varepsilon}{L}})
&\equiv& \lim_{L\rightarrow \infty}\frac{L}{2\pi}
\rho_P^{nt}(k,\sqrt{\frac{\varepsilon}{L}},L)
\label{eq:rhont}
\\
&=& \frac{\lambda_4^2}{32\pi}
\int_{-\infty}^\infty \frac{dk_2}{2\pi }
\left[\frac{1}{E_A^2} + \frac{1}{E_B^2}\right]
\frac{1}{\omega(P)}\frac{1}{\omega(k)}\frac{1}{\omega(k_2)}
\frac{1}{\omega(k_3)}
\nonumber
\end{eqnarray}
($k_3=P-k-k_2$)
plus a similar term proportional to $\lambda^6$, which will
be omitted in the following for simplicity.
The energy denominators in Eq.(\ref{eq:rhont}), corresponding
to the two time orderings,
are given by
\begin{eqnarray}
E_A &=& p_+(P)-p_+(k)-p_+(k_2)-p_+(k_3)
\nonumber\\
E_B &=& -p_+(P)-p_+(k)-p_+(k_2)-p_+(k_3).
\end{eqnarray}

The various contributions to the occupations in the presence
of a particle with momentum $P$ are shown in Fig. \ref{fig:eps}
for a number of values for $\varepsilon/L$.
The numerical values for $m$ and $P$, as well as
the coupling constants $\lambda_4$ and $\lambda_6$,
in the plots are taken to be 1.
\begin{figure}
\unitlength1.cm
\begin{picture}(14,11)(.5,1.5)
\includegraphics{eps.ps}
\end{picture}
\caption{${\cal O}(\lambda^2)$ contributions to the occupation
density $\tilde{\rho(k)}$ in the presence of an excitation
with momentum $P=1$ and mass $m=1$ for various values of
the parameter $\varepsilon/L$.
Dashed line: disconnected vacuum contribution, dotted line: tadpole
contribution, full line: non-tadpole connected (dispersive)
contribution.
}
\label{fig:eps}
\end{figure}
\begin{figure}
\unitlength1.cm
\begin{picture}(14,18)(-1,1)
\includegraphics{eps1.ps}
\end{picture}
\caption{Same as in the previous Figure but for smaller values
of $\varepsilon/L$ and plotted over a logarithmic momentum
scale.
}
\label{fig:eps1}
\end{figure}
Several effects can be observed:
\begin{itemize}
\item The non-tadpole connected (dispersive) contribution
scales in the limit $L \rightarrow \infty$,
$\varepsilon/L \rightarrow 0$. The scaling function
is the LF momentum distribution.
\item Both, the disconnected contribution as well as the
tadpole contribution, are restricted to a region
$k^2 = {\cal O}(\varepsilon/L) \Lambda^2$ near the origin, where
$\Lambda$ is some mass scale ($\Lambda=m$ to lowest nontrivial
order).
\item The integral over the disconnected vacuum contribution is independent
of $\varepsilon/L$.
\item Compared to the vacuum contribution, the tadpole term
is suppressed by one power of $\sqrt{\varepsilon/L}$ and
thus can be neglected as $\varepsilon/L \rightarrow 0$
\end{itemize}
The gap can be most easily observed by plotting the density
over a logarithmic momentum scale (Fig.\ref{fig:eps1}).

For very small values of $\varepsilon/L$, the momentum
distributions from the disconnected diagrams [momentum scale
${\cal O}(\sqrt{\varepsilon/L})$] and from the
dispersive contributions [momentum scale $0.1-1$] no
longer overlap and a gap arises. The disconnected contributions
were already present in the vacuum (ground state for $P=0$)
and are unaltered by the presence of the excitation with
momentum $P$. The only change in occupation within the small
momentum region arises from the tadpoles but its
integrated contribution vanishes as $\varepsilon/L \rightarrow 0$
and becomes negligible in that limit.
Note that $\sqrt{\varepsilon/L}$ must be extremely small
for the gap to be clearly visible.
This makes nonperturbative studies of the gap forbiddingly
difficult numerically
because one would have to cover a huge number of scales
(from $\sqrt{\varepsilon/L}$ to $1$) while keeping the
invariant volume large.

\section{Vacuum Condensates and Sum Rules}
In the previous section we explained that vacuum condensates may
enter the effective (zero-mode free) LF Hamiltonian via induced
coupling constants. The condensates cannot be calculated directly
unless one includes dynamical zero-modes. However, even without
zero-modes, it is possible to calculate at least some of the
condensates indirectly using sum rule techniques. As an example,
let us consider the two point function in a self-interacting
scalar field theory
\begin{equation}
G(x) \equiv \langle 0|\phi(0)\phi(x)|0\rangle
\end{equation}
($x^0<0$, $x^2<0$).
Inserting a complete set of states one obtains
\begin{eqnarray}
G(x) &=& \langle 0|\phi |0 \rangle^2 + \sum_n \int_0^\infty
\frac{dp_-}{2p_-}
\langle 0|\phi (0)|n,p\rangle \langle n,p|\phi (x)|0 \rangle
\\
&=& \langle 0|\phi |0 \rangle^2 + \sum_n \int_0^\infty
\frac{dp_-}{2p_-} \left| \langle 0|\phi (0)|n,p\rangle \right|^2
\exp \left( i(p_-x^-+p_+^nx^+)\right),
\nonumber
\end{eqnarray}
where the sum is over all particle states. The normalization
of the states is
$\langle n,p|m,p^\prime \rangle = 2p_-\delta (p_--p_-^{\prime})
\delta_{nm}$
and the energies are given by the on-shell dispersion relation
$p_+^n=M_n^2/2p_-$. By boost invariance (in the continuum
limit), the vacuum to ``hadron'' matrix elements are independent
of the momentum
\begin{equation}
\langle 0|\phi |n,p\rangle = \frac{g_n}{\sqrt{2\pi}},
\end{equation}
and thus
\begin{equation}
G(x) = \langle 0|\phi |0 \rangle^2 - \sum_n \frac{|g_n|^2}{4\pi}
K_0(M_n\sqrt{-x^2}),
\end{equation}
where $K_0$ is a modified Bessel function \cite{ab:mat}.
In the limit $x^2\rightarrow 0$ one thus finds
\begin{eqnarray}
\langle 0|\phi^2 |0 \rangle -\langle 0|\phi^2 |0 \rangle _{free}
&\equiv & lim_{x^2 \rightarrow 0} \left[
G(x) - G(x)_{free} \right]
\nonumber\\
&=& \langle 0|\phi |0 \rangle^2 + \sum_n \frac{|g_n|^2}{4\pi} \log
\frac{M_n}{M_{free}},
\label{eq:vevphi2}
\end{eqnarray}
where we used $\sum_n |g_n|^2 =1$ and $M_{free}$, $G_{free}(x)$
are the invariant mass and the two point function for noninteracting
fields. Eq. (\ref{eq:vevphi2}) is very interesting because it
allows to calculate $\langle 0|\phi^2 |0\rangle $ in terms of
$\langle 0|\phi |0\rangle $ and quantities ($M_n$ and $g_n$)
which are calculable in a canonical LF calculation without any
dynamical zero-modes.

A similar trick works for the cubic condensates. Of course one has
to be careful to separate the disconnected contributions first
\begin{eqnarray}
\langle 0|\phi(x) \phi(y) \phi(z)|0 \rangle &=&
\langle 0|\phi |0\rangle ^3 +
\langle 0|\phi(x) \phi(y) \phi(z)|0 \rangle ^3_C
\nonumber\\
&+& \langle 0|\phi|0 \rangle \left[
\langle 0|\phi(x) \phi(y)|0 \rangle _C+
\langle 0|\phi(x) \phi(z)|0 \rangle _C \right.
\nonumber\\
& & \quad \quad \quad \quad \quad \quad \quad \quad
+ \left. \langle 0|\phi(y) \phi(z)|0 \rangle _C \right] .
\end{eqnarray}
The connected piece is calculated similar to
$\langle 0|\phi^2|0 \rangle $ by inserting a complete
set of states. In the limit $(x-y)^2 \rightarrow 0$,
$(y-z)^2 \rightarrow 0$ one finds
\begin{eqnarray}
\langle 0|\phi^3|0\rangle
&=& \langle 0|\phi|0\rangle^3
+ 3\langle 0|\phi|0\rangle \langle 0|\phi^2|0\rangle _C
\nonumber\\
& &+\sum_n \int_0^\infty \frac{dp_-}{2p_-}
\langle 0|\phi |n,p_-\rangle \langle n,p_-|\phi^2|0 \rangle_C
\nonumber\\
&=& \langle 0|\phi|0\rangle^3
+ 3\langle 0|\phi|0\rangle \langle 0|\phi^2|0\rangle _C
\nonumber\\
& &+\sum_n \frac{g_n h_n}{4\pi}
\log M_n ,
\label{eq:vevphi3}
\end{eqnarray}
where $h_n \equiv \sqrt{2\pi}\langle 0|\phi^2|n,p_-\rangle _C$
(independent of $p_-$) and $\langle 0|\phi^2|0\rangle _C$
can be taken from above (\ref{eq:vevphi2}). Note that
$\sum_n g_n h_n=0$ because the states
$\int dp_- \exp(ip_-x^-) \phi(x^-) |0\rangle $ and
$\int dp_- \exp(ip_-x^-) \phi^2\!(x^-) |0\rangle $ are orthogonal.
Like $g_n$, $h_n$ can be calculated in a LF calculation without
dynamical zero-modes.

The generalization of these results to higher condensates
is straightforward and by recursion one can express them
in terms of $\langle 0|\phi|0 \rangle $ and matrix elements
which are accessible in a LF calculation. These
matrix elements ($g_n$ and $h_n$) depend on the states and
thus implicitly on the coupling constants in the effective
LF Hamiltonian. Since the coupling constants in the effective
LF Hamiltonian also involve the condensates (\ref{eq:leffeps2}),
this implies that it may be possible to determine the coupling
constants in the effective LF Hamiltonian self consistently.

Similar results may be derived for Yukawa theories.
In Section \ref{renf} we will relate the effective
coupling constants in the LF Hamiltonian to the
spectral densities (\ref{eq:deltam}) which are also accessible
in a LF calculation.

Extracting vacuum condensates from a canonical LF calculation
via sum rules has for example been done in Ref. \cite{zhit}
for the $m_q\rightarrow 0$ quark condensate in
$\mbox{QCD}_2 (N_C \rightarrow \infty)$. The numerical result for
$\bar{\psi}\psi$ was confirmed later in Ref. \cite{wi:vak}
in an equal time framework.
A finite quark mass calculation, based on LF wavefunctions
and sum rule techniques, was first done in Refs.
\cite{mb:phd,mb:paris}. Again the result agreed with the
result from equal time quantization \cite{th:vak}.

\chapter{Perturbative Renormalization}
\label{ren}
In practical applications of LF quantization, such as calculating
parton distributions, nonperturbative effects play a major role.
Nevertheless it makes sense to study renormalization of LF
field theories first from a perturbative point of view because
this allows to resolve some issues which would also appear
in a nonperturbative bound state equation.

Most terms in the perturbation series generated by the
LF  Hamiltonian of QED or
QCD are UV-divergent. This is not very surprising.
After all we have become used to the fact that most
quantum field theories contain divergences. However, as we
will see in the following, the structure of the divergences
in light front perturbation theory (LFPTh) is different
from the divergences in covariant perturbation theory (CPTh).
Because LF quantization is a noncovariant formulation of
field theory, different Lorentz components of a divergent
expression are not necessarily related to each other.
In addition, in many examples the degree of divergence in
LFPTh is worse than in CPTh.

On the one hand this is caused by the choice of
regulators. On a formal level, LFPTh and CPTh are
equivalent \cite{yan:sd2}. However, the ``equivalence proof''
involves steps which are ill defined in the presence
of divergences and singularities. In practice, if
one wants to demonstrate the equivalence
between LFPTh and CPTh, it is very helpful to completely
regularize the theory at the level of the Lagrangian ---
before quantizing. One possibility to do this is Pauli-Villars
regularization, where one can introduce as many regulators as
are necessary to render
the theory free of divergences and light-cone
singularities \cite{yan:sd2,bu:pv}.
Obviously, it is then not difficult to
establish the equivalence between LFPTh and CPTh.
However, for practical applications, Pauli-Villars regularization
is not very useful. On the one hand the Hamiltonian for a
Pauli-Villars regularized theory is either nonhermitian or
unbounded from below or both.
\footnote{The Pauli-Villars ghosts, must be quantized with
the ``wrong'' commutation relations in order to contribute
with opposite signs in loops, which is necessary to cancel
the divergences. The properties of the Hamiltonian then follow
from the spin statistics theorem.} On the other hand, Pauli-Villars
regulators are not very useful for nonabelian
gauge theories, because there one would have to introduce
massive vector fields, which will in general destroy the
renormalizability of nonabelian gauge theories. For these reasons,
one is not interested in employing these regulators in the
context of LF quantization.

For practical applications, it is very useful to use
regulators that are compatible with the kinematic
symmetries
\footnote{These are all Poincar\'e transformations,
which leave the $x^+=0$ initial surface invariant, such as translations,
rotations around  the $z-axis$ or longitudinal boosts.}
of the LF.\footnote{This excludes, e.g. Euclidean lattices.}
In the literature one finds for example the Brodsky-Lepage regulator
\begin{equation}
\sum_i \frac{{\vec k}_{\perp i}^2 + m_i^2}{x_i} <\Lambda^2_{BL},
\label{blcut}
\end{equation}
where the sum extends over all particles and
$x_i = k_{i-}/P_-^{tot} \in (0;1)$ are LF momentum fractions.
Other regulators are a transverse momentum cutoff
\begin{equation}
{\vec k}_{\perp }^2 < \Lambda_\perp^2
\label{perpcut}
\end{equation}
or dimensional regularization in the  transverse direction
\cite{pi:qed,mb:al1}
\begin{equation}
\int d^2k_\perp \rightarrow \int d^{2(1-\epsilon)}k_\perp       .
\label{dimperp}
\end{equation}
Very often it is in addition necessary to introduce
a cutoff for small longitudinal momenta, such as
\begin{equation}
\Theta ( x_i - \delta)
\end{equation}
and/or a cutoff in the number of particles
(Tamm-Dancoff approximation).
What all these regulators have in common is that
they are in general not compatible with Lorentz transformations
that are not kinematic symmetries of the LF (like rotations
around any axis other than the $z$-axis). Thus when using one
of these regulators, one should not be surprised if matrix
elements do not exhibit the full Lorentz invariance ---
unless one compensates for this effect by means of a more
general counterterm structure.
This last point will be the main subject for the rest of
this chapter. The Tamm-Dancoff approximation will be discussed
in more detail in Section \ref{lftd}.

\section{Scalar Fields}
\label{ren:sca}
The following observation is very helpful in analyzing
the perturbative equivalence between CPTh and LFPTh:
Hamiltonian (with $x^0$ or $x^+$ as time)
perturbation theory can be obtained from
covariant perturbation theory after integrating the
energies (i.e. the momentum variable which is canonically
conjugate to the ``time'')
first.
Thus from the mathematical point of view, the
question about equivalence between LFPTh and CPTh has been
reduced to the question whether the order of integration
plays a role in a Feynman integral.

As an example, let us consider the 1-loop self-energy $\Sigma$
in $\phi^3$-theory in 1+1 dimensions (Figure \ref{fig:1lphi3})
\begin{figure}
\unitlength1.cm
\begin{picture}(14,4)(.5,-6)
\includegraphics{1lphi3.ps}
\end{picture}
\caption{1-loop self-energy diagram in $\phi^3_{1+1}$.
}
\label{fig:1lphi3}
\end{figure}
\begin{eqnarray}
\Sigma &=& \frac{ig^2}{2} \int \frac{dk_-dk_+}{(2\pi)^2}
\frac{1}{k^2-m^2+i\varepsilon}
\frac{1}{(p-k)^2-m^2+i\varepsilon}\label{scalvacpo}\\
&=& \frac{g^2}{2}\int \frac{dk_-}{2\pi}
\frac{\Theta(k_-)}{2k_-} \frac{\Theta(p_--k_-)}{2(p_--k_-)}
\frac{1}{p_+-\frac{m^2}{2k_-}-\frac{m^2}{2(p_--k_-)}}.
\label{eq:1lphi3}
\end{eqnarray}
First, without going into the details,
it is easy to convince oneself that Eq.(\ref{eq:1lphi3}) is exactly
what one obtains in LF-Hamiltonian perturbation theory:
$p_+-\frac{m^2}{2k_-}-\frac{m^2}{2(p_--k_-)}$ is the energy
denominator and the $\Theta$-functions ensure that all momenta
are positive. The other factors arise from a vertex factor
proportional to $\left( k_-\left(p_--k_-\right)\right)^{-1/2}$
at each vertex. It is also easy to see that Eq.(\ref{eq:1lphi3})
agrees with the covariant calculation with symmetric integration.
After substituting $k_-=xp_-$ in Eq.(\ref{eq:1lphi3})
one finds
\begin{equation}
\Sigma = \frac{g^2}{2}\int_0^1 \frac{dx}{4\pi}\frac{1}{p^2x(1-x)
-\lambda^2}.
\label{svppara}
\end{equation}
In the covariant calculation one first combines the two
denominators in Eq.(\ref{scalvacpo}) with a Feynman
parameter integral and then one integrates symmetrically
over $d^2k$. This reproduces Eq.(\ref{svppara}) where the
$x$-integration corresponds to the parameter integral.

Our next example will be one where the order of integration
does matter, namely the so called simple tadpole diagram
in $\phi^4$ (for simplicity again in 1+1 dimensions)
\begin{equation}
\Sigma= \frac{ig}{2} \int \frac{d^2k}{(2\pi)^2}
\left(\frac{1}{k^2-m^2+i\varepsilon}
-\frac{1}{k^2-\Lambda^2+i\varepsilon}\right).
\end{equation}
We have already performed a subtraction because the
unregularized integral diverges logarithmically. Symmetric
integration over $d^2k$ yields
\mbox{$\Sigma = \left( g/8\pi \right) \log \Lambda^2/m^2$}.
In LFPTh (unsymmetric integration; $k_+$-integral first)
one obtains zero: for $k_- \neq 0$ one can always close
a contour integral in the complex $k_+$ plane such that
no poles are enclosed. The surface term vanishes because
of the subtraction term. The point $k_-=0$ is usually
omitted in LF quantization without zero-modes.
The mathematical reason for the difference between the
LFPTh result and the CPTh result is a term
$\propto \delta(k_-)$, which is omitted if one (as is usually,
either explicitly or implicitly done)
has a small $k_-$ cutoff, like $\Theta(k_--\varepsilon)$
at each line --- even in the limit $\varepsilon \rightarrow 0$
\cite{ma:zero}.

This result is very typical for pathologies of LFPTh with
scalar fields. Compared to CPTh, one omits certain diagrams
which are nonzero in CPTh, i.e. LFPTh yields {\it a priori}
wrong results! Fortunately (later we will see that there is
a good reason for this) the `mistake' does not depend
on the external momenta. Thus one can make up for the
mistake by means of a local counterterm in the
Lagrangian.
\begin{figure}
\unitlength1.cm
\begin{picture}(14,5)(-18,7.5)
\includegraphics{sg1.ps}
\end{picture}
\caption{Typical generalized tadpole diagrams for $\phi^4$}
\label{fig:sg1}
\end{figure}
\begin{figure}
\unitlength1.cm
\begin{picture}(14,5)(0.2,-7.)
\includegraphics{sg2n.ps}
\end{picture}
\caption{Typical tadpole diagrams arising for scalar fields
with more general polynomial self-interactions}
\label{fig:sg2}
\end{figure}
Other diagrams which suffer from the same problem
are the generalized tadpole diagrams, i.e. diagrams where
part of the diagram is connected with the rest of the
diagram only at one single point (examples are
shown in Figs.\ref{fig:sg1} and \ref{fig:sg2}).
As discussed in Ref.\cite{mb:sg},
they are all zero in LFPTh. However, because the generalized tadpoles in
these diagrams are connected to the
rest of the diagram only at one point, the covariant calculation
yields a momentum independent result for the tadpole part
(just a number), which can thus always be
replaced by a local insertion into the diagram.
In practice this means that the fact that all
generalized tadpoles are (wrongfully) zero on the LF, can
be easily compensated by appropriate redefinitions
of bare coupling constants!
Furthermore, tadpole diagrams are the only diagrams which are
treated incorrectly in naive LFPTh.

A very interesting result is
the relation between the tadpole counterterms and vacuum
condensates \cite{mb:sg}. For example, each tadpole correction to the
propagator in Fig.\ref{fig:sg1}a can be written as a mass
insertion times the free field vacuum expectation value (VEV) of
$\langle 0|\phi^2|0\rangle$. The generalized
tadpole in Fig.\ref{fig:sg1}b
corresponds to a mass insertion times a higher order correction
$\langle 0|\phi^2|0\rangle$. The higher order tadpoles
in Fig.\ref{fig:sg2} correspond to mass (a) and vertex (b) insertions
times a term that contributes to $\langle 0|\phi^4|0\rangle$.
Suppose the interaction term in the original Lagrangian is
\begin{equation}
{\cal L}_{int} = -\sum_n \frac{\lambda_n}{n!} \phi^n
\label{eq:lpol}.
\end{equation}
Then all the `missing tadpoles' are automatically taken into account if one
uses
\begin{equation}
{\cal L}_{int,eff}
= -\sum_n \lambda_n \sum_{k=0}^n \frac{\phi^{n-k}}{(n-k)!}
\frac{\langle 0 | \phi^k | 0 \rangle}{k!}
\label{eq:lpoleff}.
\end{equation}
In other words, ${\cal L}_{int,eff}$ with naive LFPTh yields
the same results as ${\cal L}_{int}$ (\ref{eq:lpol})
(\ref{eq:lpoleff}) with CPTh to all orders
in perturbation theory, if the VEVs in Eq.(\ref{eq:lpoleff})
are also given as a perturbative expansion (calculated in CPTh)
\cite{mb:sg}.

First of all this result is very useful in practice, because,
given the original interaction, it allows one immediately
to write down an ansatz for the effective LF interaction
--- even if the VEVs cannot, in general, be calculated from
the LF Hamiltonian.
Secondly, although derived perturbatively, Eq.(\ref{eq:lpoleff})
formally agrees with Eq.(\ref{eq:leffeps2}),
which was derived nonperturbatively using $\varepsilon$-coordinates.
Of course, while Eq.(\ref{eq:lpoleff}) was derived only for cases
where the VEVs can be calculated perturbatively,
Eq.(\ref{eq:leffeps2}) is valid in general.
However, the formal agreement between the two results
gives us confidence to approach other zero-mode problems
using perturbation theory as well.

As it stands, Eq.(\ref{eq:lpoleff}) is valid only for
superrenormalizable theories because we have only addressed
longitudinal divergences in the above discussion.
For renormalizable theories one must first cut off the
transverse divergences, e.g. by using a transverse lattice
(see Section \ref{lfepmc}) or dimensional regularization in
the transverse direction \cite{pi:qed,mb:al1}.
However, with such a transverse cutoff in place,
Eq.(\ref{eq:lpoleff}) is valid for field
theories in more than one spatial dimension as well.

\section{Fermions}
\label{renf}
\begin{figure}
\unitlength1.cm
\begin{picture}(14,4)(-2,.5)
\put(4,.6){\line(0,1){.2}}
\put(4,1){\line(0,1){.2}}
\put(4,1.4){\line(0,1){.2}}
\put(4,1.8){\line(0,1){.2}}
\put(4,2.2){\line(0,1){.2}}
\put(4,2.4){\line(0,1){1.8}}
\put(4,2.4){\line(1,0){2.}}
\put(5,2.4){\makebox(0,0){$/$}}
\put(6,2.4){\line(0,-1){1.8}}
\put(6,0.6){\vector(0,1){1.}}
\put(4,2.4){\vector(0,1){1.}}
\put(6,2.4){\line(0,1){.2}}
\put(6,2.8){\line(0,1){.2}}
\put(6,3.2){\line(0,1){.2}}
\put(6,3.6){\line(0,1){.2}}
\put(6,4.){\line(0,1){.2}}
\put(1,2.){\vector(0,1){1.5}}
\put(1.4,2.75){\makebox(0,0){$x^+$}}
\end{picture}
\caption{
Seagull diagram in $x^+$ ordered perturbation theory, representing
four-point interactions induced
by eliminating $\psi_{(-)}$. The dashed lines are bosons and
the full lines represent fermions.
The ``slashed'' fermion line
corresponds to instantaneous (with respect to LF-time)
fermion exchange.}
\label{fig:seagull}
\end{figure}
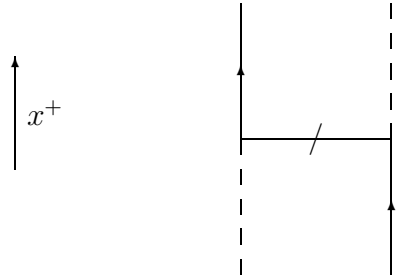

For the applications of LF quantization to DIS, we are of course
not interested in self-interacting scalar fields but rather in
theories with fermions and gauge fields. As a first step towards
this direction, let us consider fermions interacting with
pseudoscalar mesons via a Yukawa coupling (see also
Section \ref{ex:ferm})
\begin{equation}
{\cal L}_{int} = g_p\bar{\psi}i\gamma_5 \psi \chi
\end{equation}
within the framework of LFPTh. First one may be tempted to
expect that abovementioned perturbative zero-mode problem does
not occur here, because {\it a priori}
there are no tadpoles in Yukawa theory with $\gamma_5$-coupling.\\
However, after eliminating the non-dynamical
component of the fermion field ($\psi_{(-)}$) from the theory,
the canonical LF-Hamiltonian (\ref{eq:hyuk})
does contain terms which are fourth order in the fields ---
giving rise so-called ``seagull''-diagrams (Fig.\ref{fig:seagull}).
It is thus not very surprising that
the perturbative zero-mode problem arises in diagrams
which have the topology of a seagull with one vacuum contraction. (Figure
\ref{fig:rainbow}),
because this is the topology one obtains if one replaces
either $\psi_{(+)}^\dagger \partial_-^{-1}\psi_{(+)}$ or $\phi^2$
by their VEVs.
\begin{figure}
\unitlength1.cm
\begin{picture}(14,4.)(0.2,-8.)
\includegraphics{yukself.ps}
\end{picture}
\caption{
Typical self-energy diagrams in Yukawa theory, which have the
same topology as a seagull with one contraction.
The blob stands for arbitrary self energy insertions.}
\label{fig:rainbow}
\end{figure}
In practice, this works out as follows
\cite{mb:al1,mb:al2,mb:al3}: consider, for example,
the dressed
one loop self-energy diagram for a fermion
\footnote{Here we assume self-consistently that all sub-loop
counterterms have been added to the LF result, such that the
full fermion propagator is covariant.}
\begin{equation}
\Sigma(p) = g_P^2\int \frac{d^Dk}{(2\pi )^D}
\int_{0}^\infty d\mu^2
\frac{\rho(\mu^2)}{(p-k)^2-\mu^2 +i\varepsilon}
\int_{0}^{\infty}dm^2
i\gamma_5 \frac{\not \! k \rho_1(m^2) +\rho_2(m^2)}{
k^2-m^2+i\varepsilon}i\gamma_5,
\end{equation}
where the spectral functions ($\rho$, $\rho_1$, $\rho_2$)
parameterize the (unspecified) self-energy insertions.
They satisfy (follows from the canonical commutation
relations) $\int_{0}^\infty d\mu^2 \rho(\mu^2)=
\int_{0}^{\infty}dm^2\rho_1(m^2)=1$.
As far as the $k_+$ integral is concerned, the most singular
term in $\Sigma$ is the one proportional to
$\gamma^+k_+$. We thus consider
\footnote{A more detailed study shows that the other
components are free of trouble \cite{mb:al1,mb:al2}.}
\begin{eqnarray}\! \! \! \! \!
\Sigma^+&=&\frac{tr(\Sigma \gamma^-)}{4}\\
&=&
g_P^2\int \frac{d^Dk}{(2\pi )^D}
\int_{0}^\infty d\mu^2
\frac{\rho(\mu^2)}{(p-k)^2-\mu^2 +i\varepsilon}
\int_{0}^{\infty}dm^2\frac{k_+\rho_1(m^2)}{
k^2-m^2+i\varepsilon}.
\nonumber
\end{eqnarray}
To identify the troublemaker we
eliminate $k_+$ in the numerator using the algebraic identity
\begin{eqnarray}
& &\! \!\frac{k_+}{\left(k^2-m^2+i\varepsilon\right)
\left((p-k)^2-\mu^2+i\varepsilon\right)}=
\nonumber\\
& &\quad \quad \quad \quad \quad \quad \frac{1}{2p_-}
\frac{\left[ 2(p_--k_-)p_+ -{\vec p}_\perp^2+2{\vec p}_\perp
\cdot {\vec k}_\perp
+m^2-\mu^2
\right]}
{\left(k^2-m^2+i\varepsilon
\right)\left((p-k)^2-\mu^2+i\varepsilon\right)}\nonumber\\
& &\quad \quad \quad \quad \quad \quad
-\frac{1}{2p_-}\left[
\frac{1}{\left((p-k)^2-\mu^2+i\varepsilon\right)}
-\frac{1}{\left(k^2-m^2+i\varepsilon\right)}\right].
\label{eq:trouble}
\end{eqnarray}
The important point here is that the last two terms in Eq.
(\ref{eq:trouble}) give $\delta$-functions in $p_--k_-$ and
$k_-$ respectively
after the $k_+$ integration. These $\delta$-functions are missed in
the naive LF Hamiltonian without zero-modes
(this is very similar to the tadpoles in self-interacting
scalar fields). One finds (we subtract here the one-loop
result because this allows to drop the surface term
in the complex $k_-$ plane; $\rho^{free}(\mu^2)=\delta (\mu^2-\mu_0^2)$,
$\rho_1^{free}(m^2)=\delta (m^2-m_0^2)$)
\begin{eqnarray}\! \! \!
\left(\Sigma^+ - \Sigma^+_{1-loop}\right)_{covariant}
&=&
\left(\Sigma^+ - \Sigma^+_{1-loop}\right)_{canonical\ LF}
\nonumber\\
+&&\!\!\!\!\!\!\!\!\!\!\!\!\!\!\frac{g_P^2}{2p_-}
\int \frac{d^Dk}{(2\pi)^D} \int_{0}^\infty d\mu^2
\frac{\rho(\mu^2)-\rho^{free}(\mu^2)}{k^2-\mu^2+i\varepsilon}
\nonumber\\
-&&\!\!\!\!\!\!\!\!\!\!\!\!\!\!\frac{g_P^2}{2p_-}
\int \frac{d^Dk}{(2\pi)^D} \int_{0}^\infty dm^2
\frac{\rho_1(m^2)-\rho^{free}(m^2)}{k^2-m^2+i\varepsilon}.
\end{eqnarray}
Since the other component of $\Sigma$ have no problems
from zero-modes this immediately implies
\begin{eqnarray}
\left(\Sigma - \Sigma_{1-loop}\right)_{covariant}\!
&=&
\left(\Sigma - \Sigma_{1-loop}\right)_{canonical\ LF}
\nonumber\\
+&&\!\!\!\!\!\!\!\!\!\!\!\!\!\!\frac{g_P^2\gamma^+}{2p_-}
\int \frac{d^Dk}{(2\pi)^4} \int_{0}^\infty d\mu^2
\frac{\rho(\mu^2)-\rho^{free}(\mu^2)}{k^2-\mu^2+i\varepsilon}
\nonumber\\
-&&\!\!\!\!\!\!\!\!\!\!\!\!\!\!\frac{g_P^2\gamma^+}{2p_-}
\int \frac{d^Dk}{(2\pi)^4} \int_{0}^\infty dm^2
\frac{\rho_1(m^2)-\rho_1^{free}(m^2)}{k^2-m^2+i\varepsilon}.
\label{eq:deltam}
\end{eqnarray}
This result is very interesting for the following reasons:
\begin{itemize}
\item canonical LF quantization disagrees with covariant
perturbation theory
\item the mistake of canonical LF quantization can
be compensated by a counterterm to the mass term in the
kinetic energy (but not the mass term appearing in the
vertex)
\item if one adds the wrong counterterm, rotational
invariance and parity invariance for physical
observables are broken. This can
be used as a renormalization condition to ``fine-tune''
the coefficient of the counterterm.
\item the counterterm is related in a simple
way to the spectral function of fermions and
bosons which are numerically calculable in a canonical
LF-calculation!
\item The boson contribution in Eq.(\ref{eq:deltam})
can even be expressed in terms of a local VEV:
$\delta \Sigma^{boson} =
\langle 0|:\phi^2:|0\rangle g_P^2\gamma^+/2p_-$. Unfortunately this is not
possible for the term containing the fermionic spectral
density, which would read $\delta \Sigma^{fermion} =
\langle 0|:\bar{\psi} \frac{\gamma^+}{i\partial_-}\psi :|0\rangle
g^2_P\gamma^+/2p_-$.
\end{itemize}
Note that, in order to obtain the full counterterm necessary
to establish agreement between a covariant calculation and a
canonical LF calculation, one still has to add the one loop
counterterm --- but this should be obvious and can be easily
done.
Similar statements hold for fermion loops in the boson
self energy. The only difference to the above example is
that the difference between a covariant calculation and
a canonical LF calculation results in a difference in the
bare boson mass; i.e. no space time symmetries can be used
to fine-tune the counterterm. However, the difference
can still be related to the spectral density of the fermions.
Besides the ``contracted seagulls'', only disconnected
vacuum diagrams --- which are irrelevant for the dynamics
of physical states --- suffer from the zero-mode problem.
It is thus also sufficient to tune the vertex mass and the
kinetic mass independently and those masses and the boson mass
independently from the corresponding coefficients in
the covariant Lagrangian in order to recover equivalence
between covariant calculations and canonical LF calculations.
Note however, that (like in the self-interacting scalar theory)
all this holds only {\it after} rendering the transverse
momentum integrals finite (e.g. by means of dimensional
regularization \cite{pi:qed,mb:al1} or a
transverse lattice (Section \ref{lfepmc}).

It should also be noted that perturbative zero-modes
also play a role in higher-twist parton distributions.
There they can lead to violations of naive sum rules
as discussed in Refs. \cite{mb:del,mb:nag,mb:delta}.

\section{Gauge Theories}
In gauge theories the situation is much less clear
than in scalar field theories or Yukawa theories,
because of notorious infrared singularities in the
LF gauge $A^+=0$. Certain attempts have been made
to perturbatively renormalize LF-QED
\cite{pi:qed,mb:al1,mb:al3} and QCD \cite{hari:zhang}.

In the context of calculations of the electron's
anomalous magnetic moment in QED it has been shown
(up to three loops in Feynman gauge and up to two
loops in LF gauge) that all $k_-\rightarrow 0$
singularities in LFPTh cancel --- provided one
adds up all diagrams that contribute to a given
order in the coupling constant \cite{mb:al3}.
The regulators used were Pauli-Villars regulators or
dimensional regularization in the transverse direction.
Furthermore only two extra \footnote{That is beyond those
counterterms which are required in a covariant calculation.}
counterterms are necessary to render the theory
UV-finite: a kinetic mass counterterm for the electron
(similar to the one discussed in Section \ref{renf})
and a mass term for the transverse photon field.
The numerical result for $(g-2)$ thus obtained agrees
with the known result from covariant calculations.

Perturbative LF calculations of vertex functions, which
employ a Tamm-Dancoff truncation were done in Ref. \cite{pi:qed}
for QED and in Ref. \cite{hari:zhang} for QCD.
Due to incomplete cancelations of $k_-\rightarrow 0$
singularities in the Tamm-Dancoff approximation, infrared
singular counterterm functions were already in lowest
nontrivial order necessary to render the results finite.

The first calculation, relevant for asymptotic freedom,
was performed in Ref. \cite{thorn} (four gluon vertex)
and Ref. \cite{curci} (quark-gluon vertex).
Further discussions on renormalization on QCD in LF gauge (but not
LF quantization) can be found in Ref. \cite{ba:reg}.
For demonstrations of asymtotic freedom, employing both LF gauge and
LF quantization, see Ref. \cite{brasil} (and references therein).

In both types of calculation (LFPTh and LFTD)
even perturbatively the structure of the renormalized LF-Hamiltonian is not
known to higher orders.
Perhaps the cleanest way to address the problem
would be to start from the axial gauge in
$\varepsilon$-coordinates in a finite box
\cite{le:qm,le:qed,le:qcd} and to approach the LF by
carefully taking the limit $\varepsilon \rightarrow 0$
(as in Section \ref{zereps}).
While it is conceivable that this is feasible in QED,
the axial gauge Hamiltonian for $QCD_{3+1}$
in a finite box \cite{le:qcd} is perhaps too complicated
to allow one to study this limit with appropriate care.

\section{Summary}
In the renormalization of LF field theory, one can
distinguish three kinds of counterterms.
First the usual renormalizations, which can be handled
by making the bare coupling constants in the Lagrangian
cutoff dependent. In the following these will be referred
to as canonical counterterms.
Secondly, counterterms that have
to be added when one is employing a Tamm-Dancoff
cutoff. These will be discussed in Section \ref{lftd}.
Typically, one needs an {\it infinite} number of
counterterms!
Third, effects caused by an improper treatment of zero-modes
in the canonical approach. In those cases were these
effects are now understood the renormalization of
zero-mode effects can be accomplished by adding a
{\it finite} number of
counterterms that have the structure of tadpole and seagull
diagrams with some lines ``ending in the vacuum''.
In general, the zero-mode counterterms are already
included in the list of ``Tamm-Dancoff approximation
counterterms''. This means zero-mode effects become
irrelevant when one uses a Tamm-Dancoff approximation.
However, in the absence of a Tamm-Dancoff approximation,
i.e. in calculations
without or with negligible
restrictions on the Fock space (see Section
\ref{lfepmc}) or in perturbation theory if one adds all diagrams
to a given order in the coupling, abovementioned
tadpole or seagull counterterms are quite relevant because
they are the only counterterms needed besides the canonical
counterterms.

\chapter{Nonperturbative Calculations}
\label{num}
\section{Discrete Light-Cone Quantization}
\label{dlcq}
The most straightforward method for solving bound state
problems in the context of LF quantization is discrete
light-cone quantization
\footnote{Like the canonical quantization
discussed in Chapter \ref{canoni}, the quantization surface
in DLCQ is the plane $x^+=0$, i.e. a front or plane
--- and not a cone. Thus discrete light {\it front}
quantization (DLFQ) would be a more appropriate terminology.
However, because of historical reasons, the method has been named
DLCQ in the literature.}
(DLCQ) \cite{pa:dlcq}. For extensive reviews and more references
see Refs. \cite{schladming,challenge,world}.

The basic idea in DLCQ is as follows (for simplicity we
illustrate the method using the example of $\phi^4_{1+1}$).
One puts the system into an $x^-$-box of length $L$
with periodic or antiperiodic boundary conditions
\footnote{In the presence of interactions which contain
odd powers of $\phi$ one has no choice and one must use
periodic boundary conditions --- otherwise momentum
conservation is violated at the boundary!}
\begin{equation}
\phi(x^-+L,x^+)=\pm \phi(x^-,x^+).
\end{equation}
In the following, antiperiodic boundary conditions will be used,
which implies for the mode expansion
\begin{equation}
\phi(x^-) = \frac{1}{\sqrt{4\pi}} \sum_{k=1}^{\infty}
\frac{ \left[ a_k e^{-ip_-^kx^-}+a_k^\dagger
e^{ip_-^kx^-}\right]}
{\sqrt{k-\frac{1}{2}}},
\label{eq:modeex}
\end{equation}
where
\begin{equation}
p_-^k = \frac{2\pi}{L} \left( k- \frac{1}{2} \right).
\end{equation}

The main reason for choosing antiperiodic boundary conditions
is that one does not have to worry about the mode with $p_-=0$.
Another reason is that many numerical problems converge faster
when antiperiodic boundary conditions are used
(compared to periodic boundary conditions with the
\mbox{$p_-=0$} mode left out).
This can be understood in perturbation theory
because there are often non-negligible
contributions to Feynman integrals from the region near $p_-=0$.
Let $f$ be some typical function that appears as the
argument of some Feynman integral.
Then
$\ \varepsilon \sum_{n=-\infty}^{\infty} f\left( (n-\frac{1}{2})\varepsilon
\right) \ $ is usually a better approximation to
$\ \int_{-\infty}^{\infty} f(x)\ $
than
$\ \varepsilon \sum_{n=-\infty}^{-1} f\left( n\varepsilon \right)
+\varepsilon
\sum_{n=1}^{\infty}f\left( n\varepsilon \right) \ $ because in the
latter expression the point $n=0$ is missing compared
to the trapezoidal quadrature formula.

\noindent
In order for $\phi(x)$ to satisfy the canonical commutation relations
(see Chapter \ref{canoni}),
\begin{equation}
\left[ \partial_-\phi(x), \phi(y)\right]_{x^+=y^+}
= -\frac{i}{2}\delta(x^--y^-)
\end{equation}
we impose the usual commutation relations for the coefficients
$a_k$,
\begin{equation}
\left[a_k,a_q^\dagger\right] = \delta_{kq}.
\end{equation}
The above expansion is then inserted into the momentum
operator
\begin{eqnarray}
P_- &=& \int_0^L dx^- :\partial_-\phi \partial_-\phi:
\nonumber\\
&=& \frac{2\pi}{L} \sum_{k=1}^{\infty} a_k^\dagger a_k
\left( k - \frac{1}{2} \right)
\end{eqnarray}
and the Hamiltonian
\begin{eqnarray}
P_+ &=& \int_0^L dx^- \frac{m^2}{2}:\phi^2:
+ \frac{\lambda}{4!}:\phi^4:
\nonumber\\
&=& \frac{L}{2\pi}\left(T+V\right),
\end{eqnarray}
where
\begin{equation}
T=\frac{m^2}{2} \sum_{k=1}^{\infty}
\frac{a_k^\dagger a_k}{k - \frac{1}{2}}
\end{equation}
is the kinetic term and
\begin{equation}
V = \frac{\lambda \delta_{P_f P_i}}{8\pi 4!}
\sum_{k_1,k_2,k_3,k_4=1}^{\infty}
\frac{:\left(a_{k_1}^\dagger +a_{k_1}\right)}
{\sqrt{k_1-\frac{1}{2}}}
\frac{\left(a_{k_2}^\dagger +a_{k_2}\right)}
{\sqrt{k_2-\frac{1}{2}}}
\frac{\left(a_{k_3}^\dagger +a_{k_3}\right)}
{\sqrt{k_3-\frac{1}{2}}}
\frac{\left(a_{k_4}^\dagger +a_{k_4}\right):}
{\sqrt{k_4-\frac{1}{2}}}
\end{equation}
is the interaction term. $\delta_{P_fP_i}$ is a momentum conserving
Kronecker $\delta$.
Since the length of the box completely factorizes, it is useful
to work with the rescaled operators
\begin{eqnarray}
K&=&P_-\frac{L}{2\pi}\\
H&=&P_+\frac{2\pi}{L}.
\end{eqnarray}
Since the momenta of all excitations are discrete and positive,
the Fock space is finite dimensional for all $K$. Thus, at least in
principle, one can now proceed as follows: for fixed $K$ ($K$ and
$H$ commute) one diagonalizes $H$ (which is a finite matrix for
finite $K$). From the eigenvalues $E_i$ one computes the invariant
masses $M^2_i=2KE_i$ and from the eigenstates one can compute other
physical observables (like parton distributions). In general,
physical observables thus computed will of course depend on
the ``resolution'' $K$. The continuum limit is obtained by
extrapolating to $K\rightarrow \infty$. The diagonalization
is generally done using brute force matrix diagonalization
or, if one is only interested in the lowest states,
using the Lanczos algorithm \cite{lanczos}.

At this point one encounters a problem that is inherent to
Hamiltonian systems: {\it the dimension of multi-particle states
in the Fock space expansion grows exponentially with the
number of particles}. The number of particles, as well as the
number of states for a single particle are both limited by
the longitudinal momentum $K$, i.e. the dimension of the
Fock space basis shows factorial growth with $K$.
Fortunately, in $1+1$ dimensional examples, the factorial
growth sets in only rather slowly and numerical convergence
for typical observables
can be obtained before the size of the matrices becomes a problem.
DLCQ was enormously successful in many $1+1$ dimensional
field theories
\cite{pa:qed,hari,el:qed,empty,mb:deu,ho:sea,mb:phd,mb:sg,bu:dipl,mb:rb}.
In all cases, where results from other approaches to field
theories were available agreement could be shown within
numerical uncertainties
($\mbox{QED}_{1+1}$: \cite{co:bos} vs.
\cite{pa:qed,el:qed}
\footnote{However, there is still a 1\% difference in the
fundamental meson mass
for the term linear in $m_e$ in
$\mbox{QED}_{1+1}$ as calculated
from bosonization \cite{co:bos} and in LF-quantization
\cite{be:qed}. It is not clear whether this deviation is due
to the finite Fock space truncation or whether this is a real
problem \cite{qed:dipl1,qed:dipl2}.}
,$\mbox{QCD}_{1+1}$ \cite{ha:qcd}
vs. \cite{ho:sea,mb:phd}, sine-Gordon model: \cite{sg:exact}
vs: \cite{mb:sg}). Beyond reproducing known results, DLCQ
has been used to calculate new and interesting results
in $\mbox{QCD}_{1+1}$: the most notable results
are the existence of a nucleon-nucleon bound state and the
analysis of the nuclear quark distribution in comparison
with the nucleon quark distribution. Not only exhibits the
$1+1$ dimensional ``deuteron'' an EMC-effect, but it can
also be understood analytically due to the simplified
dynamics in $1+1$ dimensions \cite{mb:deu}.
Typical Euclidean lattice calculations are too ``noisy''
to even demonstrate binding of hadrons.
Another remarkable result from DLCQ calculations in
$\mbox{QCD}_{1+1}$ dimensions is ``Anti-Pauli-Blocking''
\cite{bu:dipl,mb:rb}: contrary to the naive expectation, sea quarks
in nucleons in $\mbox{QCD}_{1+1}$ tend to have the same flavor as
the majority flavor among the valence quarks (i.e. more
$\bar{u}$ than $\bar{d}$ in a nucleon $\psi_{valence} = uud$).

In $2+1$ or $3+1$ dimensions the situation changes drastically,
because there the exponential growth is much more rapid. The basic
reason is that there are now transverse degrees of freedom
besides the longitudinal degrees of freedom.
Suppose that each particle can occupy $N$ states for each
spatial dimension. Then the Fock space basis size grows like
$N^{3N_{part}}$ with the number of particles in $3+1$ dimensions,
while the corresponding growth would be only $N^{N_{part}}$
in $1+1$ dimensions. For a concrete example ($\phi^4$ with
antiperiodic boundary conditions in the longitudinal direction)
this works out as follows. For a longitudinal momentum
$K=\frac{15}{2}$ (8 longitudinal momentum states accessible)
the Fock space basis size is 27. If one has just two transverse
degrees of freedom (e.g. two points in the transverse direction)
the basis size grows to 426. For $8\times 8 =64$ degrees of
freedom in the transverse direction, that number grows to
$6 \cdot 10^{15}$. These astronomical numbers clearly demonstrate
that any direct matrix diagonalization approach or even a
Lanczos type algorithm is doomed to fail because one is not
even able to store the wavefunction in a computer \cite{mb:lfepmc}.

The most simple (and perhaps most drastic) way out of this
dilemma is to impose additional cutoffs, like restricting the
number of particles. Typically, this means restricting the
Fock space to 3 (perhaps 4) particles or less \cite{kaluza,wo:93}.
In QED, since the coupling is small, this is a good
approximation. However, to the same order in $\alpha$ within
which the 3 particle truncation is a good approximation one can
calculate the parton distributions analytically \cite{mb:pho}.
That is, even in QED there is not much point in doing numerical
DLCQ calculations with Fock space truncations to the lowest
nontrivial order! In QCD, where one faces an intrinsically
strong coupling problem, restricting the Fock space to the
lowest nontrivial component seems entirely useless. For example,
even if one allows up to 4 particles (which is about the maximum
that can be handled numerically using the Lanczos algorithm),
this means one allows at most one gluon in addition to the
three valence quarks in a proton. That is there is no chance one
can ``see'' any effects from nonlinear gluon-gluon couplings.
\footnote{A caveat to this pessimistic point of view
will be discussed in Section \ref{lftd}.}

It should be emphasized, that this  problem is not specific
for LF field theories, but occurs in many Hamiltonian approaches
to field theory --- and in many cases could be solved.
Thus there are many numerical methods available which
can potentially be useful in overcoming the difficulties
associated with exponential basis size growth.

\section{Functional Integration on a \mbox{Longitudinal Lattice}}
Functional integrals on Euclidean lattices have been
very successful in solving ground state properties
of QCD (e.g. vacuum properties, hadron masses
and ground state matrix elements).
However, since two points on a Euclidean lattice are always
separated by a space-like distance, it is only
very indirectly possible to extract information about
light-cone correlation functions from these calculations.
Of course, this is because in conventional Euclidean field
theory $\exp(-\beta P^0_E)$ is used to project on the
ground state wave function of $P^0$ at equal time.
Thus as a caveat one might be tempted to consider a similar
formalism for LF Hamiltonians.
Suppose one discretizes the $x^-$ direction,
\footnote{The transverse coordinates are irrelevant in this
argument.}
and uses a functional integral to project on the
ground state of $P_+$ \cite{mu:lat,su:lat}.
This results in an immediate problem
because the LF-energy {\it decreases} with increasing momentum
($P_+ = M^2/2P_-$ in the continuum, on a lattice
there is a minimum for $P_- = {\cal O}(1/a))$. Due to Bragg reflections,
momentum is not conserved and the particles tend to accumulate
near the minimum. However, since the momentum near that minimum
is of the order of the inverse lattice spacing, the particles
always ``see'' the lattice and no meaningful continuum limit
is obtained. It is conceivable that this problem can be cured
by adding a Lagrange multiplier proportional to the total
LF momentum to the lattice action (in the continuum limit
this amounts to minimizing
$\tilde{P}_+ = P_+ + \lambda P_-$ instead of $P_+$). However,
this idea will not be investigated here any further.
Another difficulty of the longitudinal LF-lattice is species doubling for
bosons \cite{mu:lat}!

\section{Hamiltonian Monte Carlo on a Transverse Lattice}
\label{lfepmc}
\begin{figure}
\begin{Large}
\unitlength.8cm
\begin{picture}(14,8)(1,5.5)
\put(1.5,8.5){\line(0,1){1.7}}
\put(1.,11.1){\makebox(0,0){(discrete)}}
\put(1.,12.0){\makebox(0,0){$\perp$ space}}
\put(1.5,12.6){\vector(0,1){1.6}}
\put(2.,8.){\line(3,-1){1.2}}
\put(4.5,6.9){\makebox(0,0){long. space}}
\put(4.5,6.2){\makebox(0,0){(continuous)}}
\put(7.1,6.3){\vector(3,-1){1.8}}
\put(10.4,5.8){\line(3,1){1.8}}
\put(14.5,6.2){\makebox(0,0){(continuous)}}
\put(14.5,6.9){\makebox(0,0){time}}
\put(15.8,7.6){\vector(3,1){1.8}}
\includegraphics{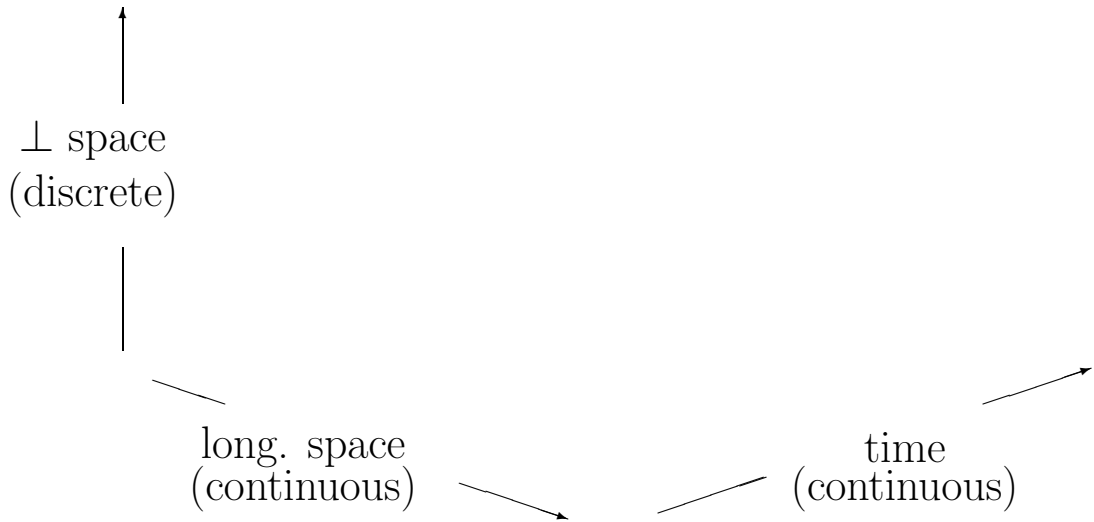}
\end{picture}
\end{Large}
\caption{Space time view of a transverse lattice}
\label{fig:perpl}
\end{figure}

While Mont Carlo calculations for
longitudinal LF-lattices seem to be plagued with
difficulties, this is not the case for
the transverse lattice \cite{mb:lfepmc}.
On a transverse lattice one keeps the longitudinal
directions ($x^+$ and $x^-$) continuous, while discretizing
the transverse coordinate (Fig. \ref{fig:perpl}) \cite{bardeen}
For simplicity, let us consider self-interacting scalar
fields in 2+1 dimensions on such a transverse lattice characterized by the
action
\begin{equation}
A^{cont.} = \int d^3x \left[
\sum_{\mu=0}^2 \frac{\partial_\mu \phi
\partial^\mu \phi}{2} - \frac{m^2}{2} \phi^2 - {\cal L}_{int}(\phi)
\right].
\end{equation}
Upon discretizing the transverse direction (spacing $a$) one thus
obtains
\begin{equation}
A^{\perp \ latt.} =
a \sum_n \int d^2x\left[ \sum_{\mu=0}^1 \frac{\partial_\mu \phi_n
\partial^\mu \phi_n}{2} -
\frac{\left(\phi_{n+1} - \phi_n\right)^2}{2a^2}-
\frac{m^2}{2} \phi^2_n - {\cal L}_{int}(\phi_n)
\right].
\label{eq:aperpl}
\end{equation}
Up to a factor of $a$ (which can be absorbed into a redefinition
of the field $\phi_n$), Eq.(\ref{eq:aperpl}) looks like the
action for a multi-flavor theory in $1+1$ dimensions
(where the site index $n$ corresponds to the ``flavor'' index).
In the next step one constructs the DLCQ Hamiltonian
for this  ``multi-flavor'' $1+1$ dimensional theory. The important
point here is that the action is local in the transverse
direction, i.e., there are only nearest neighbor interactions.
Since the DLCQ-Hamiltonian is thus also local,
\begin{equation}
H_{DLCQ} = \sum_n \left[ H_n + V_{n,n+1}\right],
\label{eq:hloc}
\end{equation}
one can apply many Monte Carlo techniques which have been
developed for other Hamiltonian systems
(see e.g. Ref. \cite{epmc2}). One technique which turns
out to be particularly useful for LF-Hamiltonians on
a transverse lattice is
the ensemble projector Monte Carlo technique
\cite{epmc1} based on the so called checkerboard
decomposition of the Hamiltonian \cite{pmc1}. Using locality
of the Hamiltonian one can write
\begin{equation}
H_{DLCQ} = H_a+H_b
\end{equation}
where
\begin{eqnarray}
H_a&=& \sum_{n=1,2,3,...} \frac{H_n}{2}
+ \sum_{n=1,3,5,...} V_{n,n+1}
=H_{1,2}+H_{3,4}+...\nonumber\\
H_b&=& \sum_{n=1,2,3,...} \frac{H_n}{2}
+ \sum_{n=2,4,6,...} V_{n,n+1}
=H_{2,3}+H_{4,5}+...
\label{eq:hab}
\end{eqnarray}
In other words, the DLCQ Hamiltonian can be written as
a sum of two terms, each of which can be written
as a direct sum of two-site-Hamiltonians.
The point to all this is that while the dimension of the
space on which $H_{DLCQ}$ acts is astronomical, the
two-site-Hamiltonians act only on a very small Hilbert space
(for our above example with $K=\frac{15}{2}$
and say 16 transverse sites:
$dim(H_{DLCQ})=7.8\cdot 10^{8}$ but $dim(H_{n,n+1})=426$).
The method is called checkerboard algorithm
because one approximates the time evolution operator
of the system by
alternating infinitesimal time evolution operators
generated by $H_a$ and $H_b$ respectively
\begin{equation}
e^{-\varepsilon H_{DLCQ}}=e^{-\frac{\varepsilon}{2} H_a}
e^{-\varepsilon H_b}e^{-\frac{\varepsilon}{2} H_a}
+{\cal O}(\varepsilon^3).
\label{eq:epsevol}
\end{equation}
If one axis of the checkerboard is the discretized
space direction and the other the time, Eq. (\ref{eq:hab})
can be interpreted as if interactions between sites occur
only across the black squares \cite{pmc1}.

Before we explain how the infinitesimal time evolution
operators are multiplied together, let us pause here
for a moment and understand the advantage of the
transverse lattice with DLCQ over the longitudinal
lattice discussed in the previous Section.
The main cause of the problem in the previous
Section was lack of longitudinal
momentum conservation. In DLCQ $P_-$, the longitudinal momentum,
is manifestly conserved. Furthermore, $P_-$ is just the
sum of momenta at each site
\begin{equation}
P_- = \sum_n P_-^n,
\end{equation}
and
the checkerboard algorithm is compatible with with longitudinal
momentum conservation ($[H_a,P_-] = [H_b,P_-]=0$). I.e.
with DLCQ on a transverse lattice,
longitudinal momentum is conserved at each step of the calculation.
Hence, one can minimize $P_+$ while keeping $P_-$
manifestly fixed and
there are no ``runaway solutions''.

In the actual calculations one uses Monte Carlo techniques
to calculate
$\left( e^{-\varepsilon H}\right)^N|\psi_i(K)\rangle $
by alternate application of
$e^{-\frac{\varepsilon}{2} H_a}$, $e^{-\varepsilon H_b}$,
$e^{-\varepsilon H_a}$,...,$e^{-\frac{\varepsilon}{2} H_a}$
to $|\psi_i(K)\rangle$. Here $|\psi_i(K)\rangle$ is an initial guess
for the ground state wave function with longitudinal
momentum $K$. For $N \rightarrow \infty$ one thus obtains
an approximation (because $\varepsilon$ is finite, the result
is not exact) to the ground state hadron with the same
good
\footnote{Of course, only those quantum numbers which are
associated with exact symmetries of DLCQ on a transverse
lattice (like C-parity or baryon number in QCD) are relevant here.}
quantum numbers as $|\psi_i(K)\rangle $.
One very useful technique is the ensemble projector Monte Carlo
method \cite{epmc2,epmc1}, which works as follows for these systems:
\begin{itemize}
\item[1)] let $|n\rangle$
be a complete set of states\\ (here product
basis of Fock state bases at each site)
\item[2)] make a good guess for $|\psi_i(K)\rangle$\\
(here a valence state with $P_\perp =0$ :
$\ |\psi_i(K)\rangle = \sum_{n_\perp} a^\dagger_{K,n_\perp}|0\rangle$)
\item[3)] start from ensemble $|i_\nu^{(0)}\rangle$ of states (from set
$|n\rangle$)
\item[4)] for each $|i_\nu^{(0)}\rangle$ select a new state
$|i_\nu^{(1)}\rangle$
with probability
$$W(i_\nu^{(1)},i_\nu^{(0)}) = \frac{
|\langle i_\nu^{(1)}| e^{-\frac{\varepsilon}{2}H_a}|i_\nu^{(0)}\rangle |}
{\sum_n
|\langle n| e^{-\frac{\varepsilon}{2}H_a}|i_\nu^{(0)}\rangle |}$$
and calculate the score
$$ S_\nu^{(1,0)}=\frac{
\langle i_\nu^{(1)}| e^{-\frac{\varepsilon}{2}H_a}|i_\nu^{(0)}\rangle }
{W(i_\nu^{(1)},i_\nu^{(0)})}$$
note: $W(i_\nu^{(1)},i_\nu^{(0)})$ factorizes into
two-site probabilities\\
$\hookrightarrow$ local (one pair of sites at a time)
``updating'' possible
\item[5)] replicate states with multiplicity:\\
$int \left[ \frac{|S_\nu^{(k,k-1)}|}{\bar{S}} +
\mbox{random number}
\in (0;1) \right]$\\[1.5ex]
($\bar{S}$: av. score)\\[1.5ex]
Thus paths with large scores
contribute with multiple weight, while paths with small scores
get eliminated.
\item[6)] repeat while alternating $H_a$ and $H_b$\\
($\frac{1}{2}$ in exponent only in $1^{st}$ and last step!)
\item[7)] observables from ensemble average, e.g., energy of
ground state hadron:
$$E_0 = \lim_{N \rightarrow \infty}
\frac{ \sum_\nu \langle \psi_f(K)| H |i_\nu^{(N)}\rangle
sign \left[ S_\nu^{(N,N-1)}
\cdot ... \cdot S_\nu^{(1,0)}\right] }
{ \sum_\nu \langle \psi_f(K)|i_\nu^{(N)}\rangle
 sign \left[S_\nu^{(N,N-1)}
\cdot ... \cdot S_\nu^{(1,0)} \right]}
$$
other observables (e.g. an observable diagonal in the basis)\\[1.5ex]
$
\langle \psi_0(K)| \hat{O} |\psi_0(K)\rangle=
$
$$
\lim_{M,N \rightarrow \infty}
\frac{ \sum_\nu \langle \psi_f(K)|i_\nu^{(N+M)}\rangle
sign \left[S_\nu^{(N+M,N+M-1)}
\cdot ... \cdot S_\nu^{(1,0)} \right]
\langle i_\nu^{(N)}|\hat{O}|i_\nu^{(N)}\rangle}
{ \sum_\nu \langle \psi_f(K)|i_\nu^{(N+M)}\rangle
sign \left[S_\nu^{(N+M,N+M-1)}
\cdot ...\cdot S_\nu^{(1,0)} \right]}
$$
\end{itemize}
In this Monte Carlo procedure, one only has to store the
ensemble of states at one ``timeslice'' plus the result of the
measurement of the observable after $N$ slices. Thus, at least
in principle, one can handle very large lattices.
The main advantages of the transverse lattice are as follows
\cite{mb:lfepmc}
\begin{itemize}
\item longitudinal momentum is manifestly conserved $\rightarrow$
no runaway solutions
\item parton distributions are diagonal in the DLCQ-basis
\item LF-vacuum is trivial $\rightarrow$ no statistical fluctuations
from updating the vacuum far away from physical states on huge
lattices.
\item species doubling for fermions occurs only for the latticized
transverse dimensions $\rightarrow$ can be easily compensated by
staggering \cite{pauldoubl}.
\item excited states are suppressed by the square of their
masses:\\
\mbox{$\exp(-N\varepsilon P_+)=\sum_n |n\rangle \exp(-N\varepsilon
M_n^2/2P_-) \langle n|$}
instead of\\
\mbox{$\exp(-N\varepsilon P^0) =
\sum_n |n\rangle \exp(-N\varepsilon M_n)\langle n|$}
which one encounters in a conventional Hamiltonian formulation.
\end{itemize}
It is interesting to see how confinement emerges on the transverse lattice
in the limit of large lattice spacing:
In this limit, the coupling between the sheets is weak and the
energy scale associated with link field excitations is high.
Therefore, when one separates two test charges in the longitudinal
direction,
the transverse lattice behaves similar to QCD$_{1+1}$ and linear
confinement results trivially. For transverse separations between the
charges, a different mechanism is at work. Gauge invariance demands
that the two charges are connected by a string of link fields.
In the limit of large spacing the link fields fluctuate only little
and the energy of such a configuration can be estimated by counting the
number of link fields needed to connect the charges, which again yields
linear confinement.

Some of the disadvantages of the transverse lattice are:
Since $x^+ \rightarrow ix^+$
is {\it not} a Wick rotation (it is just a mathematical trick
to project on the ground state of $P_+$), the metric is not
Euclidean and thus propagators oscillate. Hence,
negative scores occur already for bosons
which leads to an increase in the statistical
fluctuations. However, these negative scores turn
out to have only a small statistical weight and the
resulting ``sign-problem'' is not serious.
Very often in LF calculations, large cancellations
occur between different terms in the Hamiltonian.
For example, the instantaneous photon exchange has a
$1/q_-^2$ singularity which is canceled by vertex factors
in photon exchange. In general, it is difficult to obtain such
cancellations from a Monte Carlo calculation.
Another difficulty is that gauge invariance on a lattice can
only be maintained if one introduces link fields. On a
transverse lattice this amounts to introducing
$1+1$-dimensional gauged nonlinear sigma model fields on
each link \cite{bardeen}. Constructing a Fock space basis
out of these nonlinear degrees of freedom and calculating
appropriate matrix elements is a nontrivial task
\cite{paul,pg:zako}.

The sign problem associated with fermions is a
notorious difficulty for Monte Carlo algorithms:
due to the minus sign in exchange terms, the
infinitesimal time evolution operator tends to
contain many negative matrix elements. This very
general problem is also expected to afflict
Mont Carlo calculations on transverse lattices. However,
since the LF vacuum is trivial, there are no sign fluctuations from Z-graphs
and vacuum diagrams.
Thus one expects that the sign problem on the LF
is less severe than usual. Whether this improvement is
sufficient to render fermions tractable on
transverse lattices has not yet been investigated.

Obviously, the transverse lattice lacks manifest rotational
invariance which must be restored in the process of renormalization.
Recently, a technique has been described that allows
easy computation of the potential between infinitely
heavy quarks in a LF framework \cite{mb:zako}. Demanding
rotational invariance for this observable may prove to
be a powerful tool in such a procedure.

\section{Light-Front Tamm-Dancoff}
\label{lftd}
As we have discussed already in Section \ref{dlcq},
the dimensionality of the Fock space grows
dramatically as one includes higher Fock components.
Clearly, since $\alpha_S$ is fairly large at a low
momentum scale, a numerical solution of bound state
problems in QCD (which includes all scales) necessarily
involves many Fock components. In this chapter we will
discuss the light-front Tamm-Dancoff (LFTD) approach to
LF problems (for a comprehensive review see Ref.\cite{all:lftd}).
The basic idea is very simple \cite{phw:lftd,wi:zako}: Hadrons
are complicated objects only if one tries to build them
in terms of bare quarks and gluons whose masses and couplings
are renormalized at a scale of 1 GeV or higher. In terms
of collective excitations (constituent quarks) ground state
hadrons are
rather simple. One of the problems with the constituent quark model
is that {\it a priori} the interactions among the quarks
are {\it ad hoc}.

The goal of LFTD field theory is to systematically eliminate
higher Fock components and high energy degrees of
freedom\footnote{This procedure is explicitely demonstrated for the simple
example of $\phi^4_{3+1}$ in the two particle sector in Ref. \cite{edsel}.}.
As one goes to lower and lower scales the interaction
between the (dressed) constituents thus becomes more and more
complicated. If the whole program is successful, constituent
quarks will emerge as the quasiparticles of QCD at intermediate
energies. A major virtue of using LF quantization in this
approach is that it stays close both to physical intuition
(which may prove very helpful when it comes to developing
variational methods to analyse the Hamiltonian)
as well as to experimental observables at large
momentum transfer (useful
for phenomenological applications).

A systematic Fock space expansion, based on a
Hamiltonian formulation, for field theory was
originally developed by Tamm \cite{ta:td} and independently
by Dancoff \cite{da:td}. It turns out that such an approach
is doomed to fail if the perturbative ground state
(the starting point of the expansion) is too far from
the actual ground state. In such a situation one needs
very (or infinitely)
complicated Fock states just to build the ground state.
LF quantization is advantageous at this point, because the
vacuum of a LF Hamiltonian is trivial.

In fact, because of the vacuum, LF quantization is probably
the {\it only} framework, where such a programm can
possibly work.

In practice, even within LF quantization, it is of course not
possible to integrate out high energy states and higher
Fock states exactly. Instead one writes down a catalog of
all interaction terms that are allowed by power counting
\cite{wi:vid,osu:rg}: on the LF there are two length scales
$x^-$ and $x_\perp$. The engineering dimensions of the various
operators and terms that enter the LF-Hamiltonian
in $3+1$ dimensions can be easily derived
from free field theory \cite{osu:rg}
($\phi$ stands for a scalar field or for $A_\perp$ --- the
transverse gauge field components which have
the same engineering dimension as scalar fields;
$\psi^{(+)}$ is the dynamical component of a fermion field)
\begin{equation}
\partial_- = \left[ \frac{1}{x^-}\right],\
\partial_\perp = \left[ \frac{1}{x_\perp}\right],\
m = \left[ \frac{1}{x_\perp}\right]
\nonumber
\end{equation}
\begin{equation}
\psi^{(+)} = \left[ \frac{1}{x_\perp \sqrt{x^-}}\right],\
\phi = \left[ \frac{1}{x_\perp}\right].
\end{equation}
The Hamiltonian and the Hamiltonian density have dimensions
\begin{equation}
H = \left[ \frac{x^-}{x_\perp^2}\right],\
{\cal H} = \left[ \frac{1}{x_\perp^4}\right].
\end{equation}
Thus all allowed terms without fermion fields are \cite{osu:rg}
\footnote{See the discussion in Ref.\cite{osu:rg} why terms
with negative powers of $m$ are excluded.}
\begin{equation}
m^3\phi,\ m^2\phi^2,\ \partial_\perp^2\phi^2,\ m\phi^3,\ \phi^4.
\end{equation}
Including fermion fields one obtains \cite{osu:rg}
\begin{equation}
\frac{1}{\partial_-} \psi^{(+)\dagger} \Gamma
\left\{ m^2,\partial_\perp^2,m {\vec \gamma}_\perp {\vec \partial}_\perp,
m\phi, \phi^2 \right\} \psi^{(+)},
\end{equation}
\begin{equation}
\frac{1}{\partial_-^2}
\psi^{(+)\dagger} \Gamma_1 \psi^{(+)}\psi^{(+)\dagger} \Gamma_2 \psi^{(+)}
\label{eq:coul}
\end{equation}
($\Gamma$, $\Gamma_1$ and $\Gamma_2$ are some Dirac matrices).
Unfortunately, this is not the whole story. Already free
LF-field theory is nonlocal in the longitudinal direction
[e.g. for scalar fields because of the fundamental commutator
$\left[ \phi(x^-,x^+), \phi(y^-,x^+)\right] =
-\frac{i}{4}\varepsilon ( x^--y^-)$
and for fermions because  an inverse
derivative of $\partial_-$ appears in the kinetic energy term
(\ref{eq:hyuk})]. Thus longitudinal locality is no
longer a restriction on the functional form of the possible terms
in the Hamiltonian. As a result, any of the operators in the above
catalog may be multiplied by arbitrary functions of ratios of
longitudinal momenta! In fact, there are examples known where
complicated functions of ratios of incoming and outgoing
momenta, multiplying a four fermion counterterm,
are necessary to cancel UV divergences \cite{osu:nonloc}.

As a result of this infinitely complicated counterterm structure,
it seems one looses predictive power and
one thus might be forced to abandon LFTD as a {\it fundamental}
theory of hadrons.
It should be emphasized that, even if one does
have to abandon LFTD as a fundamental theory, it might still
have many virtues in parton phenomenology.
However, there has been recent progress towards understanding
how to apply renormalization group techniques to LFTD that
may help restore its predictive power \cite{all:lftd}
(for an excellent pedagogical review, see Ref. \cite{brasil}).

Unless one works with the full Hamiltonian, nonperturbative
bound state calculations in QCD almost inevitably violate
gauge invariance \footnote{Lattice gauge theory being
the only exception.}. Therefore, if one wants to derive
a constituent picture from QCD, one is forced to allow
that explicit gauge invariance is violated:
{\it gauge invariance becomes a hidden symmetry} \cite{all:lftd,brasil}.
In LFTD one introduces cutoffs that violate symmetries
which normally prevent a constituent picture from
arising (gauge invariance and full Lorentz invariance).
In a sense, the counterterm functions
that complicate renormalization offer a possible resolution
of apparent contradictions between the constituent picture and
QCD \cite{brasil}.

The technique to remove cutoff dependence from
physical results is renormalization: for example,
the functions of momentum fractions that
appear in the relevant and marginal operators
can be fixed by demanding covariance and gauge invariance in
physical observables. An alternative way to
fix the marginal and relevant counterterms is
{\it coupling constant coherence} (CCC): one insists
that functions appearing in
non-canonical relevant and marginal operators are not
independent functions of the cutoff, but depend on the cutoff
implicitly through their dependence on canonical couplings
\cite{ro:ann,brasil}. This automatically fixed
they way in which new variable evolve with the cutoff,
and it also fixes their value at all cutoffs
if one insists that the new counterterms vanish when
the canonical couplings are turned of \cite{brasil}.
Remarkably, in the examples studied in
Ref. \cite{ro:npb,ro:ann},
this procedure provided the precise values for the
non-canonical terms that were required to restore Lorentz
covariance for physical observables.
For an explicit example for CCC, the reader is referred to
Ref. \cite{brasil}.

Even without assuming CCC, one can employ renormalization
group techniques \cite{rengroup} to help determine
the counterterm functions: using the powerful tool
of try and error, one makes an ansatz for these
functions, which one can improve by repeatedly applying
renormalization group transformations to the
effective LF Hamiltonians with these functions included.
The fact that the QCD Hamiltonian should be an ultraviolet
stable fixed point under these transformations can be
exploited to improve the original ansatz for the
counterterm functions \cite{all:lftd}.
Probably, such transformations alone are not sufficient to
completely determine the renormalized LF Hamiltonian
for QCD, but one can improve this approach considerably
by using perturbation theory, CCC
and perhaps phenomenology to guide the ansatz functions
used in the renormalization group approach to LF Hamiltonians.

Another promising idea in the context of LFTD is
the {\it similarity transformation} \cite{gl:wi}.
Whenever one derives an effective Hamiltonian by
eliminating states above a certain cutoff perturbatively,
one faces small energy denominators, and thus
large and uncertain corrections, for states that
close to (and below) the cutoff.
This feature makes it very difficult to
repeatedly apply renormalization group
transformations because matrix elements of
states near the cutoff are large.
To resolve this problem, Glazek and Wilson have
suggested to apply a cutoff to
{\it energy differences} instead of to {\it
single particle energies}. By construction,
this resolves the problem of small energy denominators,
but it also provides a band diagonal Hamiltonian.
The {\it similarity transformation}
exploits this type of cutoff and thus provides
a way to apply renormalization group techniques
to LF Hamiltonians (and other many body problems)
\cite{gl:wi}.

\chapter{Summary, Conclusions and Outlook}
LF field theory is a very promising approach toward calculating
correlation functions along a light-like direction.
Such correlation functions appear in the theoretical analysis
of a variety of hard scattering processes, such as
deep inelastic lepton-hadron scattering and asymptotic form factors.
Probably the most intriguing and controversial property of
LF Hamiltonians is the triviality of the ground state. Recent
developments indicate that LF Hamiltonians must be regarded as
effective Hamiltonians in the sense that some of the interactions
acquire nonperturbative renormalizations with coefficients
proportional to vacuum condensates. So far one understands the
LF vacuum and is able to construct the effective LF-Hamiltonian
only in a few toy models. However, in these
examples only a finite number of condensates are necessary to
completely specify the Hamiltonian.
It would be extremely useful if one could
construct and approximately solve such an
effective LF Hamiltonian for QCD,
not only for the analysis of hard processes,
but also for our understanding of low energy
QCD: due to the triviality of the LF vacuum,
a constituent picture makes sense and an effective
LF Hamiltonian for QCD would offer the opportunity
for deriving a constituent picture as an approximation
to the QCD bound state problem.

Three main stream directions can be distinguished in the
endeavor toward constructing a LF Hamiltonian for QCD: First,
a {\it fundamental} approach where all zero-modes are
included as dynamical degrees of freedom. Second, an {\it effective}
approach where one attempts to absorb all zero-modes and associated
vacuum effects into {\it effective} interactions and coupling constants.
Third, the {\it LF Tamm Dancoff}
approach, where not only vacuum effects
but also effects from high energy and high Fock components
are ``integrated out'' and absorbed into effective interaction
terms.
In the {\it fundamental} approach
(this includes all formulations of LF field theory
where explicit zero-mode degrees of freedom are
included) the vacuum as well as the
physical particle states are complicated and one partly looses
the dynamical advantages of the LF framework.
While it seems easier to construct the Hamiltonian
than in the other two approaches, the main difficulty of the
{\it fundamental} approach lies in the fact that the
equations of motion are extremely complicated. It is not clear
whether such an approach provides any computational
advantage over a conventional Hamiltonian approach.
Nevertheless, it is very useful to pursue this
approach further in order to provide a solid theoretical basis
for other, more practical, approaches to LF field theory.
For example, studies that include zero-modes can be useful for
deriving an ansatz for the effective LF Hamiltonian in the large
volume limit.

The {\it LF Tamm Dancoff} approach corresponds to the other extreme.
The vacuum is trivial and the physical particle states are
very simple --- by construction they contain only the low
energy effective degrees of freedom.
A major virtue of this approach is that it stays
close to physical intuition and thus potentially
offers a connection between the constituent picture and QCD
\cite{brasil}. While the {\it LF Tamm Dancoff}
approach is thus very appealing from the intuitive point
of view its main disadvantage is the enormous complexity of
the effective Tamm Dancoff Hamiltonian. In principle an infinite
number of counterterms are possible.
These counterterm functions are heavily
constrained by imposing Lorentz covariance on
physical observables or by demanding cancelation
of unphysical divergences. However, so far it is not clear to
what extend one can employ renormalization group techniques to
constrain the possible interactions to the point where
only a few (instead of infinitely many) free parameters
enter the LF Tamm Dancoff
Hamiltonian of QCD.
The second ({\it effective}, in the sense of zero-mode free)
approach toward constructing the LF Hamiltonian
for QCD stands in between the other two in several respects.
The vacuum is trivial but physical particles will in general
have a complicated wavefunction. Some of the interactions
in the effective LF Hamiltonian have coefficients proportional
to vacuum condensates. Those can either be regarded as
free parameters or (in some cases) they can be determined
from self-consistency conditions.
Surprisingly, in those cases where the construction of such
an effective Hamiltonian has been accomplished,
already a finite number
of condensates is sufficient to specify the Hamiltonian.
\footnote{This approach should not be confused with the standard
QCD-sum rules approach to the strong interactions \cite{qcdsum},
where one does {\it not} solve a Hamiltonian and where practically
all the dynamics is buried in the condensates. Hence it is not
surprising that less condensates are necessary as an input in the
LF effective Hamiltonian approach than in the sum rule approach.}
This is a very encouraging
result. Perturbative calculations up to two loops indicate a
similar result for QED, where the two loop calculations do
not require any counterterms which are not already present at
the one loop level. LF perturbation theory in QCD
has so far only been performed up to one loop.

Although there has been considerable progress recently, so far
none of these three approaches has been successful to the point
where it was possible to construct a useful LF Hamiltonian for QCD.
The initial optimism about LF quantization, spurred by the very
successful application to $1+1$ dimensional field theories, was
premature. Much work remains to be done before LF quantization
can be applied to QCD.

For example, it is still not completely understood to what extend
LF Hamiltonians, with a trivial vacuum, can account for the
phenomenon of spontaneous symmetry breaking. The only examples where
this subject seems to be mostly understood are $\phi^4$ theory in
$1+1$ dimensions and field theories in the mean field approximation.
It would be interesting to study cases where the order parameter
for the symmetry breaking does not enter the Hamiltonian
--- which is for example the case in the spontaneous breakdown of
chiral symmetry in QCD.

A possibly related issue, which requires further study, concerns the
{\it non-covariant counterterms}. In the context of perturbation
theory it has been shown that a finite number of such counterterms
are necessary in the bare Hamiltonian to recover full Lorentz
covariance for physical observables. However, so far it has not been
demonstrated that the proposed counterterms are sufficient to
restore Lorentz covariance for physical observables in a
nonperturbative calculation.

Within the context of LF Tamm-Dancoff it is still necessary to
demonstrate that the renormalization
group, combined with constraints from Lorentz invariance, is
sufficient to fix the infinite number of counterterms which are
possible on general grounds.

For the transverse lattice approach to be useful, it must be shown
that the fermion sign problem, which usually limits Hamiltonian Monte
Carlo calculations with fermions considerably, is tractable.
Since vacuum fluctuations are suppressed in LF quantization, any
sign problems arising from vacuum diagrams are trivially absent.
While this is a very encouraging observation, it resolves only
part of the problem --- sign problems arising from exchange
diagrams within a hadronic state are of course still there.
Another difficulty for transverse lattice calculations occurs
because gauge invariance on such a lattice requires
the introduction of $1+1$-dimensional link fields. One must learn to
work with these ``nonlinear sigma model'' degrees of freedom in the
context of LF quantization before one can
apply the transverse lattice to QCD.

Besides QCD oriented applications of the LF formalism, it may turn
out to be very useful to consider phenomenological and/or more
nuclear physics oriented applications as well. For example, it may
be interesting to reconsider the pion contribution to nuclear
structure functions
\cite{panda} from the point of view of LF quantization.
On the one hand, this could be helpful in clarifying the role of
binding effects in such calculations. On the other hand, such works
may help to demonstrate the usefulness of LF quantization to people
who are not directly involved in the field.

LF quantization is very closely related to the
{\it infinite momentum frame} formulation of field theory.
Intuitively one would thus expect that the LF formulation
of QCD offers a new theoretical approach to
relativistic heavy ion collisions. So far, this connection
has been exploited only very little \cite{raju}.

{\bf Acknowledgments}\\
I would like to thank M. Frank for many
suggestions that helped to make this
article more ``readable''. I am also very
grateful to many colleagues and
collaborators for fruitful and
enlightening discussions over the last years, particularly with F. Lenz, S.
J. Brodsky, A. Langnau, R. J. Perry, E. Swanson and P. Griffin.
\appendix

\chapter{The Dirac Bergmann Formalism}
\label{dirac}
In this appendix, a brief introduction into the
Dirac-Bergmann quantization procedure \cite{di:db,be:db,vatican}
is given. Quite generally, it replaces the canonical
quantization procedure in the presence of constraints.
However, it can also be used to derive the correct fundamental
commutation relations for theories, where the Lagrangian
contains at most linear terms in the time derivative.
\footnote{It should be noted that, in the latter case, alternate
treatments are possible as well \cite{rj:fa}.}

This is for example the case for many field theories,
when expressed in terms of LF-variables \cite{su:82}.
For example, for a noninteracting massive scalar field
in $1+1$ dimensions one obtains
\begin{equation}
{\cal L} = \partial_+\phi \partial_- \phi - \frac{m^2}{2}
\phi^2,
\label{eq:llin}
\end{equation}
which contains no terms quadratic in
$\frac{\partial}{\partial x^+}$.
Naive canonical quantization, i.e.
\begin{eqnarray}
\pi = \frac{ \delta {\cal L} }{\delta \partial_+ \phi} &=&
\partial_-\phi
\label{eq:constr}
\\
\left[ \pi(x), \phi(y) \right]_{x^+=y^+} &=& -i\delta(x^--y^-),
\label{eq:cancomm}
\end{eqnarray}
with
\begin{equation}
P_+= \int dx^- T_{+-} = \int dx^- \frac{m^2}{2} \phi^2
\end{equation}
yields
\begin{equation}
-\partial_\mu \partial^\mu \phi =-2\partial_+ \partial_- \phi =
-i\left[ P_+, \partial_-\phi \right] = 2m^2 \phi.
\label{eq:wrongeom}
\end{equation}
Clearly, Eq.(\ref{eq:wrongeom}) differs from the (correct)
Euler-Lagrange equation $-2\partial_+ \partial_- \phi =m^2 \phi$
by a factor of two \cite{ch:73}.
This mistake arises because the kinetic term in
Eq.(\ref{eq:llin}) is only linear in the time derivative
$\partial_+$. Thus the equation relating the canonical momenta
to the fields (\ref{eq:constr}) is a constraint equation
since it contains no time derivative and therefore the phase space
variables $\pi(x)$ and $\phi(x)$
for a given time are not independent.

Quantizing a system
with constraints is a nontrivial task. Fortunately, the
Dirac-Bergmann algorithm provides a step by step prescription
for the proper quantization procedure. The basic steps of this
procedure will be illustrated in an example below. To keep
the discussion simple, zero-modes will be deliberately left
out in the discussion. A complete discussion, which includes
zero-modes, can be found in Refs.\cite{hkw:89,hkw:91,hkw:92,hkw:92b}
Furthermore, the discussion here will be restricted to a
system with a finite number of degrees of freedom (which
can, for example,
be obtained from Eq.(\ref{eq:llin}) by discretizing the
$x^-$ direction)
\begin{equation}
L(\phi_i, \dot{\phi}_i) = \sum_{i,j=1}^N A_{i,j} \dot{\phi_i}\phi_j
- \sum_i V(\phi_i)
\end{equation}
with $A_{ij}=-A_{ji}$ (the symmetric part of $A_{ij}$ corresponds
to a total time derivative and can be subtracted). The
canonical momenta are given by
\begin{equation}
\pi_i = \frac{\partial {\cal L} }{\partial \dot{\phi}_i}
= \sum_j A_{ij} \phi_j
\label{eq:clconstr}
\end{equation}
with Poisson brackets
\begin{eqnarray}
\left\{ \phi_i, \pi_j \right\} &=& \delta_{ij}\nonumber\\
\left\{ \phi_i, \phi_j \right\} &=&
\left\{ \pi_i, \pi_j \right\} =0\end{eqnarray}
Eq.(\ref{eq:clconstr}) does not contain any time derivative,
i.e. it should be considered as a constraint
\begin{equation}
\chi_i \equiv \pi_i - \sum_j A_{ij}\phi_j \approx 0
\label{eq:clconstr2}
\end{equation}
$i=1,...,N$. Here $\approx 0$ means {\it weakly vanishing}, i.e.
a constraint on the physical phase space.
The {\it canonical Hamiltonian} is constructed as usual
\begin{equation}
H_c = \sum_i \dot{\phi_i} \pi_i - L = \sum_i V(\phi_i).
\label{eq:hc}
\end{equation}
The constraints (\ref{eq:clconstr2}) have nonvanishing
Poisson brackets with $H_c$
\begin{equation}
\left\{\chi_i,H_c\right\} = -V'(\phi_i)
\end{equation}
as well as among themselves
\begin{equation}
\left\{\chi_i,\chi_j\right\} =A_{ij}-A_{ji} = 2A_{ij}.
\end{equation}
Thus if the time evolution would be generated by
Poisson brackets with $H_c$ the the theory would be
inconsistent because the constraints would not be
satisfied at all times.
To remedy the situation one adds Lagrangian multipliers
to $H_c$, yielding the {\it primary Hamiltonian}
\begin{equation}
H_p = H_c + \sum_i \lambda_i \chi_i
\label{eq:hp}
\end{equation}
and demands strong vanishing of the Poisson bracket of $H_p$ with
the constraints
\begin{equation}
0 = \left\{ \chi_i, H_p \right\}
=\left\{ \chi_i, H_c \right\}
+ \sum_k \lambda_k B_{ki}
\label{eq:lambda1}
\end{equation}
where $B_{ki}=2A_{ki}$.
To simplify the discussion, let us suppose that $B^{-1}$ exists.
\footnote{For the LF-Lagrangian this is actually not the case.
There is one zero eigenvalue --- the infamous zero-mode ---
which has to be treated separately. The resulting procedure
is known as the modified Dirac-Bergmann algorithm.}
Then one can satisfy Eq.(\ref{eq:lambda1}) by choosing
\begin{equation}
\lambda_k = \sum_i (B^{-1})_{ki}
\left\{ \chi_i, H_c \right\}.
\label{eq:lambda2}
\end{equation}
The primary Hamiltonian thus reads
\begin{equation}
H_p = H_c - \sum_{i,j} \chi_i (B^{-1})_{ij}
\left\{ \chi_j, H_c \right\}.
\end{equation}
Let us now introduce the Dirac brackets between $X$ and $Y$
\begin{equation}
\left\{X,Y\right\}_D =\left\{X,Y\right\}
-\sum_{i,j}\left\{X,\chi_i\right\}(B^{-1})_{ij}
\left\{\chi_j,Y\right\}.
\end{equation}
By construction one has
\begin{equation}
\dot{X}=
\left\{X,H_c\right\}_D =\left\{X,H_p\right\}.
\end{equation}
Actually, the Dirac bracket of any operator with any of
the constraints vanishes identically
\begin{eqnarray}
\left\{X,\chi_k\right\}_D &=& \left\{X,\chi_k\right\}
-\sum_{i,j}\left\{X,\chi_i\right\}(B^{-1})_{ij}
\left\{\chi_j,\chi_k\right\}\nonumber\\
&=&\left\{X,\chi_k\right\}
-\sum_{i,j}\left\{X,\chi_i\right\}(B^{-1})_{ij}B_{jk} =0,
\end{eqnarray}
i.e. in particular $\left\{\chi_i,\chi_j\right\}=0$.
In a sense, the Dirac brackets take all the phase space
restrictions from the constraint equations automatically
into account. This is in sharp contrast to the Poisson
brackets, which are calculated as if all the $\pi_i$'s
and $\phi_i$'s were independent
$\left\{X,Y\right\} = \sum_i \frac{\partial X}{\partial \phi_i}
\frac{\partial Y}{\partial \pi_i}
-\frac{\partial Y}{\partial \phi_i}
\frac{\partial X}{\partial \pi_i}$.
It thus seems natural to use Dirac brackets, instead of
Poisson brackets, when identifying classical brackets
with quantum commutators
\begin{equation}
\left\{X,Y\right\}_D \rightarrow i \left[ X,Y \right]
\label{eq:qrule}
\end{equation}
in the quantization process. A more thorough discussion
on this subject can for example be found in Ref.\cite{su:82}.
Here we are more interested in the consequences of
Eq.(\ref{eq:qrule}). For this purpose, let us evaluate the
fundamental Dirac brackets
\begin{eqnarray}
\left\{ \phi_i, \pi_j \right\}_D
&=&\left\{ \phi_i, \pi_j \right\}-
\sum_{k,l} \left\{ \phi_i, \chi_k \right\} (B^{-1})_{kl}
\left\{ \chi_l, \pi_j \right\}
\nonumber\\
&=& \delta_{i,j}- \sum_{k,l} \delta_{ik} (B^{-1})_{kl}
A_{lj} = \frac{1}{2} \delta_{i,j}.
\end{eqnarray}
Roughly speaking, the reduction of the number of independent
degrees of freedom by a factor of two
manifests itself in a factor $1/2$ in the Dirac bracket,
and after applying Eq.(\ref{eq:qrule})
the factor $1/2$ also appears in the quantum
commutator between $\pi_i$ and $\phi_j$.
For the LF-quantization of scalar fields this implies
\begin{equation}
\left[ \pi(x), \phi(y) \right]_{x^+=y^+} = -\frac{i}{2}
\delta(x^--y^-)
\label{eq:cancomm2}
\end{equation}
instead of Eq.(\ref{eq:cancomm}). Clearly, this remedies the
abovementioned (\ref{eq:wrongeom})
problem with the extra factor of 2 in the LF equation of
motion generated by $P_+$ (\ref{eq:wrongeom}).

{}From the physics point of view \cite{pa:db}, the whole
difficulty in quantization with constraints could be avoided if
it were possible to choose degrees of freedom which are
compatible with the constraints. For the above example this
is actually possible, since the constraint
(\ref{eq:clconstr2}) is linear in the fields. Let us thus make
the ansatz
\begin{equation}
\phi_i^D = \alpha \left[ \phi_i + \sum_j \beta_{ij}\pi_j \right],
\end{equation}
where $\alpha$ is a constant and the $\beta_{ij}$ are
determined by requiring a vanishing Poisson bracket
between $\phi_i^D$ and the constraints
\begin{equation}
0 = \left\{ \phi_i^D, \chi_j \right\}
=\alpha \left[ \delta_{ij} + \sum_k \beta_{ik}A_{kj} \right],
\end{equation}
i.e. $\beta_{ij} = \left(A^{-1}\right)_{ij}$. The normalization
$\alpha$ is fixed by demanding that $\phi_i^D \approx \phi_i$, i.e.
\begin{equation}
\phi^D_i = \alpha \left[ \phi_i + \sum_j\left(A^{-1}\right)_{ij}
\pi_j \right] \approx \alpha\left[ \phi_i + \sum_j\left(A^{-1}\right)_{ij}
\sum_k A_{jk} \pi_k \right] = 2 \alpha \phi_i,
\end{equation}
yielding $\alpha = \frac{1}{2}$. By construction, $\phi^D_i$ does
not ``see'' the constraint and one can apply canonical quantization
rules directly
\begin{equation}
\left[ \pi_i, \phi^D_j \right] = -i\delta_{ij}
\end{equation}
and thus
\begin{equation}
\left[ \pi_i, \phi_j \right] = -i\alpha \delta_{ij} = - \frac{i}{2}
\delta_{ij}.
\end{equation}
In the continuum limit this results in Eq.(\ref{eq:cancomm2}).

As already indicated above, there are more points that need
to be discussed before the Dirac-Bergmann procedure
for constructing the LF Hamiltonian is complete. In particular,
one has to address the issue of zero-modes. In the LF example,
the analog of the matrix $A_{ij}$ is the differential
operator $\partial_-$. Fields which are independent of $x^-$
are annihilated by $\partial_-$ and thus correspond to eigenvectors
with eigenvalue zero
(on a finite interval, with periodic boundary conditions,
these zero-modes have to be considered for a complete
formulation of the theory). In such a situation one first has to
project on the Hilbert space orthogonal to the zero-modes
before the simplified procedure above can be applied.
The resulting modified Dirac-Bergmann procedure is quite involved
and has been discussed extensively in the literature
\cite{hkw:89,hkw:91,hkw:92,hkw:92b}. The basic difficulty arises
because the constraint equation for the zero-mode is nonlinear.
For example, in $\phi^4$ theory in a ``box'' with
periodic boundary conditions in the
$x^-$-direction, integrating the Euler Lagrange equation
$-2\partial_-\partial_+\phi=m^2\phi + \frac{\lambda}{3!}\phi^3$
over $x^-$ yields \cite{hksw:92,pisa:1,pisa:2}
\begin{equation}
m^2 \int dx^- \phi +
\frac{\lambda}{3!} \int dx^- \phi^3
=0.
\label{eq:0constr}
\end{equation}
In the first term in Eq.(\ref{eq:0constr})
only the zero-mode is projected out but
in the second term higher modes contribute
as well. Because the constraint equation (\ref{eq:0constr})
is nonlinear, the resulting quantum theory is as
complicated as the formulation in usual coordinates.
So far, it is not clear whether any dynamical
simplifications (like ``freezing out'' of the zero-mode)
arise in the infinite volume limit.


\begin{thebibliography}{9}
\bibitem[Am 94]{edsel} E. A. Ammons, Phys. Rev. {\bf D50} (1994) 980.
\bibitem[AS 70]{ab:mat} M. Abramowitz and I. A. Stegun (Eds.), ``Handbook of
Mathematical Functions'', (Dover, New York, 1970).
\bibitem[BB 91]{mb:rb} M. Burkardt and R. Busch, contributed to the Lake
Louise
 Winter Institute: Particle Physics --- The Factory Era, Lake Louise,
 Canada, February 17-23, 1991, (World Scientific, Singapore, 1991), eds.:
B. A. Campbell, A. N. Kamal, P. Kitching and F. C. Khanna.
\bibitem[BD 65]{bj:rel} J. D. Bjorken and S. D. Drell: ``Relativistic
Quantum Fields'',
(Mc Graw Hill, New York, 1965).
\bibitem[Pa 93]{challenge} H.-C. Pauli,
proceedings of Leipzig Workshop on Quantum
Field Theory Theoretical Aspects of High Energy Physics, Bad Frankenhausen,
Germany, 20-24 Sep 1993.
\bibitem[BD+ 85]{ba:reg} A. Bassetto, M. Dalbesco, I. Lazzizzera and
R. Soldati, Phys. Rev. {\bf D31} (1985) 2012; A. Basetto,
G. Nardelli and R. Soldati, ``Yang-Mills Theories in Algebraic
Non-Covariant Gauges'', (World Scientific, Singapore, 1991).
\bibitem[BD+ 92]{bj:hq} J. D. Bjorken, I. Dunietz and J. Taron,
Nucl. Phys. {\bf B371} (1992) 111.
\bibitem[BD+ 94]{ba:2d}  A. Bassetto, F. De Biasio, L. Griguolo, Phys. Rev.
Lett.
{\bf 72} (1994) 3141.
\bibitem[Be 56]{be:db}
P. G. Bergmann, Helv. Phys. Acta Suppl. {\bf 4} (1956) 79.
\bibitem[Be 77]{be:qed} H. Bergknoff, Nucl. Phys. {\bf B122} (1977) 215.
\bibitem[BF+ 72]{bu:pv}
C. Buchiat, P. Fayet and N. Sourlas, Nuovo Cim. Lett. {\bf 4} (1972) 9.
\bibitem[BG 78]{ba:qcd} I. Bars and M. B. Green, Phys. Rev. {\bf D17} (1978)
537.
\bibitem[BK+ 71]{bj:71}
J. D. Bjorken, J. B. Kogut and D. E. Soper,
Phys. Rev. {\bf D3} (1971) 1382.
\bibitem[BL 80]{br:lep} S. J. Brodsky and G. P. Lepage,
Phys. Rev. {\bf D22} (1980) 2157.
\bibitem[BL 91a]{mb:al1}
M. Burkardt and A. Langnau, Phys. Rev. {\bf D44} (1991) 1187.
\bibitem[BL 91b]{mb:al2}
M. Burkardt and A. Langnau, Phys. Rev. {\bf D44} (1991) 3857.
\bibitem[BM+ 93]{world} S. J. Brodsky, G. McCartor, H.-C. Pauli, S. S.
Pinsky; Part. World {\bf 3}(1993) 109.
\bibitem[BP 76]{bardeen} W. A. Bardeen and R. B. Pearson, Phys.\ Rev.\
 {\bf D14}  (1976) 547;\\ W. A. Bardeen, R. B. Pearson and
E. Rabinovici, {\it ibid} {\bf D21} (1980) 1037.
\bibitem[BR+ 73]{brs:73}
S. Brodsky, R. Roskies and R. Suaya, Phys. Rev. {\bf D8} (1973) 4574.
\bibitem[BP 91]{schladming} S. J. Brodsky and H.-C. Pauli,
Invited lectures given at 30th Schladming
Winter School in Particle Physics:
Field Theory, Schladming, Austria, Feb 27 - Mar
8, 1991.
\bibitem[Bu 89a]{mb:phd} M. Burkardt, Doktorarbeit, Erlangen, 1989.
\bibitem[Bu 89b]{mb:deu} M. Burkardt, Nucl. Phys. {\bf A504} (1989) 762.
\bibitem[Bu 89c]{bu:dipl} R. Busch, Diplomarbeit, Erlangen 1989.
\bibitem[Bu 92a]{mb:pho}
M. Burkardt, Nucl. Phys. {\bf B373} (1992) 371.
\bibitem[Bu 92b]{mb:del}
M. Burkardt, Nucl. Phys. {\bf B373} (1992) 613.
\bibitem[Bu 92c]{mb:nag} M. Burkardt,
in Proceedings to SPIN 92 (Nagoya, Japan, Nov 92).
\bibitem[Bu 92d]{mb:hq} M. Burkardt, Phys. Rev. {\bf D46} (1992) R1924;
R2751; M. Burkardt and E. Swanson, Phys. Rev. {\bf D46} (1992) 5083.
\bibitem[Bu 93]{mb:sg} M. Burkardt, Phys. Rev. {\bf D47} (1993) 4628.
\bibitem[Bu 94a]{mb:lfepmc} M. Burkardt, Phys. Rev. {\bf D49}
(1994) 5446.
\bibitem[Bu 94b]{mb:paris} M. Burkardt,
Workshop on ``Quantum Infrared Physics'', Paris, France, 1994,
hep-ph/9409333.
\bibitem[Bu 94c]{mb:zako} M. Burkardt, proceedings to 'Theory of Hadrons and
Light-Front QCD',
Zakopane, August 1994, hep-ph/9410219.
\bibitem[Bu 95]{mb:delta} M. Burkardt, submitted to
Phys. Rev. D., hep-ph/9505226.
\bibitem[CC+ 76]{ca:qcd} C. G. Callan, N. Coote and D. J. Gross, Phys. Rev.
{\bf D13} (1976) 1649.
\bibitem[CF+ 80]{curci} G. Curci, W. Furmanski and R. Petronzio, Nucl. Phys.
{\bf B175} (1980) 27.
\bibitem[CL 84]{ch:gau} T. P. Cheng and L. F. Li: ``Gauge theory of
elementary
particle physics'', (Oxford Univ. Press, Oxford, 1984).
\bibitem[CM 69]{ma:zero}
S.-J. Chang and S. K. Ma, Phys.\ Rev.\ {\bf 180} (1969) 1506.
\bibitem[Co 76]{co:bos} S. Coleman, Ann. Phys. (N.Y.) {\bf 101} (1976) 239.
\bibitem[CR+ 73]{ch:73} S. Chang, R. Root and T. Yan, Phys. Rev. {\bf D7}
(1973) 1133.
\bibitem[CY 72]{yan:sd2}
S.-J. Chang and T.-M. Yan, Phys.\ Rev.\ {\bf D7} (1972) 1147.
\bibitem[CZ+ 95]{wz:hq} C.-Y. Cheung, W.-M. Zhang and G.-L. Lin,
hep-ph/9505232.
\bibitem[Da 50]{da:td} S. M. Dancoff, Phys. Rev. {\bf 78} (1950) 382.
\bibitem[DH+ 75]{sg:exact} R. F. Dashen, B. Hasslacher and A. Neveu,
Phys. Rev. {\bf D11} (1975) 3424; A. Zamolodchikov and A. Zamolodchikov,
Ann. Phys. (N.Y.) {\bf 120} (1979) 253.
\bibitem[Di 49]{di:49} P. A. M. Dirac, Rev. Mod. Phys. {\bf 21} (1949)
392.
\bibitem[Di 50]{di:db}
P. A. M. Dirac, Canad. J. Math. {\bf 1} (1950) 1;
`Lectures on Quantum Mechanics', (Benjamin, N.Y., 1964).
\bibitem[DP 85]{epmc1} T. A. DeGrand and J. Potvin, Phys. Rev. {\bf D31}
(1985) 871.
\bibitem[Ei 76]{ei:str} M. B. Einhorn, Phys. Rev. {\bf D14} (1976) 3451;
ibid. {\bf D15} (1977) 1649, 3037.
\bibitem[EP+ 87]{pa:qed} T. Eller, H.-C. Pauli and S. J. Brodsky, Phys. Rev.
{\bf D35} (1987) 1493.
\bibitem[EP 89]{el:qed} T. Eller and H.-C. Pauli, Z. Phys. {\bf C42} (1989)
59.
\bibitem[ES 95]{be:eng}
M. Engelhardt and B. Schreiber, Z. Phys. {\bf A 351} (1995) 71.
\bibitem[FF 65]{fu:inf} S. Fubini and G. Furlan,
Physics {\bf 229} (1965).
\bibitem[FJ 88]{rj:fa} L. Fadeev and R. Jackiw, Phys. Rev. Lett. {\bf 60}
(1988) 1962.
\bibitem[FP 83]{panda} B. L. Friman, V. R. Pandharipande and R. B. Wiringa,
Phys. Rev. Lett. {\bf 51} ( 1983) 763.
\bibitem[GH+ 93]{osu:nonloc}
S. D. Glazek et al., Phys. Rev. {\bf D47} (1993) 1599.
\bibitem[GP 92]{osu:rg}
S. D. Glazek and R. J. Perry, Phys. Rev. {\bf D45} (1992) 3740.
\bibitem[Gr 92a]{gr:sg} P. A. Griffin,
Phys.\ Rev.\  {\bf D46} (1992) 3538.
\bibitem[Gr 92b]{paul} P. A. Griffin, Mod.\ Phys.\ Lett. {\bf A7}
(1992) 601; P. A. Griffin, Nucl.\ Phys.\ B\ {\bf 372} (1992) 270.
\bibitem[Gr 93]{pauldoubl} P. A. Griffin, Phys.\ Rev.
{\bf D47} (1993) 1530.
\bibitem[Gr 94a]{pa:db} P. Griffin, private communications.
\bibitem[Gr 94b]{pg:zako} P. Griffin,
proceedings to 'Theory of Hadrons and Light-Front QCD',
Zakopane, August 1994, hep-ph/9410243.
\bibitem[GW 93]{gl:wi} S. D. Glazek and K. G. Wilson,
Phys. Rev. {\bf D48} (1993) 5863; {\it ibid.} {\bf D49} (1994) 4214.
\bibitem[Ha 82]{ha:qcd} C. J. Hamer, Nucl. Phys. {\bf B195} (1982) 503.
\bibitem[Hi 91]{lanczos} J. R. Hiller, Phys. Rev. {\bf D43} (1991) 2418;
Phys. Rev. {\bf D44} (1991) 2504;
J. J. Wivoda and J. R. Hiller, Phys. Rev. {\bf D47} (1993) 4647.
\bibitem[HK+ 89]{hkw:89}
T. Heinzl, S. Krusche, E. Werner, Z. Phys. {\bf A334} (1989) 443.
\bibitem[HK+ 91a]{hkw:92b}
T. Heinzl, S. Krusche, E. Werner, Phys. Lett. {\bf B256} (1991) 55.
\bibitem[HK+ 91b]{hkw:91}
T. Heinzl, S. Krusche, E. Werner, Phys. Lett. {\bf B272} (1991) 54.
\bibitem[HK+ 92a]{hksw:92}
T. Heinzl, S. Krusche, S. Simburger, E. Werner, Z. Phys.
{\bf C56} (1992) 415.
\bibitem[HK+ 92b]{hkw:92}
T. Heinzl, S. Krusche, E. Werner, Phys. Lett. {\bf B275} (1992) 410.
\bibitem[HL 93]{su:lat} S. Huang and W. Lin, Ann. Phys. (N.Y.) {\bf 226}
(1993) 248.
\bibitem[HN+ 88]{ne:nor} S. Huang, J. W. Negele and J. Polonyi, Nucl. Phys.
{\bf B307} (1988) 669.
\bibitem[Ho 74]{th:qcd} G.'t Hooft, Nucl. Phys. {\bf B72} (1974) 461; ibid.
{\bf B75} (1974) 461.
\bibitem[Ho 91]{kent} K. Hornbostel, in ``From Fundamental Fields
to Nuclear Phenomena'', eds. J. A. McNeil and C. E. Price,
(World Scientific, Singapore, 1991).
\bibitem[Ho 92]{ho:vac} K. Hornbostel,
Phys. Rev. {\bf D45} (1992) 3781.
\bibitem[HP+ 90]{ho:sea} K.Hornbostel, H.-C.Pauli and S.J.Brodsky, Proc. of
the
Ohio State Workshop on Relativistic Many-Body Physics, World Scientific,
1988;
\\
S.J.Brodsky, Invited talk presented at the Third Lake Louise Winter
Institute
on QCD: Theory and Experiment, Chateau Lake Louise, Alberta, Canada, March
6-12,
1988;\\
K. Hornbostel, Ph.D.Thesis, SLAC-Report-333;\\
K. Hornbostel, S. J. Brodsky and H.-C. Pauli, Phys. Rev.
{\bf D41} (1990) 3814.
\bibitem[HP 91]{hari:yuk} A. Harindranath and R. J. Perry,
Phys.\ Rev. {\bf D43} (1991) 492.
\bibitem[HR 76]{vatican}
A. Hanson, T. Regge and C. Teitelboim,
``Constrained Hamiltonian Systems'',
Academia Nazionale dei Lincei, (Rome, 1976).
\bibitem[HS+ 82]{pmc1} J. E. Hirsch, R. L. Sugar, D. J. Scalapino and
R. Blankenbecler, Phys.\ Rev.\ {\bf B26} (1982) 5033;
J. Potvin and T. A. DeGrand, Phys.\ Rev.\ {\bf D30} (1984) 1285.
\bibitem[HV 87]{hari} A. Harindranath and J. P. Vary, Phys.\ Rev.\
{\bf D36} (1987) 1141; {\bf D37} (1988) 1064; 1076; 3010;\\
W.-M. Zhang, Chin. J. Phys. {\bf 32} (1994) 717.
C. J. Benesh and J. P. Vary, Z. Phys. {\bf C49} (1991) 411.
\bibitem[HZ 93]{hari:zhang}
A. Harindranath and W.-M. Zhang,
Phys. Rev. {\bf D 48} (1993) 4868; 4881; 4903.
\bibitem[Ja 72]{rj:70} R. Jackiw, {\it Springer Tracts in Modern Physics}
{\bf 62}, ed. G. H\"ohler (Springer, 1972).
\bibitem[Ja 85]{jaffe} R. L. Jaffe, in ``Los Alamos School on
Relativistic Dynamics and Quark-Nuclear Physics'',
ed. M. B. Jackson and A. Picklesimer, (Wiley, New York,
1985).
\bibitem[Ji 93]{ji:com}
X. Ji, Comments Nucl. Part. Phys. {\bf 21} (1993) 123.
\bibitem[KP 92]{kaluza} M. Kalu\v{z}a and H.-C. Pauli, Phys. Rev. {\bf D45}
(1992) 2968;\\
M. Krautg\"artner, H.-C. Pauli and F. W\"olz, Phys. Rev. {\bf D45}
(1992) 3755; L. C. L. Hollenberg, K. Higashijima, R. C. Warner and
B. H. J. McKellar, Prog. Theor. Phys. {\bf 87} (1992) 441.
\bibitem[KP 93]{kalli}
A. C. Kalloniatis and H. C. Pauli,Z. Phys. {\bf C60} (1993) 255;
 Z.Phys. {\bf C63} (1994) 161;\\
A. C. Kalloniatis and D. G. Robertson, Phys. Rev. {\bf D50} (1994) 5262.
\bibitem[KP+ 94]{kpp94} A. C. Kalloniatis, H.-C. Pauli and S. S. Pinsky,
Phys. Rev. {\bf D50} (1994) 6633.
\bibitem[KS 70]{ks:qed}
J. B. Kogut and D. E. Soper, Phys. Rev. {\bf D1} (1970) 2901.
\bibitem[LB 93]{mb:al3} A. Langnau and M. Burkardt,
Phys. Rev. {\bf D47} (1993) 3452.
\bibitem[Li 86]{wi:vak} M. Li, Phys.Rev. {\bf D34} (1986) 3888;\\
M. Li, L. Wilets and M. C. Birse, J.Phys. {\bf G13} (1987) 915.
\bibitem[LN+ 94a]{le:qm}
F. Lenz, H. W. L. Naus, K. Ohta and M. Thies, Ann. Phys. (N.Y.) {\bf 233}
(1994) 17.
\bibitem[LN+ 94b]{le:qed}
F. Lenz, H. W. L. Naus, K. Ohta and M. Thies, Ann. Phys. (N.Y.) {\bf 233}
(1994) 51.
\bibitem[LN+ 94c]{le:qcd}
F. Lenz, H. W. L. Naus and M. Thies, Ann. Phys. (N.Y.) {\bf 233} (1994) 317.
\bibitem[LS 93]{so:bs} H.-H. Lin and D. Soper, Phys. Rev.
{\bf D48} (1993) 1841.
\bibitem[LT 89]{th:vak} F.Lenz and M.Thies, private communications.
\bibitem[LT 91]{le:ap} F. Lenz, S. Levit, M. Thies and K. Yazaki,
Ann. Phys.(N.Y.) {\bf 208} (1991) 1; F. Lenz, Proceedings of
NATO Advanced Study Institute on ``Hadrons and Hadronic Matter'',
eds. Vautherin et al. (Plenum, New York, 1990).
\bibitem[Ma 83]{ml:reg} S. Mandelstam, Nucl. Phys. {\bf B213} (1983) 149;\\
G. Leibbrandt, Phys. Rev. {\bf D29} (1984) 1699; Rev. Mod. Phys. {\bf 59}
(1987) 1067.
\bibitem[Ma 85]{manton} N. S. Manton, Ann. Phys. (N.Y.) {\bf 159} (1985)
220.
\bibitem[Ma 92]{qed:dipl1} K. Mahrt, Diplomarbeit, Erlangen, 1992.
\bibitem[Mc 88]{empty} G. McCartor, Z. Phys. {\bf C41} (1988) 271;
{\bf C52} (1991) 611; {\bf C64} (1994) 349.
\bibitem[MP+ 91]{pi:qed} D. Mustaki, S. Pinsky, J. Shigemitsu and
K. G. Wilson, Phys. Rev. {\bf D43} (1991) 3411.
\bibitem[Mu 88]{mu:lat} D. Mustaki, Phys. Rev. {\bf D38} (1988) 1260.
\bibitem[MV 94]{raju} L. McLerran and R. Venugopalan,
Phys. Rev. {\bf D49} (1994) 2233; 3352; {\it ibid.}
{\bf D50} (1994) 2225.
\bibitem[NO 87]{epmc2} J. W. Negele and H. Orland, ``Quantum Many-Particle
Systems'', (Addison Wesley, Redwood City, 1987).
\bibitem[NS+ 81]{qcdsum}
V. A. Novikov, M. A. Shifman, A. I. Vainshtein
and V. I. Zakharov,
Nucl. Phys. {\bf B191} (1981) 301.
\bibitem[Pa 93]{challange}
H.-C. Pauli, proceedings of Leipzig Workshop on Quantum
Field Theory Theoretical Aspects of High Energy Physics, Bad Frankenhausen,
Germany, 20-24 Sep 1993.
\bibitem[PB 85]{pa:dlcq} H.-C. Pauli and S. J. Brodsky, Phys. Rev. {\bf D32}
(1985) 1993, 2001.
\bibitem[Pe 94a]{ro:ann} R. J. Perry, Ann. Phys. (N.Y.) {\bf 232}
(1994) 116.
\bibitem[Pe 94b]{brasil} R. J. Perry, invited lectures presented at
'Hadrons 94', Gramado, Brasil, April, 1994.
\bibitem[PF 89]{fr:eps} E. V. Prokhvatilov and V. A. Franke,
Sov. J. Nucl. Phys. {\bf 49} (1989) 688.
\bibitem[PH+ 90]{phw:lftd}
R. J. Perry, A. Harindranath and K. G. Wilson,
Phys. Rev. Lett. {\bf 65} (1990) 2959.
\bibitem[PH 91]{lftd} R. J. Perry and
A. Harindranath, Phys.\ Rev. {\bf D43} (1991) 4051.
\bibitem[PN+ 95]{pnp} E. V. Prokhvatilov, H.W.L. Naus and
H.-J. Pirner, Phys. Rev. {\bf D51} (1995) 2933.
\bibitem[PV 93]{pisa:1}
S. S. Pinsky and B. Van de Sande, Phys. Rev. {\bf D48} (1993) 816.
\bibitem[PV 94]{pisa:2}
S. S. Pinsky and B. Van de Sande, Phys. Rev. {\bf D49} (1994) 2001.
\bibitem[PV+ 95]{pisa:3} S. S. Pinsky, B. Van de Sande and
J. R. Hiller, to appear in Phys. Rev. D
\bibitem[PW 93]{ro:npb} R. J. Perry and K. G. Wilson, Nucl. Phys.
{\bf B403} (1993) 403.
\bibitem[RM 92]{smu} D. G. Robertson and G. McCartor, Z. Phys. {\bf C53}
(1992) 661;
G. McCartor and D. G. Robertson, Z. Phys. {\bf C53} (1992) 679.
\bibitem[RM 94]{smu2} G. McCartor and D. G. Robertson, Z. Phys.{\bf C62}
(1994) 349;
hep-ph/9501107, to appear in Z. Phys. {\bf C}.
\bibitem[Ro 70]{ro:ini} F. R\"ohrlich, Acta Phys. Aust. {\bf 32} (1970) 87.
\bibitem[Ro 93]{dave:symm}
D. G. Robertson, Phys. Rev. {\bf D47} (1993) 2549.
\bibitem[St 92]{qed:dipl2} S. Stampfer, Diplomarbeit, Erlangen, 1992.
\bibitem[Su 68]{su:inf} L. Susskind, Phys. Rev.
{\bf 165} (1968) 1535.
\bibitem[Su 82]{su:82}
K. Sundermayer, `Constrained Dynamics', Lecture Notes in
Physics 169, (Springer, Berlin, 1982).
\bibitem[Ta 45]{ta:td} I. Tamm, J. Phys. (USSR) {\bf 9} (1945) 449.
\bibitem[TB 91]{tang} A. C. Tang, S. J. Brodsky and H.-C. Pauli, Phys. Rev.
{\bf D44} (1991) 1842.
\bibitem[Th 79]{thorn} C. B. Thorn, Phys. Rev. {\bf D19} (1979) 639; {\it
ibid}
{\bf D20} (1979) 1934.
\bibitem[VF 94]{fields} J. Vary and T. J. Fields,
proceedings to 'Theory of Hadrons and Light-Front QCD',
Zakopane, August 1994, hep-ph/9411263.
\bibitem[We 69]{we:69} S. Weinberg,
Phys. Rev. {\bf 150} (1966) 1313.
\bibitem[Wi 75]{rengroup} K. G. Wilson, Rev. Mod. Phys. {\bf 47}
(1975) 773.
\bibitem[Wi 91]{wi:vid} K. G. Wilson, talk at Aspen Center for Physics,
Aspen, CO, 1991\\
(a video tape of this lecture is available from S. Pinsky).
\bibitem[Wo 93]{wo:93}
P. M. Wort, Phys. Rev. {\bf D47} (1993) 608.
\bibitem[WR 94]{wi:zako} K. G. Wilson and D. G. Robertson,
proceedings to 'Theory of Hadrons and Light-Front QCD',
Zakopane, August 1994, hep-ph/9411007.
\bibitem[Wu 77]{wu:int} T. T. Wu, Phys. Lett. {\bf 61B} (1977) 142;
Phys. Rep. {\bf 49} (1979) 245.
\bibitem[WW+ 94]{all:lftd}
K. G. Wilson et al., Phys. Rev. {\bf D49} (1994) 6720.
\bibitem[Yn 83]{yn:qcd} F. J. Yndurain: ``Quantum Chromodynamics'',
(Springer, New York, 1983).
\bibitem[Zh 85]{zhit} A. R. Zhitnitsky, Phys. Lett. {\bf 165B} (1985) 405;
Sov. J. Nucl. Phys. {\bf 43} (1986) 999;
Sov. J. Nucl. Phys. {\bf 44} (1986) 139.
\end{thebibliography}
\end{document}